\documentclass[12pt,preprint]{aastex}

\usepackage{url}
\usepackage[caption=false]{subfig}
\usepackage[table]{xcolor}% http://ctan.org/pkg/xcolor
\usepackage{longtable}
\usepackage{bbm}
\usepackage{graphicx}
\usepackage{subfig}
\usepackage[showframe=false]{geometry}
\usepackage{changepage}
\usepackage{tablefootnote}
\usepackage{amsmath}
\begin{document}
\title{A Search for Ringed Exoplanets using Kepler Photometry}
\shortauthors{Heising, Marcy, and Schlichting}
\shorttitle{Search for Extrasolar Rings}
\author{Matthew Z. Heising$^1$, Geoffrey W. Marcy$^2$, and Hilke E. Schlichting$^1$}
\affil{$^1$Department of Earth, Atmostpheric, and Planetary Sciences, Massachusetts Institute of Technology, Cambridge, MA 02139}
\affil{$^2$University of California, Berkeley, CA 94720}
\email{hazmatt@mit.edu}

\begin{abstract}
Models are developed to simulate lightcurves of stars dimmed by transiting exoplanets with and without rings.  These models are then applied to \textit{Kepler} photometry to search for planetary rings in a sample of 21 exoplanets, mostly hot Jupiters, chosen to offer the best observational opportunity for discovering potential rings.  We also examine what kinds of rings might be expected for these planets, in terms of both size and orientation, based on arguments involving the host planet's equilibrium temperature, its likely obliquities, and the formation and stability of possible ring systems.  Finding no evidence for rings, for each of the 21 studied planets it is determined on an observational basis which potential rings can be rejected out of a representative set of fiducial rings, varying in both size and orientation.  For 12 of the 21 planets, we determined that Saturn-like rings could be ruled out for at least certain orientations.  Additionally, the detectability of rings is studied, and it is found that ringed planets with small obliquities (roughly $5^{\circ}-10^{\circ}$) can yield large signals, which is encouraging for future work, since such small obliquities are expected for hot Jupiters.
\end{abstract}


\section{INTRODUCTION}

Planetary rings are known to be common in our Solar System, with all four outer planets having at least one ring each (albeit some of them very tenuous).  Saturn's main rings are of course the most prominent, and were discovered the earliest, in 1610 by Galileo (who, mystified, called them ``ears" of Saturn).  More recently, the Centaur 10199 Chariklo was also determined to have a ring \citep{chariklo}, bringing the total number of astronomical bodies with confirmed rings up to 5.  However, when we turn our thoughts to outside our Solar System, we know essentially nothing about rings, and the questions of their occurence and nature remain unanswered.  There has been one claim of a discovery of an extrasolar ring-like structure \citep{mamajek} yet it is unlike any object in our Solar System; it has an outer edge extending to 0.6 AU \citep{km} and is thought not to be a stable planetary ring but rather a disk in which either planets or satellites are forming, depending on whether the disk's host is a brown dwarf or a giant planet.  In this paper, we present a search for planetary rings of a more familiar nature through detailed study of transit lightcurve data taken with the \textit{Kepler} satellite \citep{borucki}.

Rings can potentially have large surface areas compared to the size of their host planet.  Saturn's main rings, for example, extend out to more than twice Saturn's equatorial radius.  As such, certain ringed exoplanets' transit lightcurves might be significantly different from that of any possible ringless planets.  If this is the case, such a ring would be detectable through transit photometry \citep{bf}.  However, it is also possible that a ring, while blocking out a significant fraction of the stellar light, might change the lightcurve in a non-detectable manner, if the ring's orientation and size is such that the ringed planet's profile, as observed from Earth, resembles that of a spherical ringless planet.  \textit{Kepler}, a NASA mission that has taken detailed measurements of lightcurves of thousands of transiting exoplanets, currently provides the ideal opportunity to search for these possibly altered lightcurves due to the presence of planetary rings.  It should be noted that the planets \textit{Kepler} has probed are all different from the ring-bearing planets we know of; the longest period of any confirmed \textit{Kepler} planet is 705 days (\textit{Kepler}-421 b [\citealt{k14}]) with the typical \textit{Kepler} planet being closer to 10 days in orbital period.  By contrast, Jupiter has a period of 11.9 years, the shortest of all the outer planets, while Saturn's is 29.5 years.  However, stable rings around planets of shorter periods, with semimajor axes less than 1 AU and down to (and possibly below) 0.1 AU are thought to be theoretically possible \citep{warm}, although they would need to be composed of rocky material, unlike the icy rings around the outer planets.

\section{RING SIGNAL}

\subsection{Ringless Lightcurves}
\label{sec:ringless}

The signal from a hypothetical transiting ringed planet is determined by the difference between its lightcurve and the best fit ringless lightcurve \citep{bf}.  Thus, it is necessary to understand ringless transits in order to understand the ring signal.  All planets were assumed spherical for this paper.  A quadratic limb darkening model for the star was considered, characterized by two parameters, $\gamma_1$ and $\gamma_2$, such that the profile of the stellar flux is given by:
\begin{equation}
I(r)=1-\gamma_1(1-\mu)-\gamma_2(1-\mu)^2
\end{equation}
where $r$ is the normalized radial coordinate on the stellar disk such that $0\leq r \leq1$; $I(r)$ is the specific intensity, normalized so that $I(0)=1$; $\mu=\cos\,\theta=\sqrt{1-r^2}$, where $\theta$ is the angle between the observer's line of sight and the normal to the stellar surface.  For this to be a physical model, two conditions must be met: $dI(r)/dr\leq0$ for $0\leq r\leq1$ and $I(1)\geq0$ (which together imply $I(r)\geq0$ for $0\leq r\leq1$).  These conditions yield the following constraints \citep{kip}:
\begin{subequations}
\label{eq:ld}
\begin{equation}
\gamma_1+\gamma_2\leq1
\end{equation}
\begin{equation}
\gamma_1\geq0
\end{equation}
\begin{equation}
\gamma_1+2\gamma_2\geq0.
\end{equation}
\end{subequations}

Lightcurves were calculated following the small planet approximation described by \cite{ma}.  This approximation is valid for planet sizes of $\lesssim0.1$ times the star size, roughly the size limit of observed exoplanets.  While this approximation may introduce a small source of error to the calculated lightcurves, such an error should not confuse any potential ring signal, as the discrepancy between a lightcurve calculated with the small planet approximation and a lightcurve calculated without this approximation is very different in shape from a ring signal (see Section 2.3).  In particular, the ring signal is concentrated almost entirely during ingress/egress, while the small planet approximation can introduce a small error over an entire transit \citep{ma}.

These small planet lightcurves were then modified to account for the integration time of \textit{Kepler} photometry.  In this paper, we examine only the "short cadence" \textit{Kepler} data, taken with 54.18-second exposure times \citep{gilliland}.  Therefore, to calculate the observed lightcurve at a given time $t_0$ during the transit, the theoretical lightcurve was calculated at 10 different times evenly spaced throughout 60 seconds with $t_0$ at its center; the observed lightcurve was computed as the mean of these 10 values.

To fit a set of photometric data to such a lightcurve, there were 7 parameters that were allowed to float.  The first was the size of the planet, measured in units of stellar radii.  The second and third were orbital parameters: the semimajor axis (also in units of stellar radii) and the inclination of the orbit to the line of sight, which together give the position of the planet as a function of orbital phase for a circular orbit.  The fourth and fifth parameters were $\gamma_1$ and $\gamma_2$, which were allowed to float over all values subject to the constraints derived by physical limb darkening, as described by Equations (2a-2c).  In addition, there was a sixth free parameter to allow for an overall multiplicative rescaling of the lightcurve, and a seventh to allow for an overall shift of the folded lightcurve in orbital phase.  Planets were assumed to be on circular orbits and to have constant periods (no transit timing variations), characteristics that were selected for in the set of \textit{Kepler} planets studied in this paper (see Section 3.2).

\subsection{Ring Model}
\label{sec:model}

Because the sizes of the \textit{Kepler} planets are determined solely by their transit lightcurves, the signal a ring would produce is not the difference between the ringed planet's lightcurve and the lightcurve of the same planet without a ring, which is unknown; rather, it is the difference between the ringed planet's lightcurve and the lightcurve of the ringless planet that best approximates the ringed planet's transit \citep{bf}.  Thus, the detectability of a potential ring decreases with the ringed planet's capacity to masquerade as a ringless planet.  This can lead to some counter-intuitive results: rings that are very large in projected surface area may in fact yield small ring signals, appearing through transit photometry instead as a large ringless planet.  However, if such a planet's mass were known (through radial velocity measurements), the deduced mean density of the planet would be lower than its true value, due to the overestimation of the planet's size caused by the additional stellar light blocked by its ring.  In theory, as noted by \cite{z15}, the presence of such rings could therefore explain the anomalously low measured densities of several ``inflated hot Jupiters" that have been observed - see, for example, \cite{hotjup}.

For this paper, all rings are assumed to be optically thick two-dimensional circular annuli.  There are four parameters that determine the profile of such a ring.  The first two are the ring's physical size: its inner radius, $R_1$, and its outer radius, $R_2$.  The third parameter is the ring's obliquity, $\Omega$, defined as:
\begin{equation}
\Omega=\varepsilon+i-90^{\circ}
\end{equation}
where $\varepsilon$ is the ring's axial tilt with respect to the planet's orbital axis and $i$ is the planet's orbital inclination with respect to the sky plane.  An obliquity of $0^{\circ}$ implies an edge-on ring, which would therefore be unobservable, assuming negligible thickness.  The fourth parameter, referred to here as the "season", or $\phi$, represents an azimuthal angle.  For a ring with some positive obliquity that lies in its planet's equatorial plane, a season of $0^{\circ}$ degrees implies that the planet is being observed transiting during its northern hemisphere's winter solstice, while a season of $90^{\circ}$ implies that it is the planet's northern hemisphere's vernal equinox.  Generally speaking, a ring will appear in profile as an elliptical annulus, with inner and outer semimajor axes given by the inner and outer radii of the ring, respectively, and an eccentricity and rotation about the line of sight given by the combination of obliquity and season.  Figure 1 shows visualizations of ringed planets with six different seasons for a given obliquity of $20^{\circ}$, illustrating the effect that the season has on a ringed planet's projection onto the sky.  A season of $\pm90^{\circ}$ implies a ring that is viewed edge-on.

%\end{figure}
\begin{figure}
\plotone{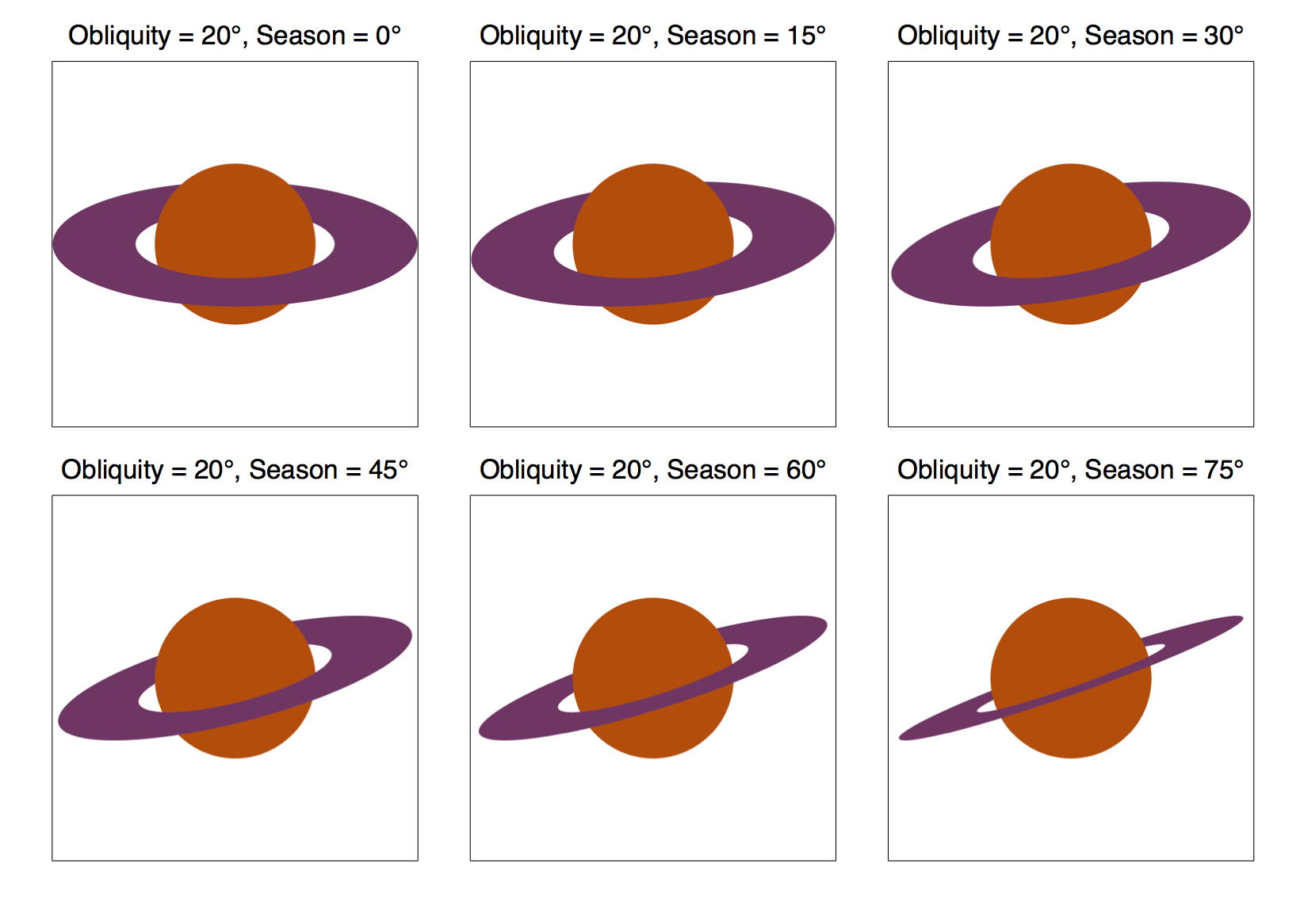}
\caption{Visualizations of six ringed planets, each with an obliquity of $20^{\circ}$, and with seasons of $0^{\circ}$, $15^{\circ}$, $30^{\circ}$, $45^{\circ}$, $60^{\circ}$, and $75^{\circ}$.  A season of $90^{\circ}$ corresponds to an edge-on ring.}
\label{seasonal}
\end{figure}

It is important to note that although we assume completely opaque rings, exoplanetary rings very well might not be optically thick.  For example, when viewed face-on, Saturn's B Ring has an optical depth of order unity.  The less opaque a potential ring might be, the smaller its transit ring signal would be, though the ring signal wouldn't change much in shape.  Rings below a certain optical depth would be undectable, depending on the level of noise in the data.  However, when a ring is viewed at an angle, its optical depth is enhanced by a factor of $\sec\,\eta$, where $\eta$ is the angle between the ring's normal axis and the line of sight (this enhancement factor holds because planetary rings, to a very good approximation, are 2-dimensional).  Therefore, a ring with a modest face-on optical depth might be optically thick when viewed close to edge-on, as is expected to be the case for most potential ringed planets (see Section 4.1).

Our assumption of optically thick rings implies that the lightcurve will not be altered by scattering effects \citep{bf}.  Thus, transit lightcurves were calculated considering only the extinction of the stellar light, integrating the limb-darkened star's flux over the area blocked by the ring and planet.  The flux blocked by just the planet was calculated as described in Section~\ref{sec:ringless}.  

A two-step algorithm was devised to numerically integrate the additional flux blocked by the ring with the necessary speed to be useful for fitting data.  The first step consisted of semi-analytically finding the area of overlap between the ring's profile and the stellar profile outside of the region defined by the planet's profile.  The only part of this step that was performed numerically was finding the points of intersection between the ring (both inner and outer edge) and the stellar disk.  These intersection points were found by approximating each of the ring's inner and outer ellipses as a 200-sided polygon with vertices defined as points on the ellipse spaced uniformly in angle from $0$ to $2\pi$, as was done by \cite{cw} to find the points of intersection between the profile of an oblate planet and a star.  The intersections of the polygon and the stellar disk were then calculated analytically and were used to approximate the intersection points between the ring and star profiles.  The intersection points between the planet and the star were calculated analytically as the intersection of two circles, and the intersection points between the ring and the planet were also calculated analytically, as made possible due to their common geometrical center.  The total area of overlap was then calculated analytically as sums and differences of circular and elliptical segments.

\begin{figure}
\plotone{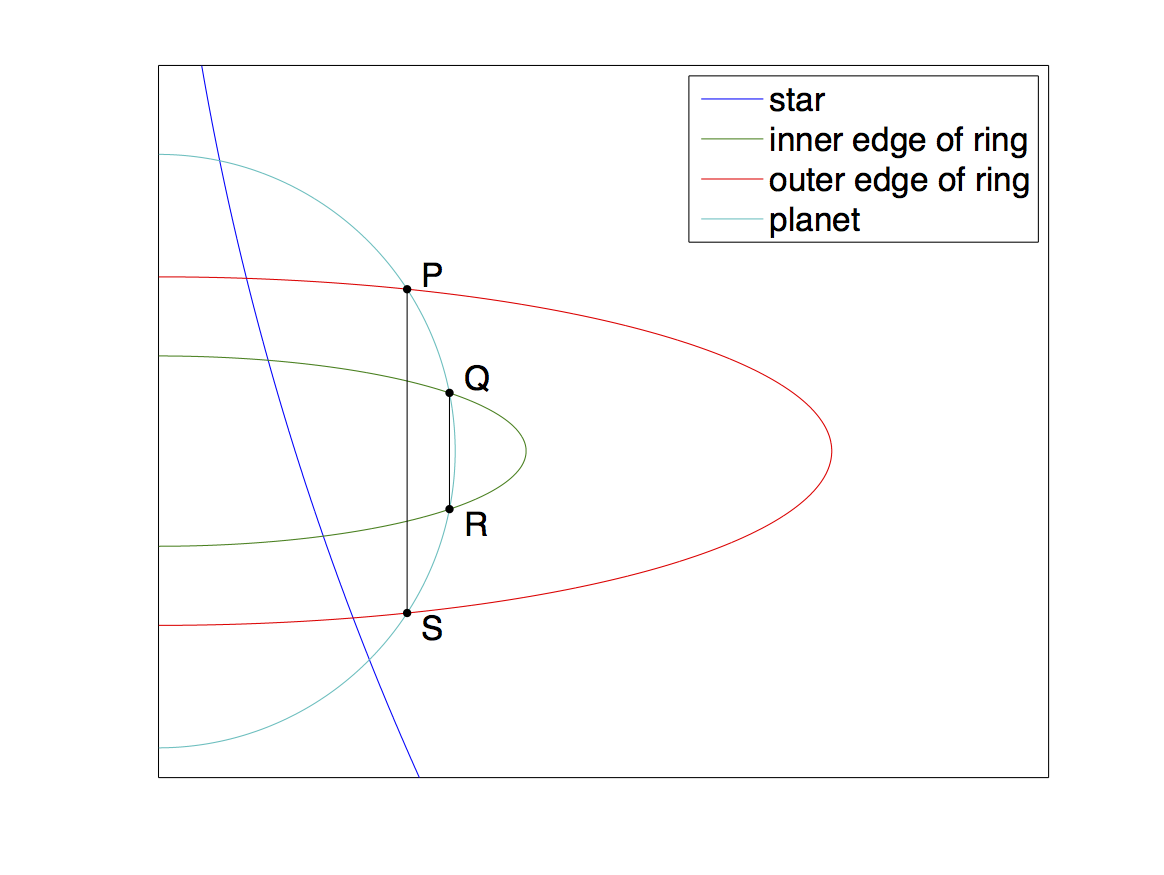}
\caption{Geometry of a ringed exoplanet partially obscuring its host star.  The intersection points between the projected ring and planet are shown.  The area of the overlap between the projected ring and star, outside of the planet, was calculated as sums and differences of circular and elliptical segments.}
\label{rog}
\end{figure}

An illustrative example of calculating such an area of overlap is provided in detail.  The setup is shown in Figure \ref{rog}: we consider a system consisting of a star and a ringed planet.  The ringed planet is a scaled version of Saturn and its main rings, and the planet's radius is 0.1 stellar radii.  It is observed with an obliquity of $15^{\circ}$ and a season of $0^{\circ}$.  We define the $x$-axis to be parallel to the projected ring's major axis, and the $y$-axis to be parallel to the projected ring's minor axis, with an origin located at the star's center, such that the planet's center is located at $x$ = -1 stellar radii and $y$ = -0.3 stellar radii.  We aim to calculate the area of projected overlap between the ring and the star, outside of the planet, and approach the problem as such: first, the projected ring's inner and outer edges are calculated as ellipses with known semimajor and semiminor axes by applying rotational matrices to account for the ringed planet's obliquity and season.  Next, the points of intersection between the projection of the ring's edges and the planet - $P$, $Q$, $R$, and $S$ - are found analytically by considering a circle and an ellipse with a common center.  We then define four areas, all of which can be calculated analytically given these four points of intersection.  We define $W$ as the area of the circular sector defined by the planet and the chord $\overline{PS}$, and $X$ as the area of the elliptical sector defined by the ring's outer edge and the chord $\overline{PS}$.  We then define $Y$ as the area of the circular sector defined by the planet and the chord $\overline{QR}$, and $Z$ as the area of the elliptical sector defined by the ring's inner edge and the chord $\overline{QR}$.  With these four areas, the problem is easily solved:
\begin{equation}
\textrm{Area of overlap}=(X-W)-(Z-Y).
\end{equation}
This example covers just one of many cases, depending on where the intersection points between the edges of the projected planet, ring, and star all lie.  Each case requires a different calculation, but the strategy is the same, and for each case, the area of overlap can be calculated purely as the sums and differences of various circular and elliptical segments.

%The second step consisted of finding the average stellar intensity over the area blocked by the ring, again excluding the region also blocked by the planet.  An exact expression for this average flux could not be derived analytically, due to the complicated region of integration.  Instead, the region of integration was approximated for this purpose as an annular sector (a region defined as an annulus cut off by two angles) of the stellar profile, or the sum of four annular sectors in the case that the rectangle that circumscribed the ring's outer ellipse contained the stellar center.  The average intensity of this simpler region was then calculated analytically by integrating the quadratically limb darkened stellar intensity.  The average intensity was then applied to the area of overlap, yielding an expression for the deficit in stellar flux due to the presence of the ring that was calculated semi-analytically.

The second step consisted of finding the average stellar intensity over the region blocked by the ring.  An exact expression for this average flux could not be derived analytically, due to the complicated region of integration.  Instead, the region of integration was approximated for this purpose as an annular sector, representing the ring's outer edge, with a "hole" in the shape of another annular sector, representing the ring's inner edge.  The dimensions of the annular sectors were chosen such that they would circumscribe the rectangle that circumscribes the ellipse that was being approximated (this leads to a region of integration that extends somewhat beyond the rings).   In the event that the rectangle that circumscribed the ellipse contained the stellar center, four annular sectors, as opposed to one, were used to approximate the ellipse.  Once this region had been defined, the average intensity was then calculated analytically by integrating the quadratically limb darkened stellar intensity.  The average intensity was then applied to the area of overlap, yielding an expression for the deficit in stellar flux due to the presence of the ring that was calculated semi-analytically.

\begin{figure}
\plotone{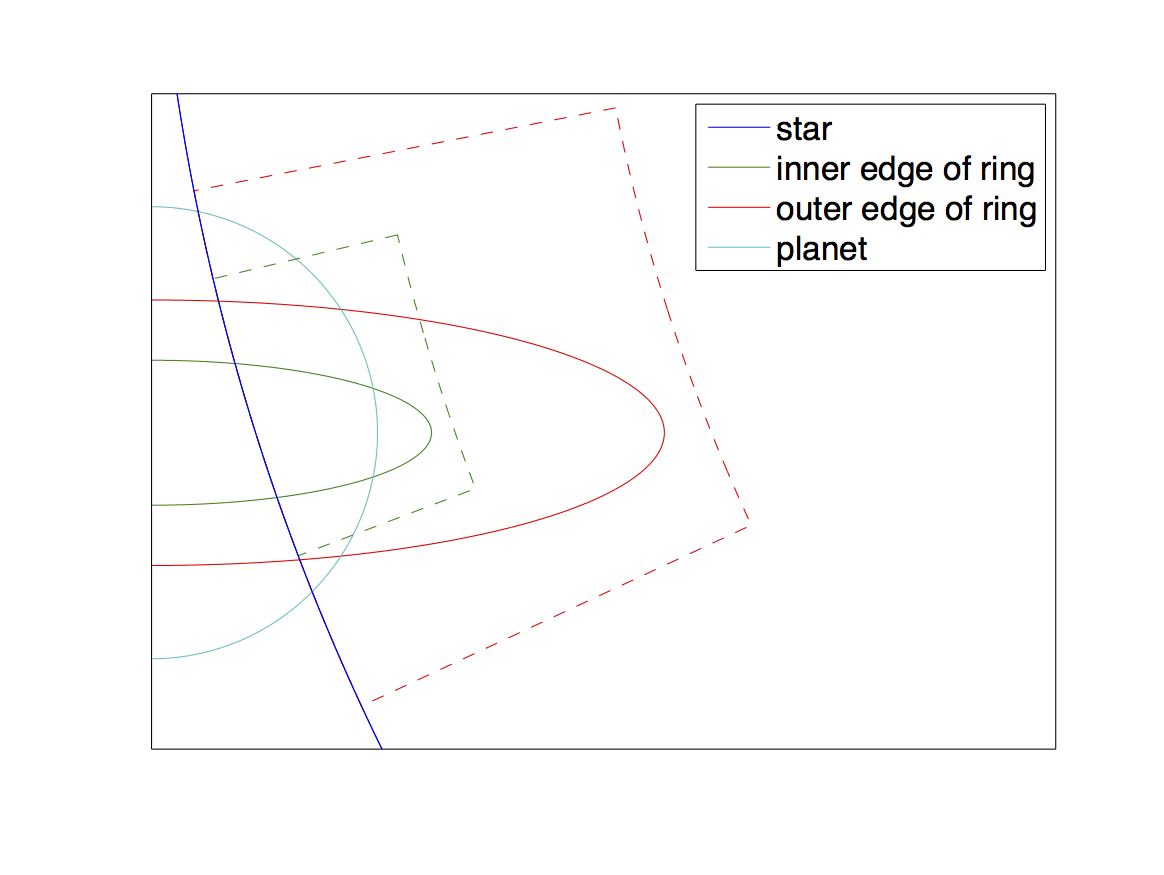}
\caption{Geometry of a ringed exoplanet partially obscuring its host star, with the same setup as in Figure 2.  The average stellar flux blocked by the ring could not be calculated analytically.  Instead, it was approximated by considering a simpler region, that defined by the dashed red curve (representing the ring's outer edge), the dashed green curve (representing the ring's inner edge), and the stellar border, over which the average stellar flux could be integrated analytically.}
\label{roi}
\end{figure}

A visualization of what this modified region of integration looked like is shown in Figure \ref{roi}, which uses the same star-planet-ring setup as described for Figure \ref{rog}.  The region of integration is the area enclosed by the dashed red curve, the dashed green curve, and the star's border; the region enclosed by the planet is not specifically excluded in this approximation, even though an exact expression would not include this region.  Note that while the approximation of the region of integration may at first look very poor, it is not the area blocked by the ring, but only the average stellar flux the ring is blocking, that is being calculated.  By comparing this semi-analytic approach to a more computationally intensive method that calculates the flux deficit via a pixel-by-pixel integration, it was determined that the semi-analytic method introduced a level of error well within the uncertainty of the Kepler data.  For example, for the geometry shown in Figures \ref{rog} and \ref{roi}, with limb darkening parameters $\gamma_1=\gamma_2=0.2$, the algorithm described above yields an error of $1.19 \times 10^{-4}$, in units of normalized stellar flux, while the smallest mean uncertainity for the transit lightcurves of the planets studied in this paper was $3.63 \times 10^{-4}$ (for \textit{Kepler}-14b), yielding an error-to-uncertainty ratio of 0.33.  The error-to-uncertainty ratio was generally smaller than this, for three reasons: the typical \textit{Kepler} lightcurve in this paper had a mean uncertainty closer to $10^{-3}$; the ringed planet in this example is large enough to yield a transit depth at the very upper limit of the observed transit depths in this paper; and the approximation is more accurate away from ingress/egress, where the stellar intensity profile is closer to constant.  Future work that requires more precise calculations could likely benefit from improving the algorithm to calculate the average stellar flux blocked by the ring.

\subsection{Ring Detectability}
%\subsubsection{Single-Transit Lightcurves}
\label{sec:detectability}

\begin{figure}
\plotone{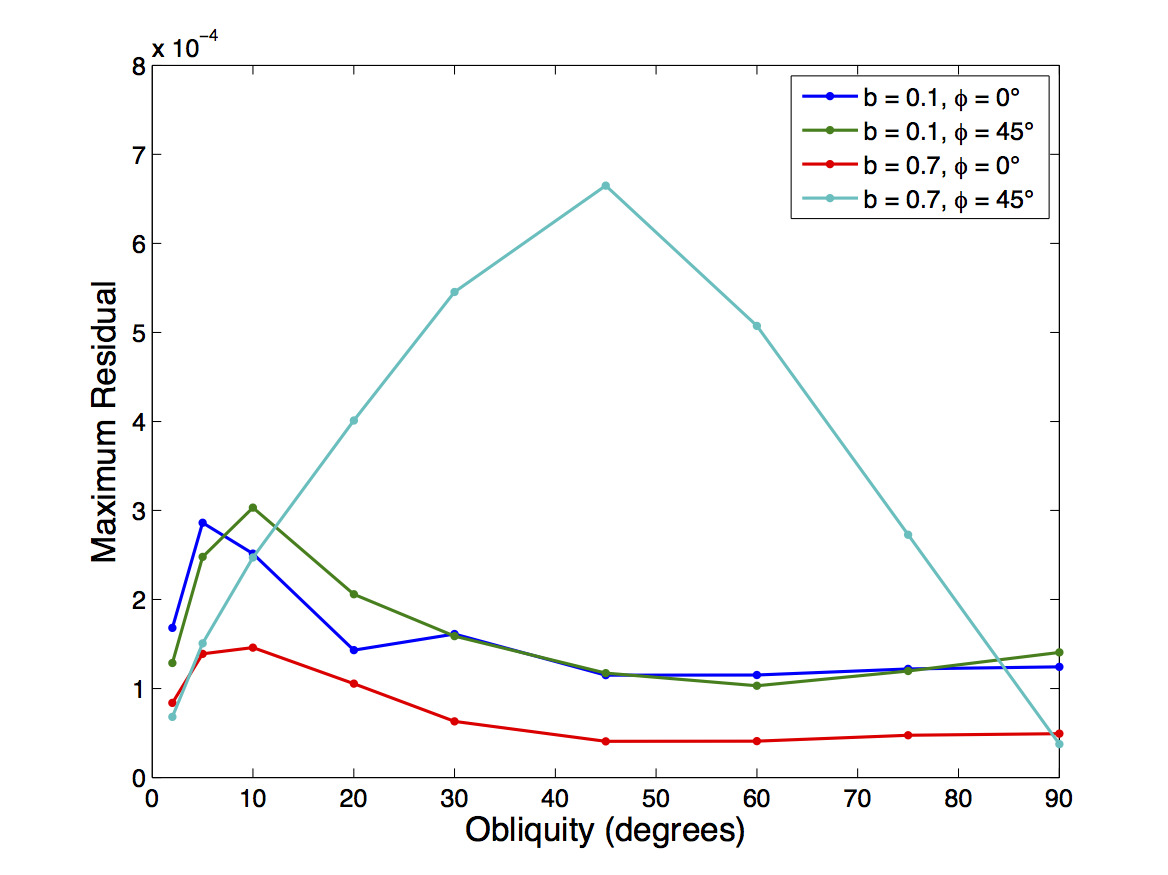}
\caption{Maximum absolute residual of a ringless fit to a simulated ringed lightcurve, in units of normalized flux, as a function of obliquity for ringed planets of varying season and impact parameter.  The ringed planet for each simulated lightcurve was a scaled Saturn, and was chosen to have a total area (the area of the ring and planet together projected onto the sky) equal to that of a spherical planet of radius 0.0964 stellar radii.}
\label{od}
\end{figure}

To characterize the amplitude of a signal left by a ring, ringed lightcurves were calculated for several different obliquities.  Each lightcurve was then fit to a ringless model, and the maximum absolute residual was calculated and plotted for each fit, the result of which can be seen in Figure \ref{od}.  Ringed lightcurves were calculated for different values of season ($0^{\circ}$ and $45^{\circ}$) as well as different values of the impact parameter, $b$ (0.1 and 0.7), defined as the minimum projected distance between the planet center and the stellar center, in units of stellar radii.  The host star was assumed to have limb darkening parameters of $\gamma_1=0.4$ and $\gamma_2=0.3$, values that were chosen to be roughly in the range of observed limb darkening parameters for stars in the \textit{Kepler} filter \citep{stagger}.  In addition, the nominal ring was assumed to have proportions given by $R_1/R_p=1.239$ and $R_2/R_p=2.269$, where $R_p$ represents the radius of the planet, $R_1$ the ring's inner radius, and $R_2$ its outer radius.  These ratios were taken from Saturn's main rings, so that the simulated ringed planets represent scaled Saturns.  The overall size of the planet and ring were rescaled for each transit so that the total area blocked by the planet and ring was equal to the area blocked by a ringless, spherical planet of radius 0.0964 (in units of stellar radii), the radius that was estimated in this paper for \textit{Kepler}-8b (see Section 3).  This rescaling was performed so that each simulated lightcurve would have approximately the same transit depth, so that the quality of the ringless fit would indicate how similar a particular ringed planet's transit lightcurve is to that of a ringless planet for a given transit depth; i.e., the amplitude of the ring signal was thus dependent only on the shape of the ringed planet's profile, and not on its size, which increases significantly with obliquity.  There were no fixed parameters when deriving the best ringless transit fit.

The most obvious feature of Figure \ref{od} is the large maximum that occurs around an obliquity of $45^{\circ}$ for an impact parameter of 0.7 and a season of $45^{\circ}$.  This is due to an asymmetric transit lightcurve, with different durations of ingress and egress, made possible by the combination of nonzero impact parameters and nonzero seasons.  The geometry can be seen in Figure 5a, which shows a ringed planet characterized by $\Omega=45^{\circ}, b=0.7,$ and $\phi=45^{\circ}$ at mid-transit, moving from left to right across the limb-darkened star.  The resulting simulated lightcurve, best ringless fit, and residual can be seen in Figure 5b.

Another prominent feature of Figure \ref{od} is the peak that occurs around an obliquity of $5^{\circ}$ for other impact parameters and seasons.  While the ring signal must go to zero as the obliquity approaches $0^{\circ}$, some of the largest signals occur for relatively small obliquities, despite the small area of the sky rings take up at such low obliquities.  This large signal is due to the fact that at small obliquities, there is a sharp division between the narrow ring profile and the circular planet profile, an effect which is smoothed out at larger obliquities.  The fact that such low obliquities can yield relatively large signals is promising for those who want to search for rings, as close-in ringed planets are expected to have low axial tilts (see Section 4.1).  For very close-in transiting planets, obliquities exceeding $5^{\circ}$ can be achieved even for rings that lie perfectly in their planets' orbital planes, due to the larger allowable range of inclinations.  The profile, lightcurve, and fit for a transit consisting of a ringed planet with $\Omega=5^{\circ}, b=0.1,$ and $\phi=0^{\circ}$ is shown in Figure 6.  The slight discontuinity in the simulated ringed lightcurve near midtransit in Figures 5b and 6b is due to the algorithm used to efficiently compute ringed transit lightcurves, as described earlier in this section; however, the algorithm is accurate to well within the level of noise in the Kepler data.  Furthermore, this discontinuity only occurs near midtransit, and the vast majority of signal from a transiting ringed planet occurs roughly during ingress and egress, as Figures 5b and 6b show.

\begin{figure}[htb]
\centering
  \subfloat[Stellar profile]{%
    \includegraphics[width=.5\textwidth]{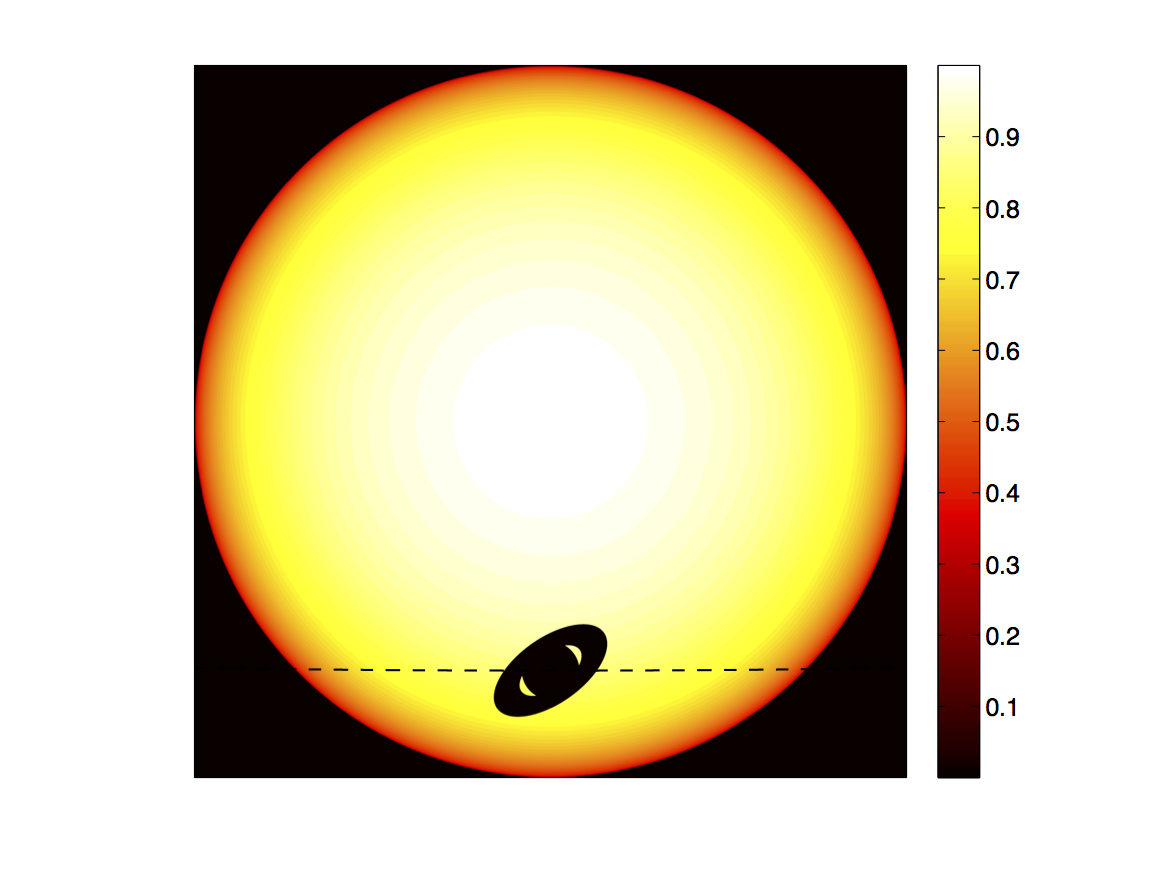}}\hfill
  \subfloat[Lightcurve and ringless fit]{%
    \includegraphics[width=.5\textwidth]{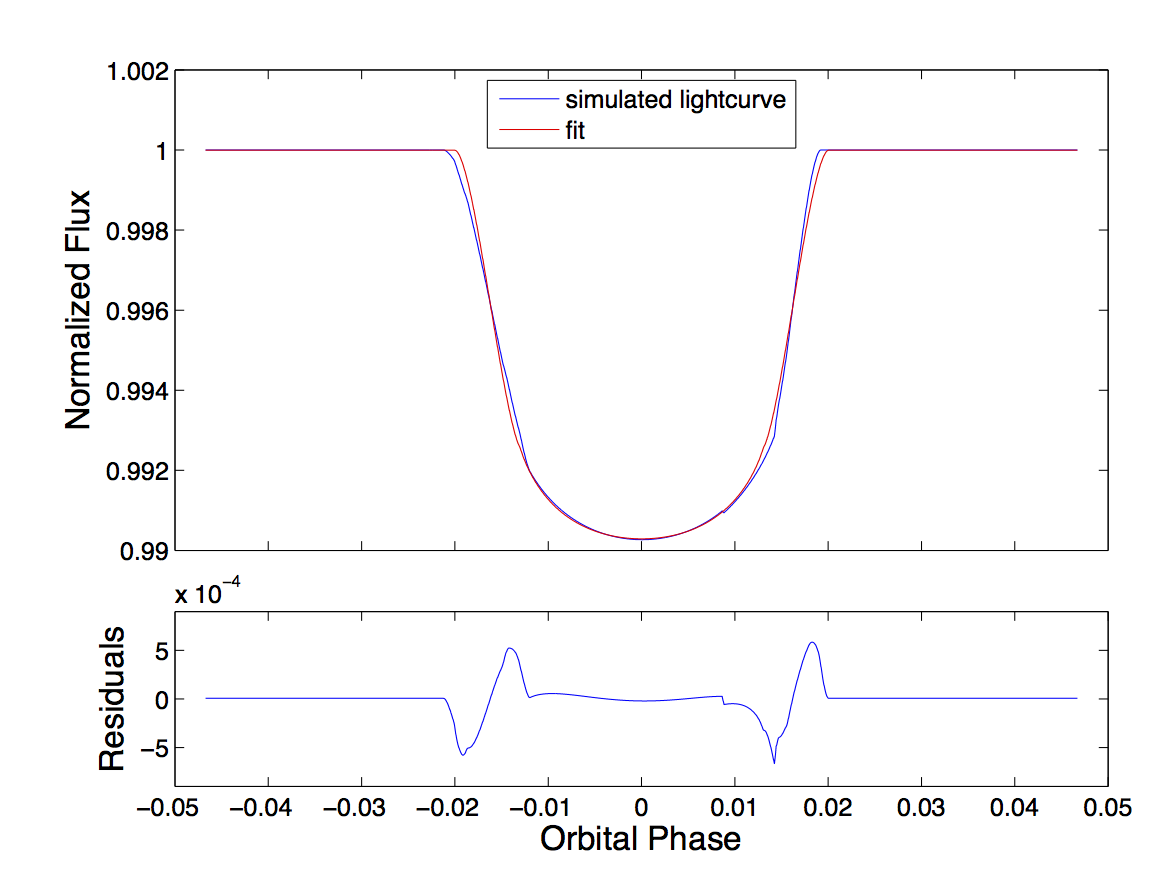}}
  \caption{Stellar profile indicating $I(\vec{r})$ on the stellar disk for a simulated ringed planet at midtransit characterized by $\Omega=45^{\circ}, b=0.7,$ and $\phi=45^{\circ}$.  The ringed planet moves along the dashed line from left to right.  At right is shown the simulated transit lightcurve and the best fit ringless lightcurve.  The residuals (data minus fit) are shown below, reaching a maximum of $6.65\times10^{-4}$ in absolute magnitude, easily detectable with Kepler short cadence photometry.}
\label{fig:visig1}
\end{figure}

\begin{figure}[htb]
\centering
  \subfloat[Stellar profile]{%
    \includegraphics[width=.5\textwidth]{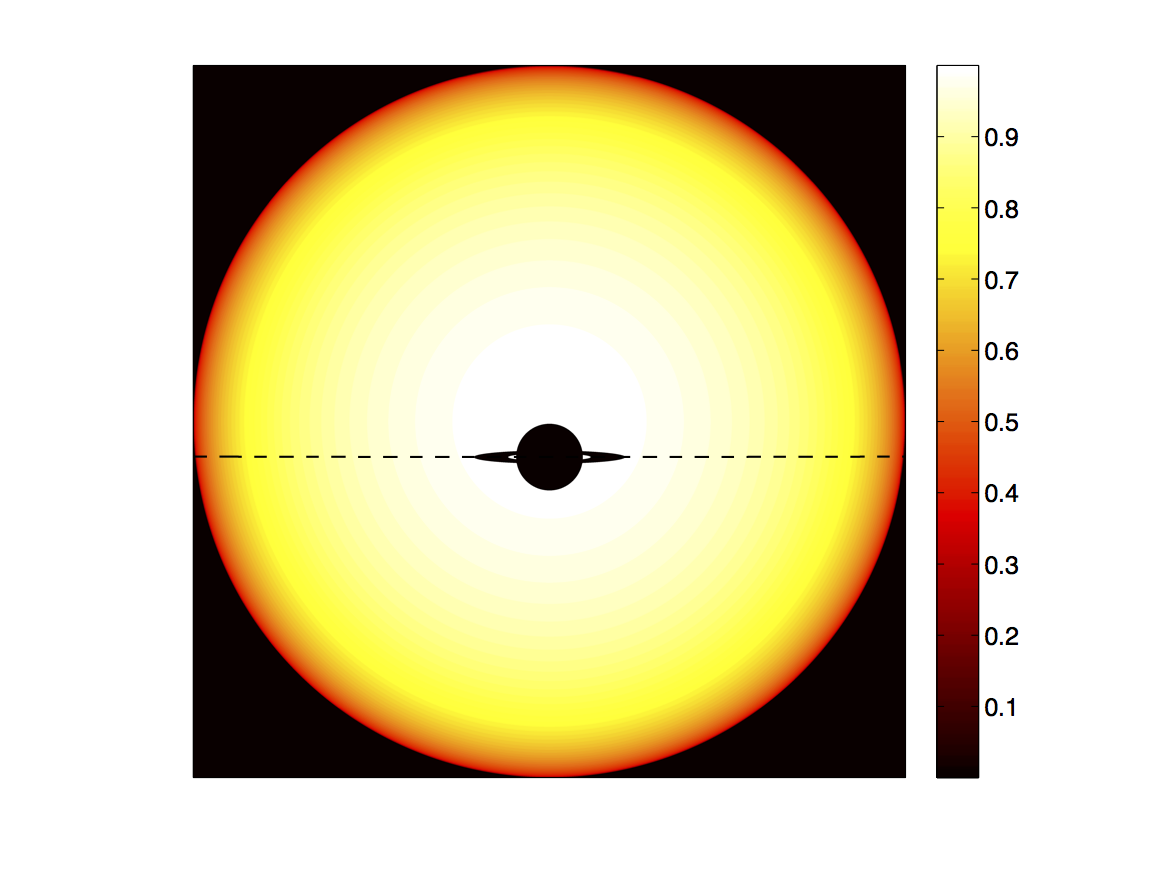}}\hfill
  \subfloat[Lightcurve and ringless fit]{%
    \includegraphics[width=.5\textwidth]{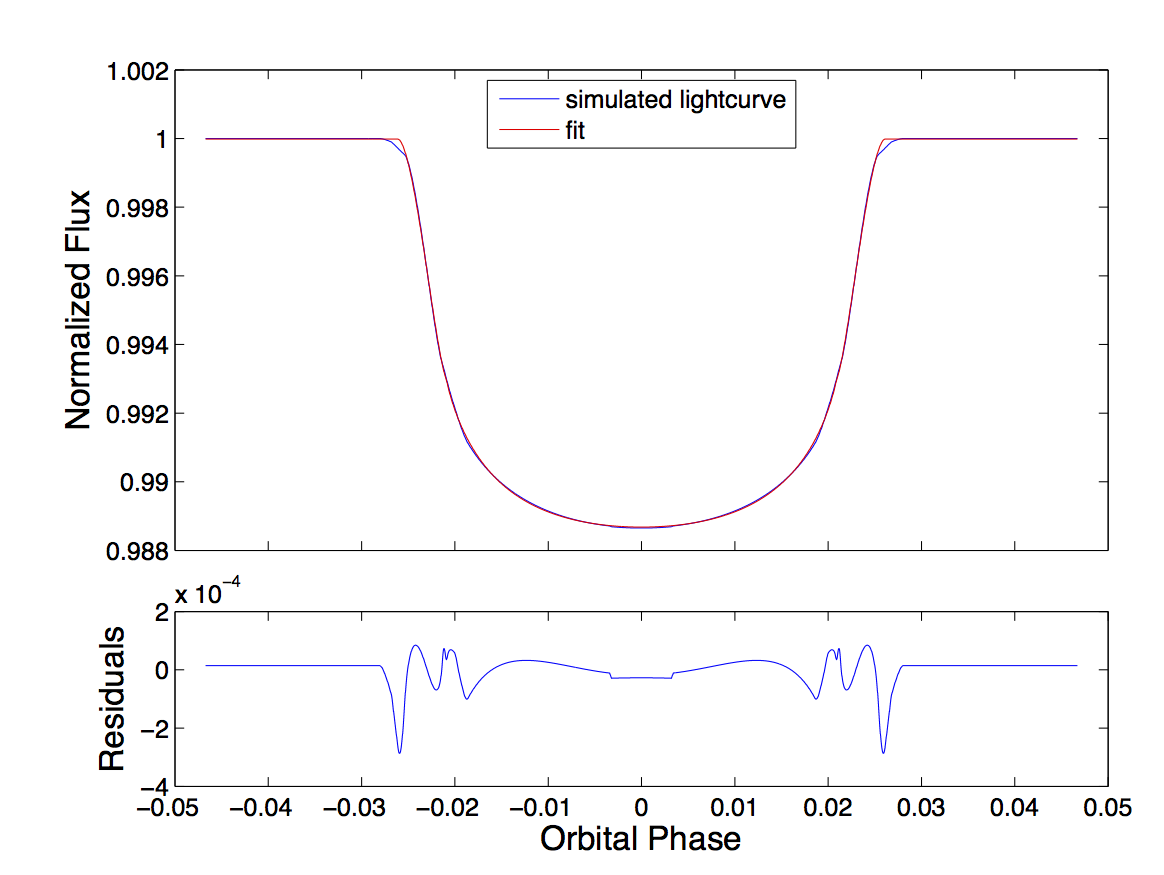}}
  \caption{Stellar profile indicating $I(\vec{r})$ on the stellar disk for a simulated ringed planet at midtransit characterized by $\Omega=5^{\circ}, b=0.1,$ and $\phi=0^{\circ}$.  The ringed planet moves along the dashed line from left to right.  At right is shown the simulated transit lightcurve and the best fit ringless lightcurve.  The residuals (data minus fit) are shown below, reaching a maximum of $2.86\times10^{-4}$ in absolute magnitude, likely detectable with \textit{Kepler} short cadence photometry.}
\label{fig:visig2}
\end{figure}

An alternate version of Figure 4 was also prepared, in which the size of the ring/planet was not rescaled to have the same total area for each ring orientation; rather, the radius of the planet was fixed at 0.05 stellar radii for every fit, and the ring was of Saturn's proportions.  The resulting plot is shown in Figure 7.  The largest difference between the two plots is that the amplitude of the ring signal at larger obliquities is enhanced in Figure 7, due to the larger areas of overlap between the star and the ringed planet, and therefore larger transit depths, at larger obliquities.  However, the largest signal is still due to asymmetric transit lightcurves, at $\Omega\approx45^{\circ}-60^{\circ}, b=0.7,$ and $\phi=45^{\circ}$.

\begin{figure}
\plotone{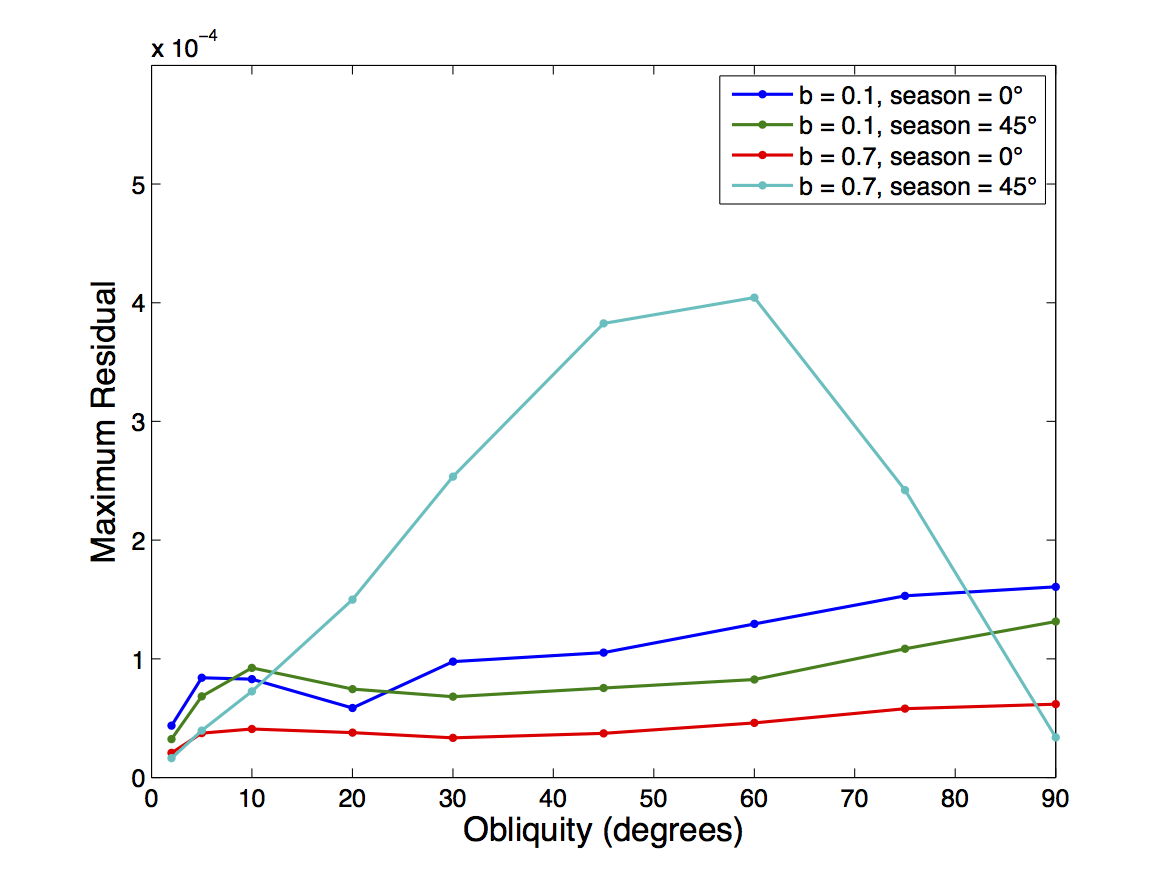}
\caption{Maximum absolute residual of a ringless fit to a simulated ringed lightcurve, in units of normalized flux, as a function of obliquity for ringed planets of varying season and impact parameter.  The ringed planet for each simulated lightcurve was a scaled Saturn with a radius of 0.05 stellar radii.}
\label{od2}
\end{figure}

It is likely that there is some degeneracy in the ring signal.  For example, the signal from a transiting ringed planet might be similar to that of an oblate planet in certain cases.  For the profile of a ringed planet to be similar to that of a realistic oblate planet, it would need to be nearly face-on, with an obliquity close to $\pm90^{\circ}$.  Furthermore, a phase-folded transit lightcurve of an exoplanet with a moon could yield a signal similar to a ring signal, provided the moon's semimajor axis is close to a typical ring radius and enough transits have been observed to sample the moon at many different phases \citep{h14}.  However, such a scenario could potentially be distinguished from a ringed exoplanet by comparing the individual (unfolded) transit lightcurves, which would vary in the case of an exomoon, but would be constant for a ring.
%
%The method used for searching for rings in this paper is examining residuals to transit lightcurve fits for effects caused by extinction of stellar light by the potential rings.  However, other methods for searching for exoplanetary rings have been proposed.

It should be noted that there are other possible ways of detecting potential rings around extrasolar planets.  These include: photometric studies that examine forward-scattering due to diffraction of stellar light caused by ring particles \citep{bf}; ensemble analyses that probe for anomalously large transit depths and/or anomalous estimations of lightcurve-derived stellar denisities {\citep{z15}; and direct imaging techniques, from which the presence of rings can be argued from photometry that is anomalously bright at certain wavelengths \citep{k08}.

\subsubsection{Axial precession}

Over multiple transits, the signature of a ringed planet is complicated by the fact that the spin axis of the planet is expected to precess about the orbit normal due to the planet's rotationally induced oblateness (for how a close-in planet might attain a spin axis significantly offset from its orbit normal, see Section 4.1).  This oblateness manifests itself as an equatorial bulge around the planet, on which its host star applies a graviational torque, driving the precession.  The effect of this precession for a planet with $i=90^{\circ}$ is a gradual change of a planet's season from transit to transit.  For a circular orbit, the period of this precession, $P_{prec}$, is given by
\begin{equation}
P_{prec}=\frac{2}{3}\frac{P_{orb}^2}{P_{rot}}\frac{\mathbbm{C}}{J_2}\frac{1}{\textrm{cos}(\varepsilon)}
\end{equation}
where $P_{orb}$ is the planet's orbital peiod, $P_{rot}$ is its rotational period, $\mathbbm{C}$ is its dimensionless moment of inertia\footnote{defined such that the planet's angular momentum about its spin axis is given by $\mathbbm{C}M_pR_{eq}^2$, where $M_P$ is the planet's mass and $R_{eq}$ is its equatorial radius; a spherical planet of uniform density will have $\mathbbm{C}=0.4$, while more centrally concentrated planets will have smaller values for $\mathbbm{C}$.}, and $J_2$ is its quadrupole gravitational moment \citep{g59, cw}.

While the rotational period of any exoplanet has yet to be measured, a planet will synchronize its spin to its rotational period to its orbital period on a timescale $\tau$ given by
\begin{equation}
%\tau\equiv\left\lvert\frac{\omega}{d\omega/dt}\right\rvert\approx\frac{4}{9}Q\mathbbm{C}\left(\frac{R_{eq}^3}{GM_p}\right)\omega\left(\frac{M_p}{M_*}\right)^2\left(\frac{a}{R_{eq}}\right)^6
\tau\equiv\left\lvert\frac{\omega}{d\omega/dt}\right\rvert\approx\frac{4}{9}Q\mathbbm{C}\left(\frac{M_p}{M_*}\right)\left(\frac{a}{R_{eq}}\right)^3\left(\frac{\omega}{n}\right)n^{-1}
\end{equation}
where $\omega=2\pi/P_{rot}$ is the planet's rotational angular frequency, $n=2\pi/P_{orb}$ is its mean motion, $Q$ is its tidal dissipation factor, $a$ is its semimajor axis, $R_{eq}$ is its equatorial radius, $M_p$ is its mass, $M_*$ is the mass of the host star, and $G$ is the gravitational constant \citep{gs,cw}.  Taking $Q=10^5$, typical for the Solar System gas giants \citep{gs}, and an initial $\omega=1.7\times10^{-4} s^{-1}$, also typical for the Solar System gas giants (which are too far-out to have rotational periods that have been affected by the Sun's tidal gravity in any specific way), one finds that hot Jupiters ($a\approx 0.05 AU$), which comprise the majority of planets studied in this paper (see Section 3.2), will synchronize their rotation on a timescale of $10^5$ years around Sun-like stars (and smaller planets will synchronize even more quickly), indicating that $P_{rot}=P_{orb}$ should be expected for close-in planets.

Assuming synchronous rotation, then, an exoplanet's $J_2$ can be roughly approximated knowing only its size and mass, from the following relations \citep{md99}:
\begin{subequations}
\begin{equation}
f=\frac{3}{2}J_2+\frac{1}{2}\frac{\omega^2R_{eq}^3}{GM_p}
\end{equation}
\begin{equation}
\frac{J_2}{f}\approx -\frac{3}{10}+\frac{5}{2}\mathbbm{C}-\frac{15}{8}\mathbbm{C}^2
\end{equation}
\end{subequations}
where $f$ is the planet's oblateness, defined as $f=\frac{R_{eq}-R_{pol}}{R_{eq}}$, $R_{pol}$ being the planet's polar radius \citep{md99}.  The first equation is derived by considering that the surface of a rotating planet must lie on an equipotential, and the second relation is the Darwin-Radau approximation.  The Darwin-Radau relation holds precisely only for a solid body with uniform density.  However, it predicts $\mathbbm{C}$ for each of the Solar System giant planets to within a few percent \citep{bf03}, and so we adopt it for our purposes here.

The quantity which is the most poorly constrained is the planet's dimensionless moment of inertia, $\mathbbm{C}$.  Jupiter and Saturn have $\mathbbm{C}=0.26401$ and $\mathbbm{C}=0.22037$, respectively \citep{hm89}.  Adopting $\mathbbm{C}=0.23$, we find that the spin axis of a Jupiter-sized body on a 3.5-day synchronously rotating orbit around a Sun-like star will precess with a period of $(9.9 \textrm{yr})/\textrm{cos}(\varepsilon)$.  Photometric data from planets studied in this paper cover up to 3 years, which is long enough compared to 9.9 years that precession may significantly change a ringed planet's lightcurve over this time, and phase-folding all of the transits could potentially hide some of the signal.  However, it was found that under the Darwin-Radau approximation, $P_{prec}$ was highly dependent on $\mathbbm{C}$; taking $\mathbbm{C}=0.15$, for example, leads to a precession period of 45 years, long enough to make the photometric effects of such precession negligible.  To put such a value for $\mathbbm{C}$ in context, \citet{bodenheimer01} calculate $\mathbbm{C}$ ranging from $0.133$ to $0.266$ for various models of hot Jupiters.  Note that while planets with lower axial tilts precess more quickly, the amplitude of their precession is less; a planet with $\varepsilon=0$, of course, will not precess at all.

To determine how precession might affect a phase-folded transit lightcurve, we simulated transits of a scaled Saturn at 6 different seasons, equally spaced from $-75^{\circ}$ to $75^{\circ}$ (the ring orientations with seasons from $-90^{\circ}$ to $90^{\circ}$ are the same as those with seasons from $90^{\circ}$ to $270^{\circ}$).  The planet was chosen to have $R_p=0.05$, $b=0.7$, and $\Omega=45^{\circ}$, while the host star was chosen to have  $\gamma_1=0.4$ and $\gamma_2=0.3$.  An average lightcurve was then compiled out of the 6 lightcurves, and was fit with a ringless model.  The season-averaged lightcurve, its fit, and the residuals can be seen in Figure 8a.  For comparison, a lightcurve with the season fixed at $45^{\circ}$ is shown in Figure 8b, also with a ringless fit and residuals.  This season was chosen to rouhgly maximize the ring signal, due to the effect of an asymmetric lightcurve.  As Figure 8 demonstrates, the largest residual from the season-averaged lightcurve is roughly 4 times smaller than the largest residual from the lightcurve with season = $45^{\circ}$.  However, the easiest way to detect a transiting precessing ringed planet would likely be through transit depth variations.  For a modest obliquity of $5^{\circ}$, a ringed planet's total projected area will change by a factor of 1.09 as its season changes from $0^{\circ}$ to $90^{\circ}$, and larger obliquities will yield larger variations in transit depth.

The above analysis assumes that the ring always lies flat in the planet's equatorial plane.  Because of the planet's equatorial bulge and the frequent collisions of ring particles, this is a safe assumption for all slowly precessing planets.  However, for a more quickly rotating planet, the ring might not be able to adapt quickly enough to stay in the planet's constantly changing equatorial plane.  The mechanism for exchanging angular momentum between the planet and its ring is the nodal precession of the ring particles caused by the planet's equatorial bulge.  This angular momentum exchange is what allows for an initially offset ring to align itself with the planet's equator.  The period of nodal precession, $P_{nodal}$, for a ring particle on a circular orbit about its host planet is approximated as
\begin{equation}
P_{nodal}=\frac{2}{3}P_r\left(\frac{a_r}{R_p}\right)^2\frac{1}{J_2}\frac{1}{\textrm{cos}(i_r)}
\end{equation}
where $P_{r}$ is the orbital period of the ring particle, $a_r$ is its semimajor axis, and $i_r$ is its inclination \citep{boc75}.

We define $\lambda$ as the ratio of the period of nodal precession of a ring particle to the period of spin precession of the planet, and find (assuming a synchronously rotating planet)
%\begin{equation}
\begin{multline}
\lambda\equiv P_{nodal}/P_{prec} = \left(\frac{a_r}{R_p}\right)^2\frac{P_r}{P_{orb}}\frac{1}{\mathbbm{C}}\frac{\textrm{cos}(\varepsilon)}{\textrm{cos}(i_r)}\\
=\left(\frac{a_r/R_p}{1.72}\right)^{7/2}\left(\frac{R_p}{R_J}\right)^{3/2}\left(\frac{M_p}{M_J}\right)^{-1/2}\left(\frac{P_{orb}}{3.5\textrm{ days}}\right)^{-1}\left(\frac{\mathbbm{C}}{0.23}\right)^{-1}\frac{\textrm{cos}(\varepsilon)}{\textrm{cos}(i_r)}
\end{multline}
%\end{equation}
where $M_J$ is Jupiter's mass and $R_J$ is Jupiter's mean radius.  For $\varepsilon\approx i_r$, then, which is a reasonable assumption for a ring particle that is offset from its host planet's equatorial plane due to the planet's axial precession, ring particles further out than roughly 1.73 planetary radii from a hot Jupiter will yield $\lambda>1$, and will therefore have a longer nodal precession period than the axial precession period of the planet.  In the limit of $\lambda\gg1$, a ring particle will only feel an equatorial bulge from the planet that has been averaged over many axial precessional periods.  This averaging out will cause an excess of mass that lies in the planet's orbital plane, not its instantaneous equatorial plane.  Therefore, we argue that ring particles existing beyond roughly 1.73 planetary radii may lie in the planet's orbital plane, while closer-in ring particles will lie in the planet's equatorial plane, potentially causing a ``kinked" ring.  Furthermore, the high frequency of collisions between ring particles, which causes individual ring particles to quickly lose any out-of-plane momentum component they may have, implies that all rings we are considering should be very flat throughout: \citet{dpl01} argue that for rings with optical depth of order unity or greater, the flattening time is a few orbits of the ring particles.  Since a typical ring orbit around a giant planet is of the order of a few hours, the flattening timescale is much shorter than the planet's precession period, and the ring will stay flat.

However, the analysis of the ring's shape so far has ignored the direct effects of the stellar tide on the ring particles.  The competing effects between the planet's equatorial bulge and the host star's tides define the Laplace plane, the "plane" (actually curved surface) in which a satellite orbiting the host planet will not precess, and thus the plane in which a ring is expected to lie.  At short distances from the planet, the planetary bulge dominates, and the Laplace plane coincides with the planet's equatorial plane, while at larger distances from the planet, the host star's tides dominate, and the Laplace plane coincides with the planet's orbital plane.  The transition between these two regimes is continuous, but happens roughly at the Laplace radius, $R_L$, which for a circular planetary orbit is given by
\begin{equation}
\frac{R_L}{R_p}=\left(J_2\left(\frac{a}{R_p}\right)^3\left(\frac{M_p}{M_*}\right)\right)^{1/5}=1.45\left(\frac{J_2}{0.005}\right)^{1/5}\left(\frac{a/0.05 \textrm{ AU}}{R_p/R_J}\right)^{3/5}\left(\frac{M_p/M_*}{0.001}\right)^{1/5}
\end{equation}
\citep{l1805,ttn09}.  The shape of the ring, then, can be described as thus: at a certain distance from the planet, the ring will transition from lying in the planet's equatorial plane to lying in its orbital plane.  This transition will occur roughly at either the Laplace radius or the radius at which the nodal precession of the ring particles is slower than the axial precession of the planet, whichever is smaller.

Because transit depth variations were not observed in the planets studied, and because the precession periods for these planets could not be calculated from known quantities with much accuracy, the rest of this paper assumes zero axial precession over the time period of observations.  Furthermore, because a planet's $J_2$ cannot be estimated with great certainty, rings are assumed to lie solely in their host planets' equatorial planes.  It should be noted that the planets studied in this paper are generally expected to have their spin axes and orbital axes aligned (see Section 4.1), in which case both of the above assumptions are automatically met.

\begin{figure}[htb]
\centering
  \subfloat[Season-averaged]{%
    \includegraphics[width=.5\textwidth]{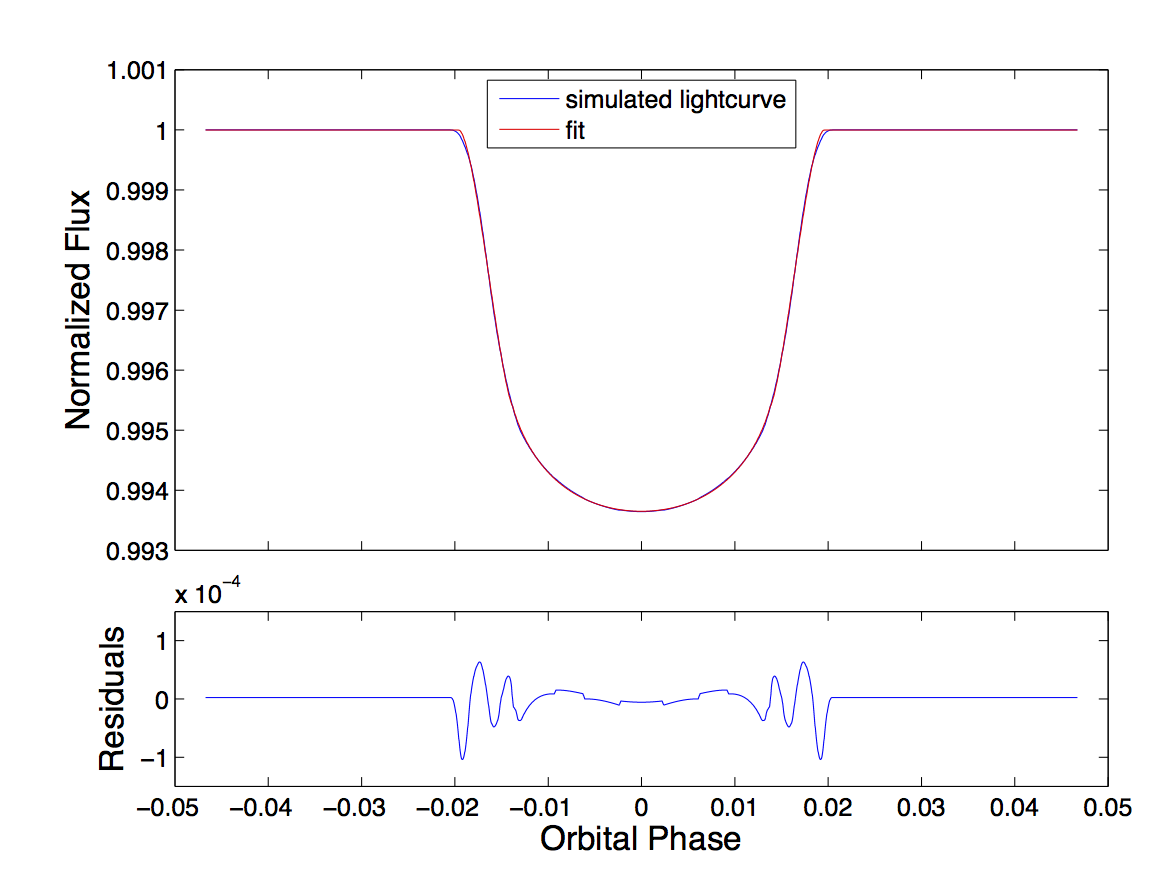}}\hfill
  \subfloat[Season = $45^{\circ}$]{%
    \includegraphics[width=.5\textwidth]{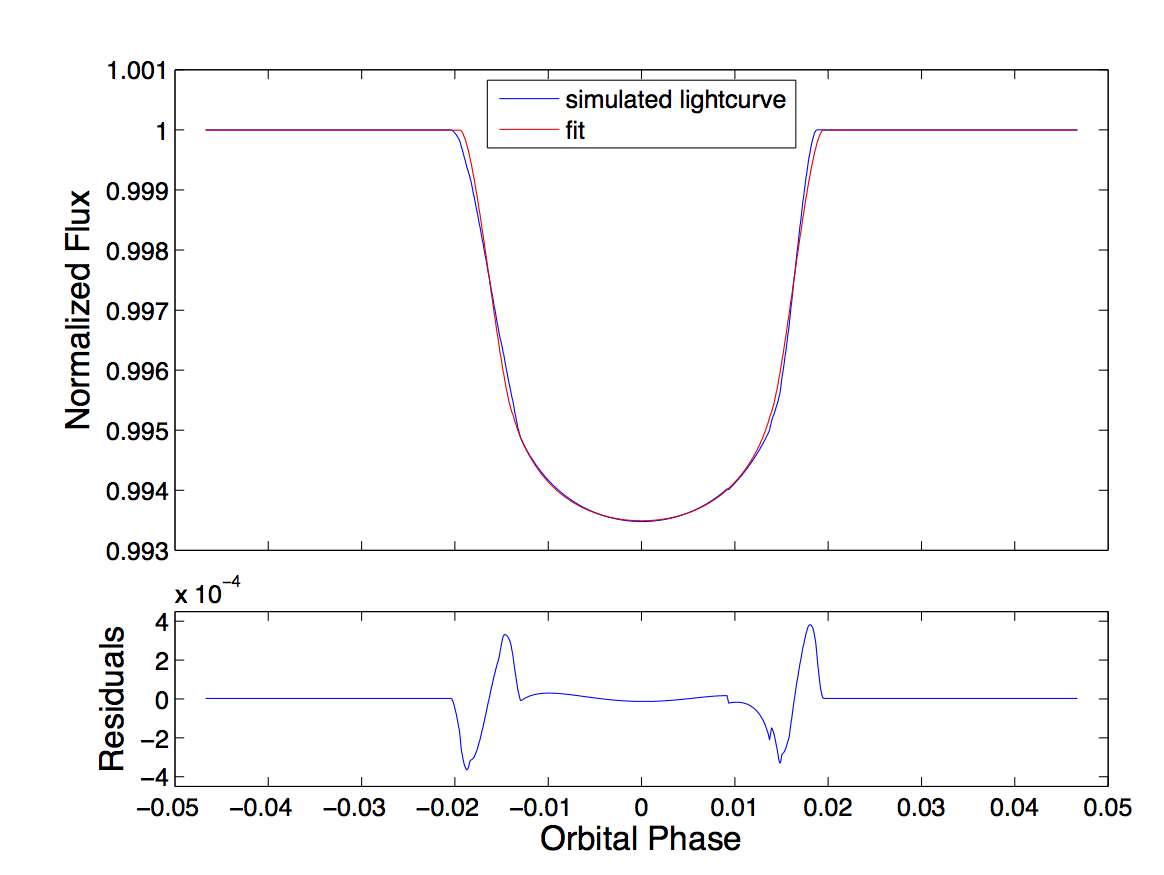}}
  \caption{Comparison of season-averaged lightcurve (a) to one with season = $45^{\circ}$ (b).  Both lightcurves were created by simulating the transit of a scaled Saturn with $R_p=0.05$, $b=0.7$, and $\Omega=45^{\circ}$ across a star with $\gamma_1=0.4$ and $\gamma_2=0.3$.  A ringless fit and the resulting residuals are also shown for each simulated lightcurve.  Note the difference in scale for the residuals.  The maximum residual for the fixed-season lightcurve is approximately 4 times larger than the maximum residual for the season-averaged lightcurve, indicating that precession of a ringed planet can remove the largest ring signals, caused by asymmetric lightcurves.}
\label{fig:prec}
\end{figure}

\section{THE SEARCH FOR RINGS}

\subsection{Treatment of Data}
\label{sec:data}

The existence of rings around exoplanets was tested exclusively using \textit{Kepler}'s short cadence data, photometric measurements collected with 54.18-second exposure times, publicly available from the Mikulski Archive for Space Telescopes (MAST)\footnote{\url{http://archive.stsci.edu/missions/kepler/lightcurves/}}.  In particular, the Pre-search Data Conditioning (PDC) lightcurves were used.  These lightcurves have been generated by processing the raw photometry through the \textit{Kepler} data analysis pipeline, removing effects caused by sources of systematic error during observation, such as pointing drift and changes in focus \citep{jc}.  Each file of a given lightcurve was independently normalized so that the average out-of-transit flux received was equal to 1.  The out-of-transit flux was selected for by excluding all photometric data below 1/5 of the transit depth subtracted from the average flux.  Of the 21 planets examined in this paper, two of them (\textit{Kepler}-17b and \textit{Kepler}-75b) had transit lightcurves that were not adequately detrended by the \textit{Kepler} pipeline, characterized by non-constant out-of-transit fluxes.  Their lightcurve files were therefore further detrended by computing running means, calculated as the average flux received within 4.8 hours of a given time, excluding data received during transits, and then normalizing the files by these running means.

The \textit{Kepler} satellite collected many orbits worth of short cadence data for a typical planet of interest in this paper.  These were folded into a single transit lightcurve using the planet's best fit orbital period from the cumulative list of \textit{Kepler} Objects of Interest from the NASA Exoplanet Archive.  Folded transits were cropped in orbital phase so that the primary transit occupied roughly half of the lightcurve, with the other half being ``flat."  Outlying points were then rejected on the condition that they were at least five local standard deviations removed from the local mean (``locally" being defined by breaking up the lightcurve by orbital phase into ``local" groups of 100 points each).

Uncertainties were calculated by normalizing the reported \textit{Kepler} PDC photometry uncertainties as the corresponding lightcurves were normalized.  While we believe these uncertainties adequately account for instrumental sources of error as well as photon-counting statistics, there are other sources of variation in the received flux.  This extra source of noise is termed ``stellar flicker" and accounts for variation within the emitted light due to various stellar activity phenomena \citep{flicker}.  The true uncertainty in a measurement is more accurately the flicker added in quadrature to the reported uncertainty.  For each folded lightcurve, the amount of flicker was assumed to be constant with time, and was calculated by requiring that during the flat part of the transit, the average uncertainty was equal to the standard deviation of the flux (unless the reported uncertainties already were greater than the standard deviation, in which case the flicker was assumed to be 0).  Adding stellar flicker typically increased the uncertainty by a factor of roughly 1.05 only, indicating that the reported Kepler uncertainties account for the majority of the noise.

\subsection{Ringless Fits to 21 \textit{Kepler} Planets}

A total of 21 confirmed \textit{Kepler} planets were probed for rings by determining whether their lightcurves were adequately described by transits of ringless circular planets on circular orbits; these planets are listed by name and properties in Table \ref{props}.  Orbital periods, transit depths, planetary radii, and equilibrium temperatures (calculated using a Bond albedo of 0.3) were obtained from the cumulative list of \textit{Kepler} Objects of Interest, while estimates for the planetary masses (from which mean planetary densities were calculated) were obtained from the list of Confirmed Planets, both from the NASA Exoplanet Archive.  The most up-to-date values are used for the planets' transit depths, radii, and equilibrium temperatures.  The orbital periods were taken from the cumulative list of \textit{Kepler} Objects of Interest in August 2014.  The planets were chosen based on the criteria that they have relatively large transit depths, a large amount of short cadence data, and no known orbital eccentricities.  Furthermore, for the sake of simplicity, they were chosen to not have any known transit timing variations.  The resulting list consisted mainly of hot Jupiters, as there exists much less short cadence data for longer-period planets.  Although such planets are too hot to host rings composed of water ice particles, they are all cool enough for rings composed of at least some rocky materials; SiO$_2$, for example, has a melting point of approximately 1870 K, hotter than the equilibrium temperature of each planet studied.

\begin{table}[ht]
\begin{adjustwidth}{-1.45cm}{}
\captionsetup{justification=centering}
\centering
%\caption{Properties of surveyed \textit{Kepler} planets}
\caption{Properties of surveyed \textit{Kepler} planets (from the NASA Exoplanet Archive)\label{props}}
\begin{tabular}{p{1.4cm}|p{4cm}|p{3cm}|p{2.7cm}|p{2.4cm}|p{2.0cm}}
\hline\hline
\textit{Kepler} name&Period (days)&Transit depth (ppm)&Radius (Jupiter radii)&Density (g/cm$^3$)&Equilibrium temp. (K)\\
\hline
4b&3.2136641 $\pm$ 4.2e-06&727.7 $\pm$ 2.6&0.377 $\pm$ 0.010&1.90 $\pm$ 0.33&1507\\[0.1cm]
5b&3.54846566 $\pm$ 7.3e-7&7451.9 $\pm$ 3.6&$1.39\:^{+0.11}_{-0.12}$&$1.04\:^{+0.26}_{-0.24}$&1640\\[0.1cm]
6b&3.23469955 $\pm$ 4.2e-7&10814.1 $\pm$ 4.0&$1.222\:^{+0.088}_{-0.094}$&$0.49 \pm 0.11$&1355\\[0.1cm]
7b&4.88548917 $\pm$ 8.6e-7&7569.3 $\pm$ 3.4&$1.576\:^{+0.053}_{-0.054}$&0.147 $\pm$ 0.020&1499\\[0.1cm]
8b&3.52249913 $\pm$ 7.3e-7&9379.3 $\pm$ 6.6&$1.35\:^{+0.11}_{-0.12}$&$0.32\:^{+0.11}_{-0.13}$&1521\\[0.1cm]
12b&4.43796291 $\pm$ 5.8e-7&16721.1 $\pm$ 5.2&$1.66\:^{+0.36}_{-0.29}$&$0.125\:^{+0.065}_{-0.082}$&1338\\[0.1cm]
14b&6.7901235 $\pm$ 3.6e-6&2252.3 $\pm$ 2.4&1.113 $\pm$ 0.037&$8.09 \pm 0.88$&1445\\[0.1cm]
15b&4.94278327 $\pm$ 9.3e-7&11498.1 $\pm$ 6.6&$0.98\:^{+0.16}_{-0.06}$&$0.93\:^{+0.21}_{-0.48}$&1008\\[0.1cm]
17b&1.48571127 $\pm$ 1.6e-7&21802 $\pm$ 11&$1.349\:^{+0.078}_{-0.076}$&$1.32\:^{+0.23}_{-0.24}$&1561\\[0.1cm]
41b&1.85555773 $\pm$ 2.7e-7&11037.3 $\pm$ 7.0&$0.933\:^{+0.062}_{-0.056}$&$0.90\:^{+0.22}_{-0.23}$&1436\\[0.1cm]
43b&3.02409489 $\pm$ 6.2e-7&8004.8 $\pm$ 7.3&$1.103\:^{+0.081}_{-0.089}$&$3.20\:^{+0.80}_{-0.73}$&1475\\[0.1cm]
44b&3.246732 $\pm$ 1.6e-6&7414 $\pm$ 14&$1.158\:^{+0.096}_{-0.091}$&$0.87\:^{+0.21}_{-0.22}$&1460\\[0.1cm]
75b&8.8849176 $\pm$ 4.1e-6&17670 $\pm$ 47&$1.004\:^{+0.061}_{-0.053}$&$13.0\:^{+2.1}_{-2.5}$&770\\[0.1cm]
77b&3.57878272 $\pm$ 5.4e-7&11741.5 $\pm$ 6.8&$0.95\:^{+0.12}_{-0.04}$&$0.66\:^{+0.10}_{-0.26}$&1144\\[0.1cm]
94b&2.5080585 $\pm$ 1.6e-6&1548.2 $\pm$ 7.7&$0.227\:^{+0.030}_{-0.029}$&3.9 $\pm$ 1.6&828\\[0.1cm]
101b&3.4876878 $\pm$ 4.2e-6&1340 $\pm$ 4.4&$0.397\:^{+0.092}_{-0.050}$&$3.4\:^{+1.3}_{-2.4}$&1291\\[0.1cm]
108b&49.18431 $\pm$ 3.8e-5&1645.7 $\pm$ 5.9&$0.817\:^{+0.045}_{-0.046}$&------------&706\\[0.1cm]
117c&50.790563 $\pm$ 7e-5&5890 $\pm$ 12&$0.83\:^{+0.51}_{-0.10}$&------------&580\\[0.1cm]
122c&12.466071 $\pm$ 1.4e-5&2323.6 $\pm$ 8.9&$0.42\:^{+0.15}_{-0.04}$&------------&804\\[0.1cm]
148c&4.1800659 $\pm$ 8.7e-6&1880.5 $\pm$ 15&$0.324\:^{+0.098}_{-0.024}$&------------&962\\[0.1cm]
412b&1.72086037 $\pm$ 2.5e-7&10541.3 $\pm$ 7.5&$1.14\:^{+0.73}_{-0.12}$&$0.84\:^{+0.28}_{-1.60}$&1681\\
\end{tabular}
%\label{table:props}
\end{adjustwidth}
\end{table}

Each of these planets was then fit to a ringless transit model, as described in section 2.1, allowing the planet's radius ($R_p$), semimajor axis ($a$), and orbital inclination ($i$), the star's limb darkening parameters ($\gamma_1$ and $\gamma_2$), a phase offset, and an overall normalization factor to float.  The resulting fits are shown in Figure \ref{fr}, and the resulting physical parameters from each fit are shown in Table \ref{params}.  For each fit, a $\chi^2$ statistic was calculated over the set of phases during which the planet was near or in either ingress or egress, where the majority of a potential ring's signal is expected.  Values for  $\chi^2$, normalized by the number of data points, $N$, are shown for each fit in Figure \ref{fr}.  Residuals and their uncertainties were also calculated for each fit, in non-overlapping bins, usually of 3 minutes in width, and are plotted in Figure \ref{fr}.  The bin size used for the residuals and the root-mean-square (RMS) of the residuals are shown for each fit.  Bin sizes were chosen to be as large as possible, so as to maximize the signal-to-noise ratio of any potential ring signal, while being small enough that the ring signal could still be resolved on that timescale.

A good fit to a set of data would imply a normalized $\chi^2$ close to 1.  The fits for the 21 \textit{Kepler} planets pass this test, loosely speaking - the highest $\chi^2$ statistic is 1.13 (for \textit{Kepler}-117c).  To determine exactly how far from unity $\chi^2$ can deviate before the discrepancy is significant, a more complicated statistical procedure, such as bootstrapping, is required.  However, such an analysis would yield little insight as to the nature of such a discrepancy.  While a ringed planet would cause an anomalously high $\chi^2$ value, so would a number of other possible factors, such as an error in the calculation of the uncertainties for each data point, a somewhat oblate planet, a host star that cannot be adequately described by a quadratic limb darkening model, etc.  Instead, we note that the residuals for each fit lack any trends that would be characteristic of a ring signal (as discussed in Section~\ref{sec:detectability}, and more thouroghly in \citealt{bf}), and conclude that the data are consistent with no optically thick rings being present among the 21 planets.

\begin{figure}
\begin{adjustwidth}{-1.8cm}{}
\captionsetup[subfigure]{labelformat=empty}
\centering
%\vspace{-6mm}
\subfloat[]{
  \includegraphics[width=59mm]{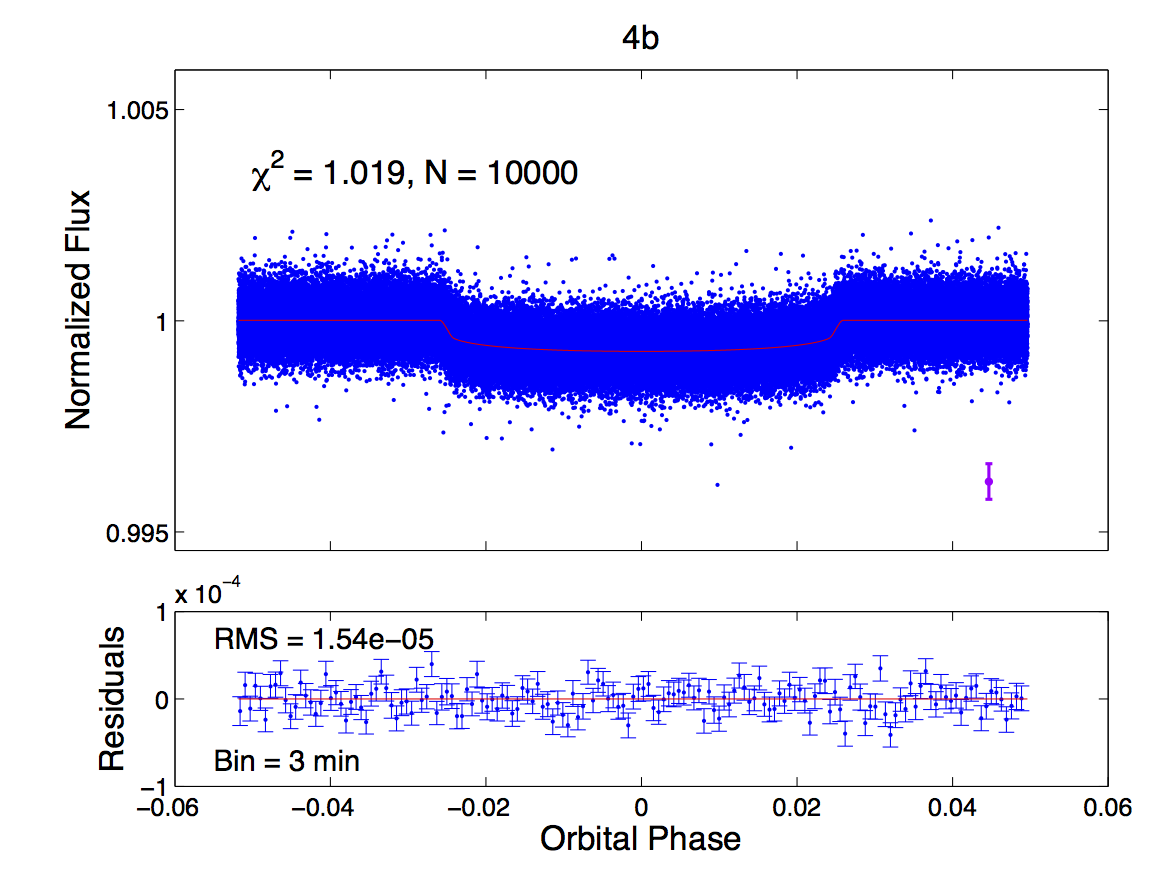}
}
\subfloat[]{
  \includegraphics[width=59mm]{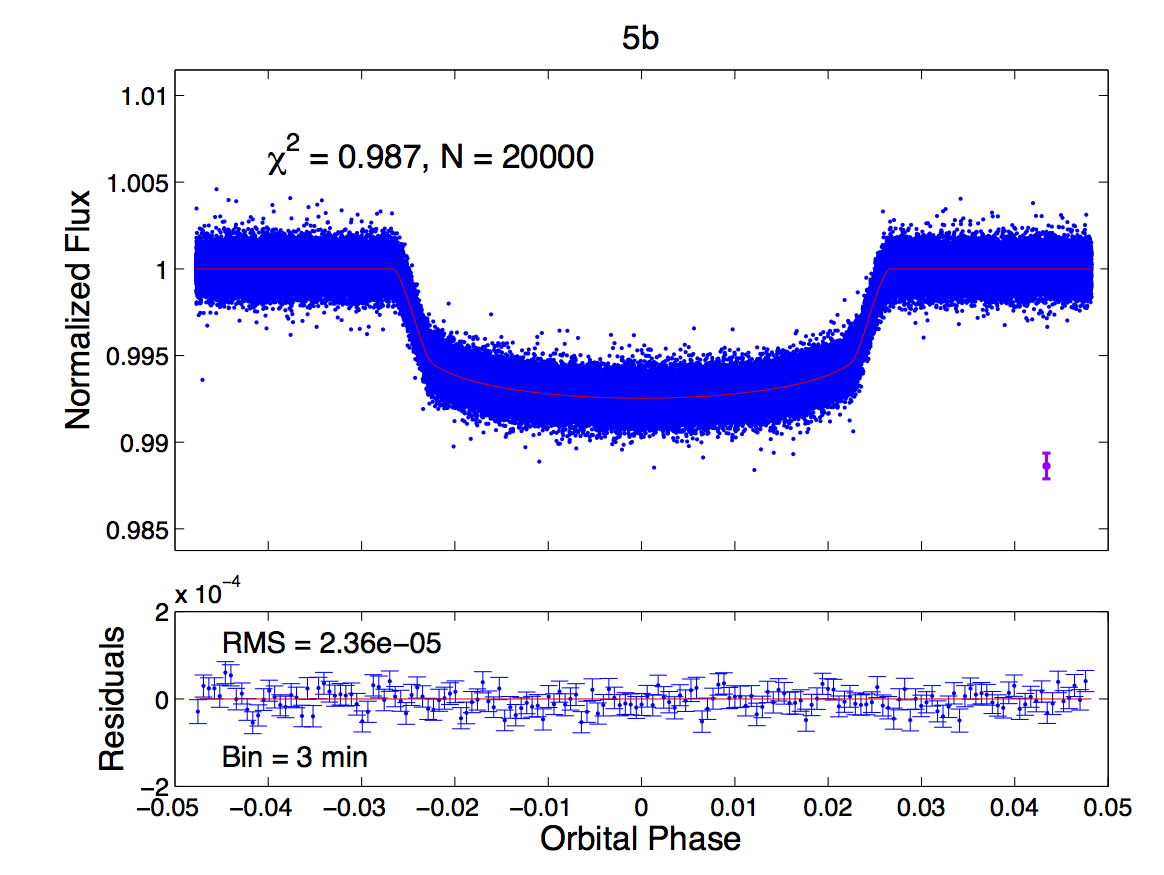}
}
\subfloat[]{
  \includegraphics[width=59mm]{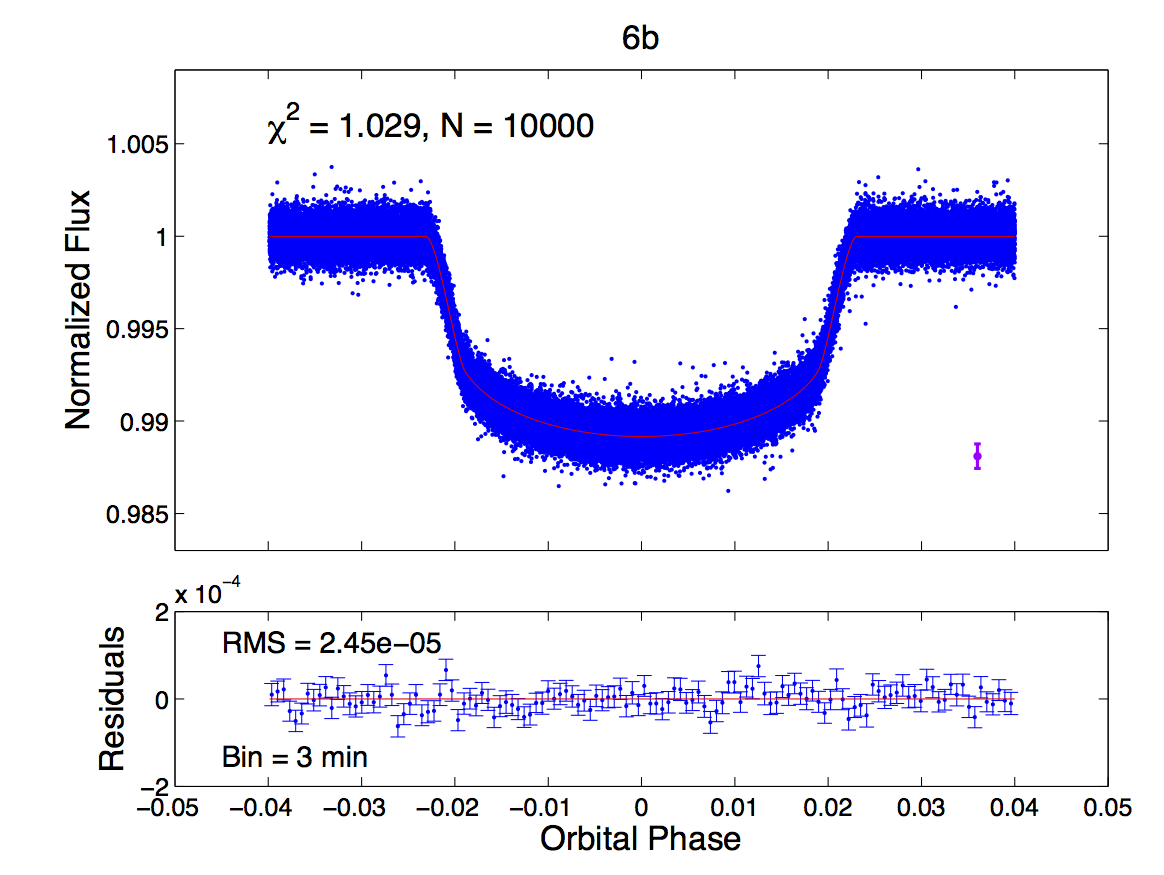}
}\\
\subfloat[]{
  \includegraphics[width=59mm]{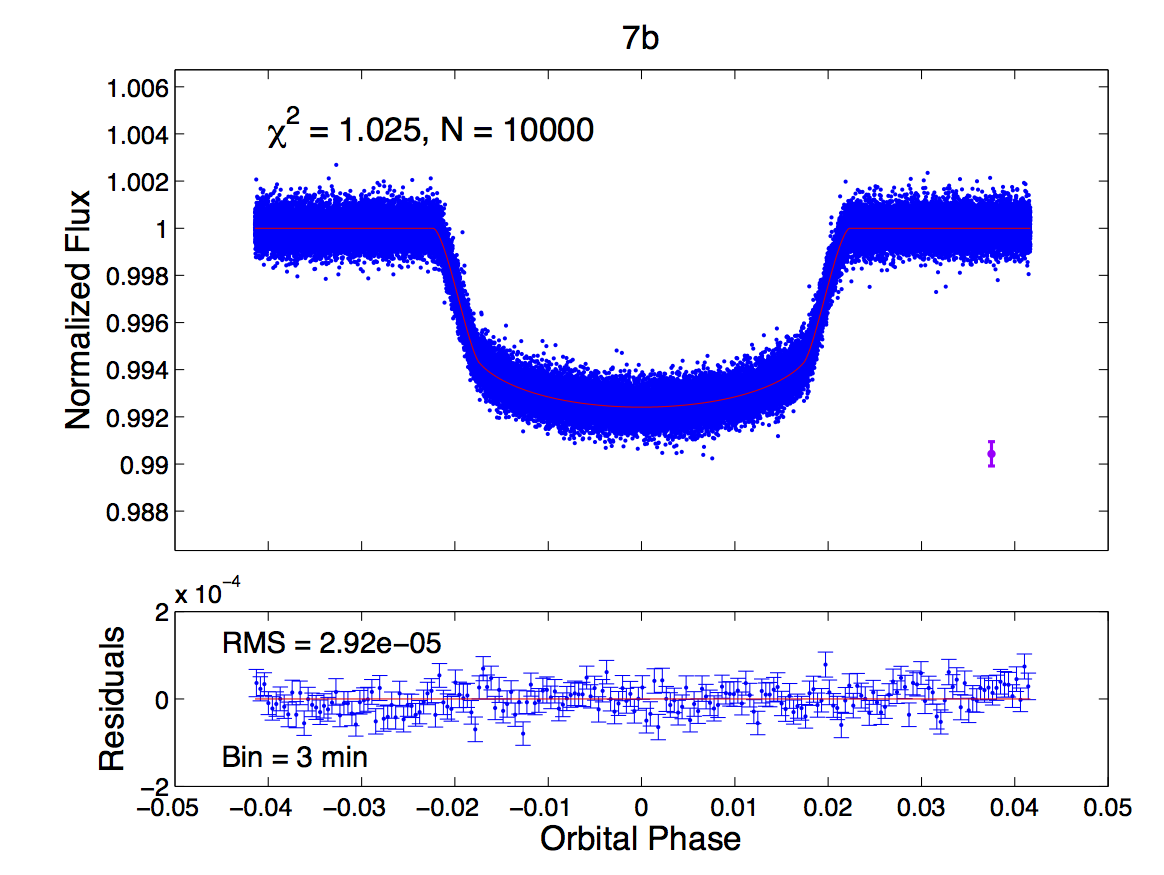}
}
\subfloat[]{
  \includegraphics[width=59mm]{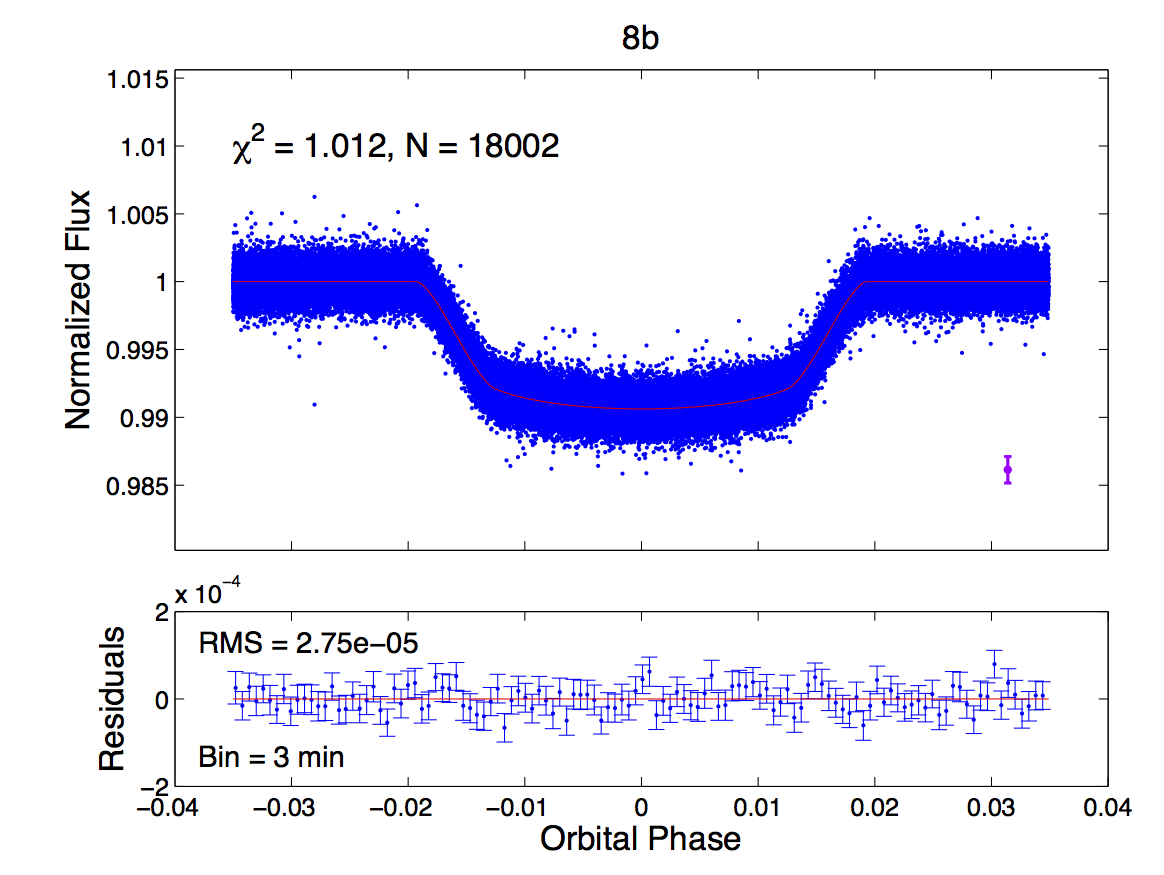}
}
\subfloat[]{
  \includegraphics[width=59mm]{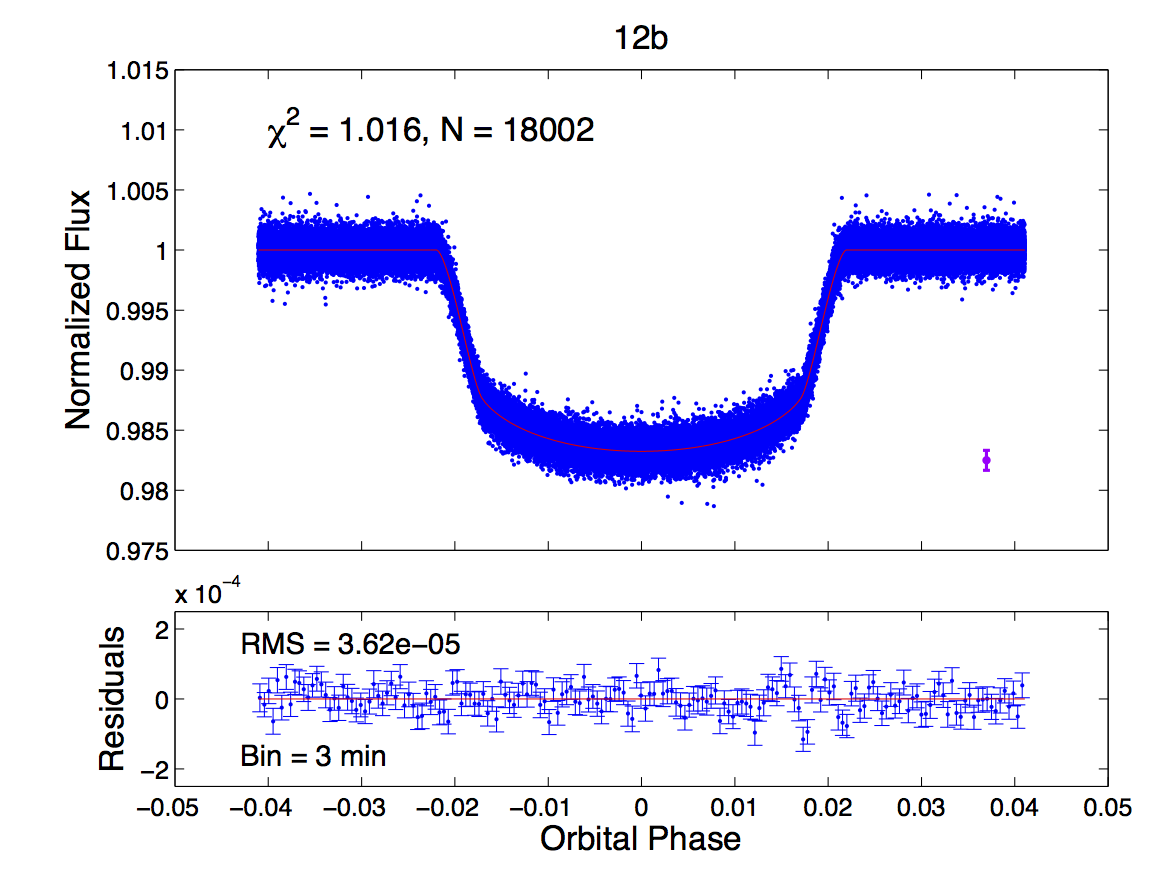}
}\\
\subfloat[]{
  \includegraphics[width=59mm]{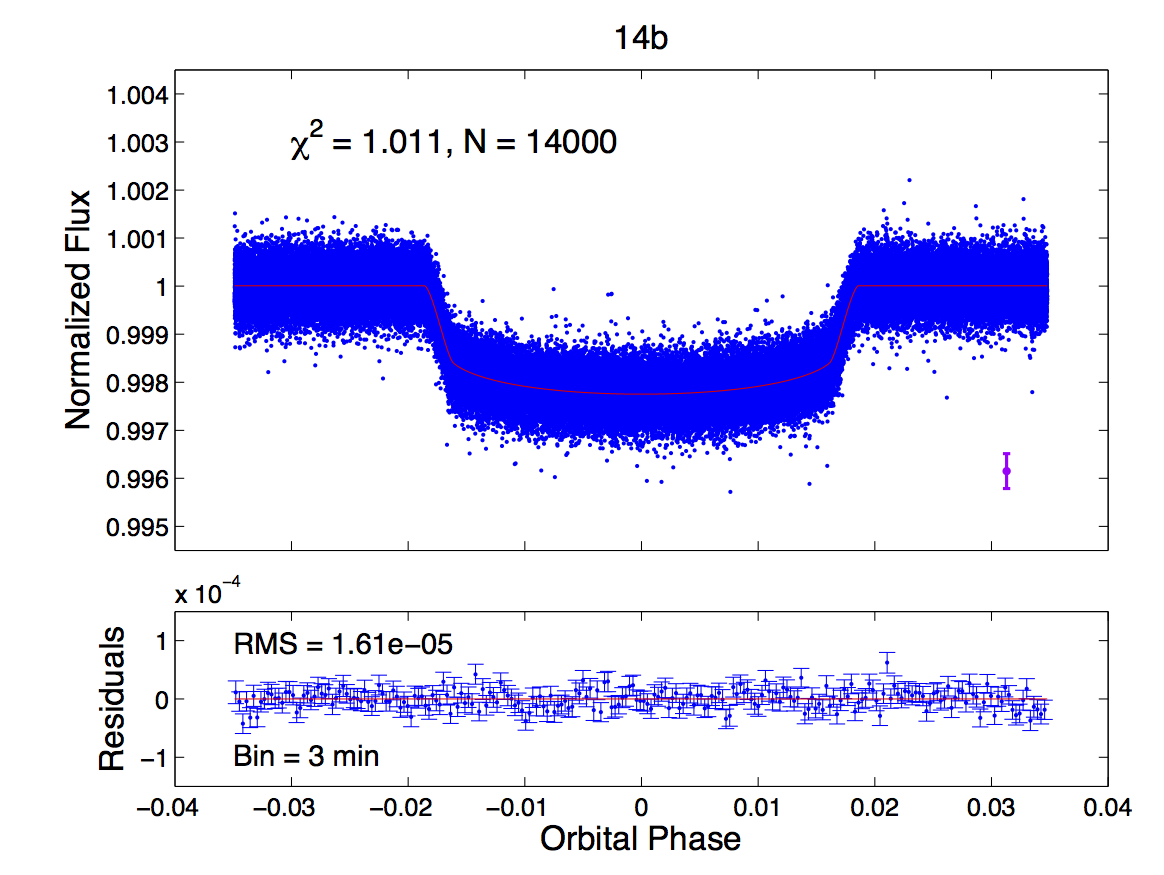}
}
\subfloat[]{
  \includegraphics[width=59mm]{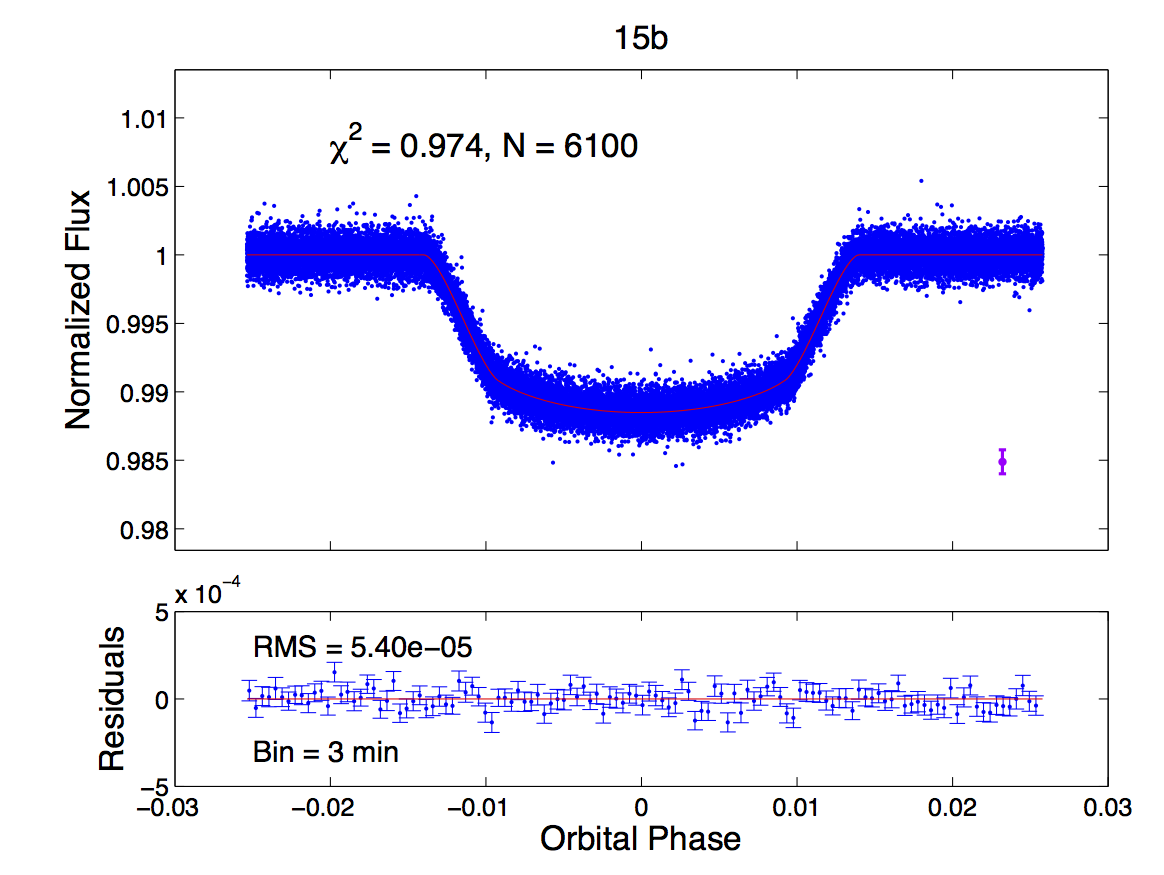}
}
\subfloat[]{
  \includegraphics[width=59mm]{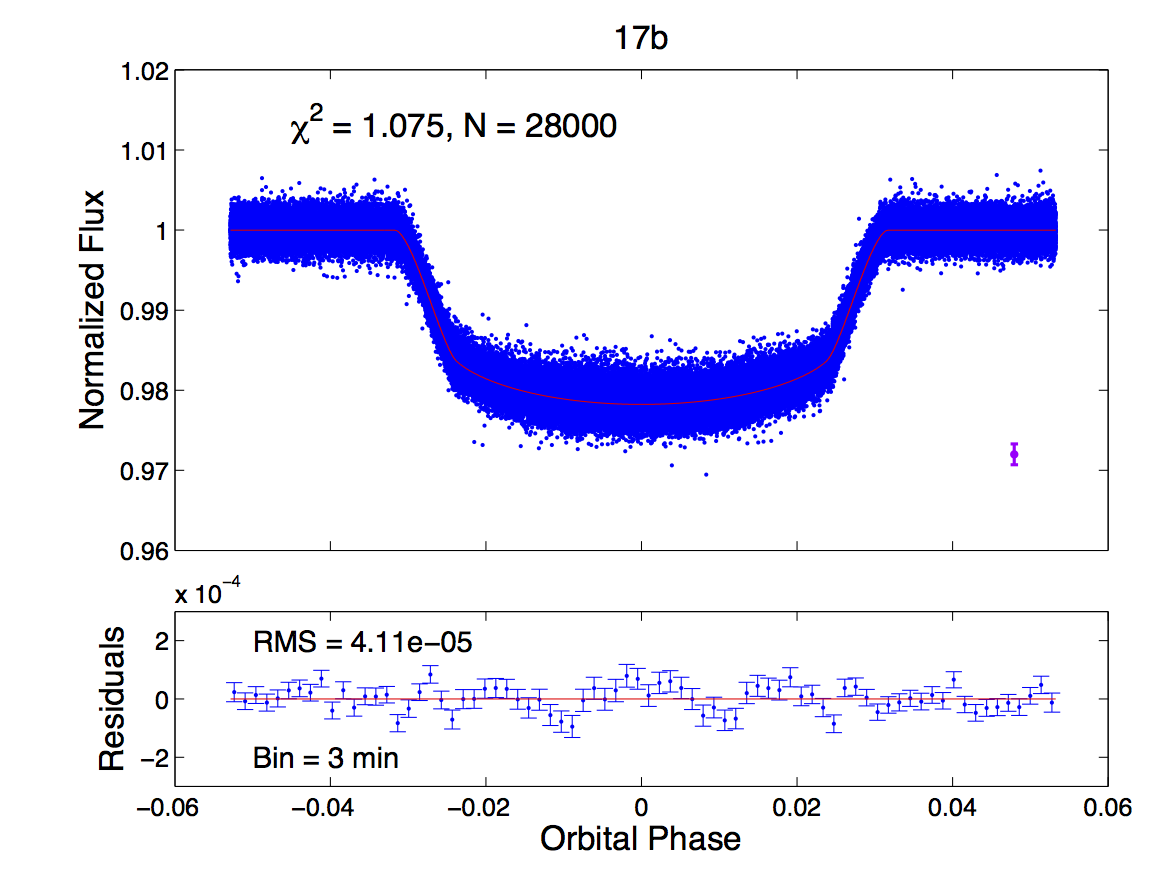}
}
\end{adjustwidth}
%\vspace{-4mm}
\caption{(Continued on next page) Ringless fits to 21 \textit{Kepler} planets.  In the upper plot of each panel, the blue points represent the phase-folded lightcurve, and the red curve represents the fit.  Individual error bars are omitted for clarity, and the purple error bar in the lower right indicates the average error about an artificial data point.  In the lower plot of each panel is a plot of the residuals in blue, with bin sizes given by the text, and a red line at $y=0$.\label{fr}}
\end{figure}

\begin{figure}
\setcounter{figure}{8}
\begin{adjustwidth}{-1.8cm}{}
\captionsetup[subfigure]{labelformat=empty}
\centering
\subfloat[]{
  \includegraphics[width=59mm]{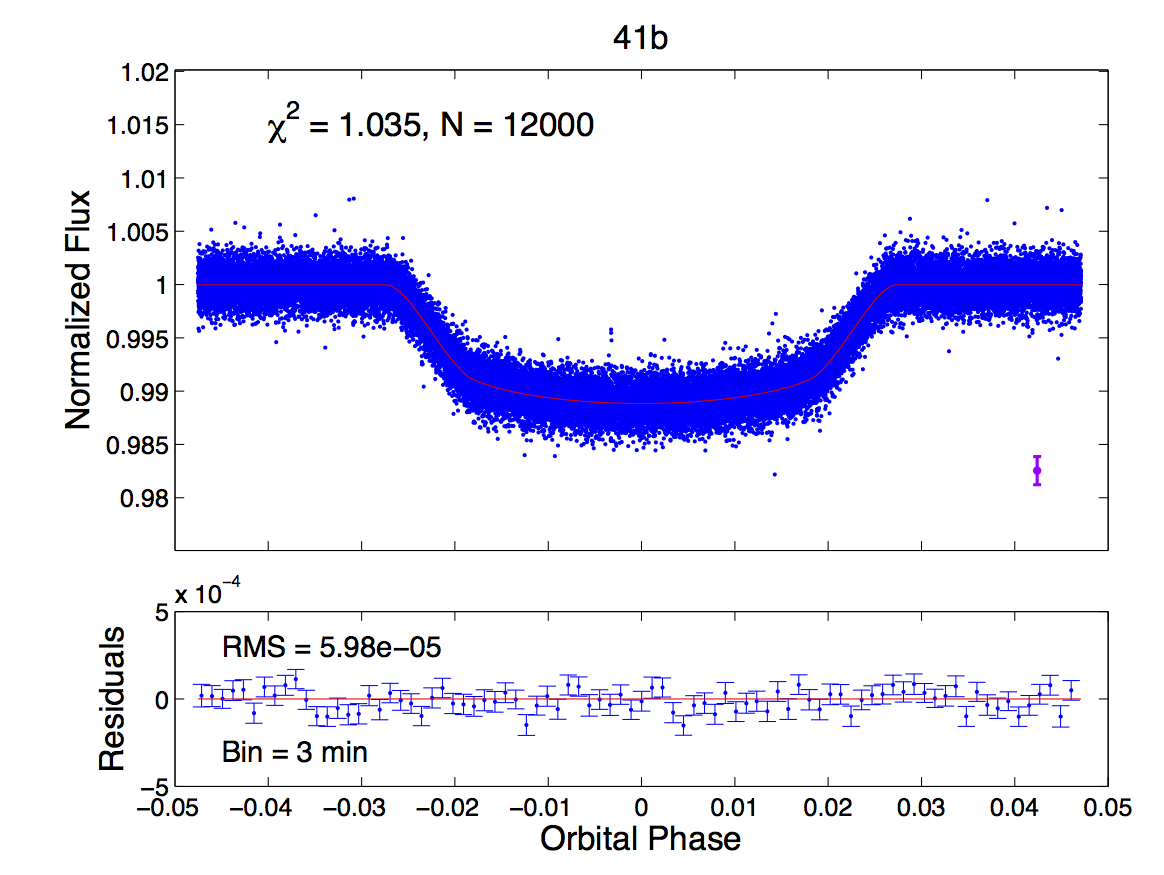}
}
\subfloat[]{
  \includegraphics[width=59mm]{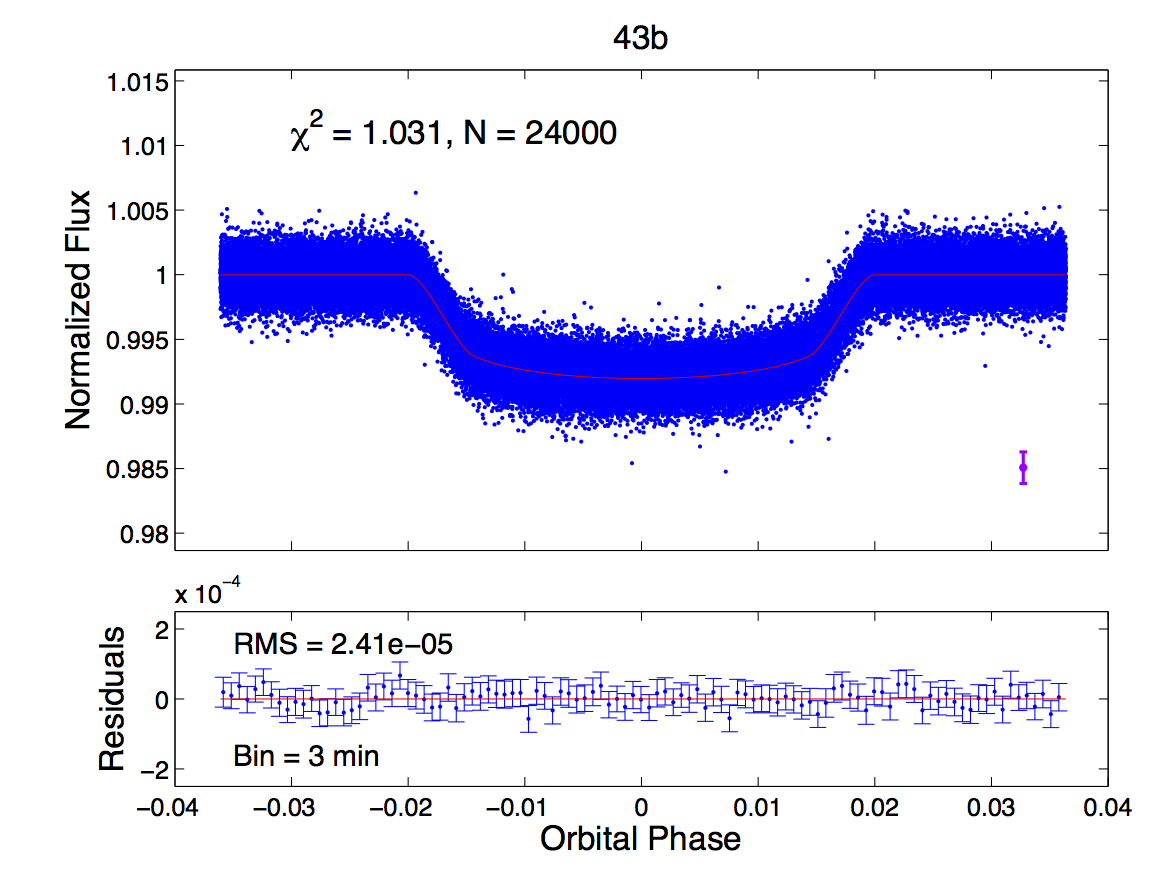}
}
\subfloat[]{
  \includegraphics[width=59mm]{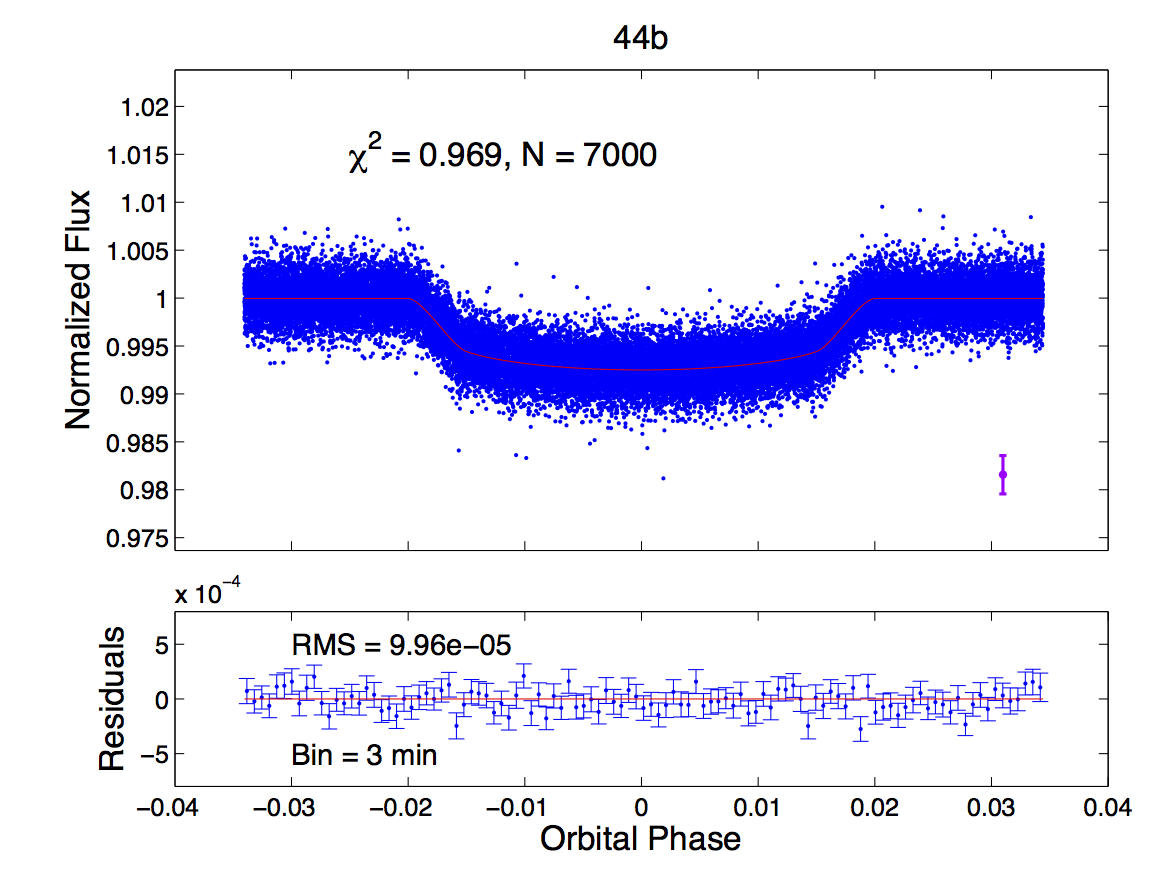}
}\\
\subfloat[]{
  \includegraphics[width=59mm]{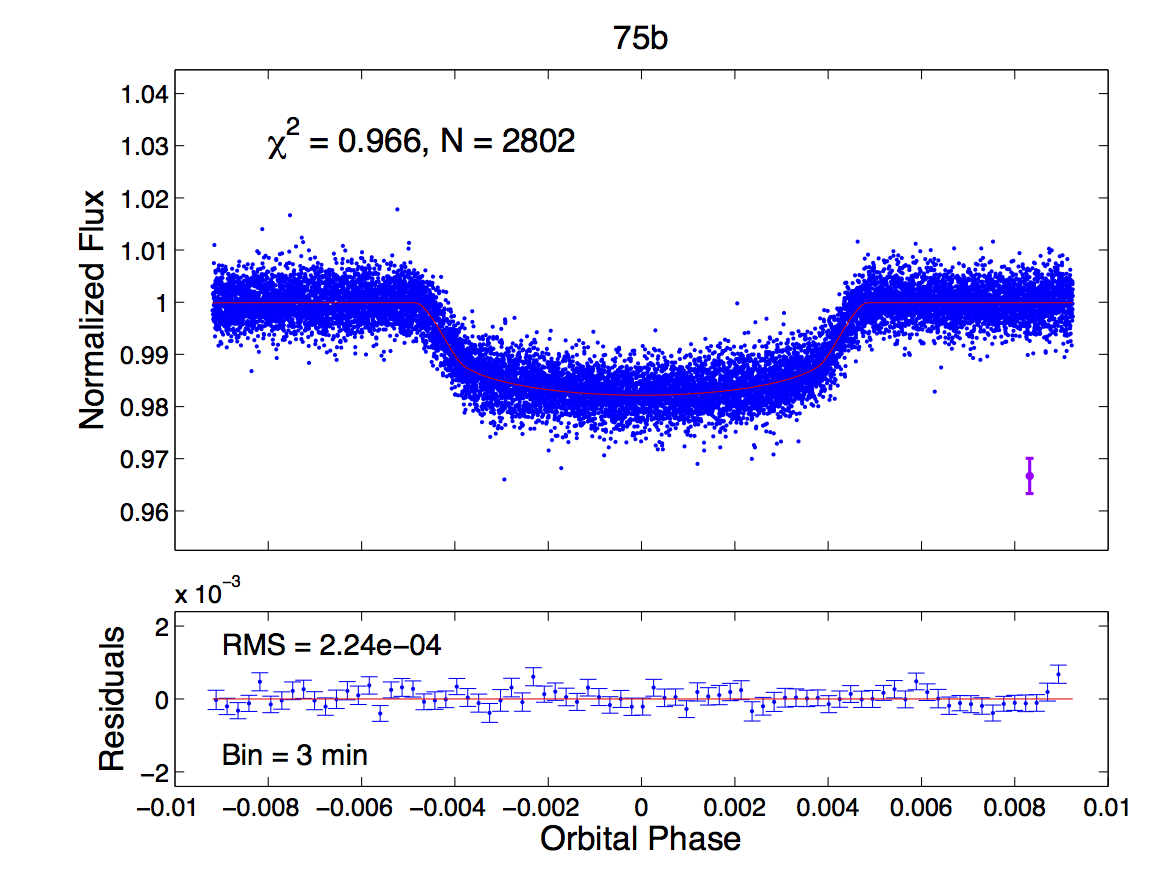}
}
\subfloat[]{
  \includegraphics[width=59mm]{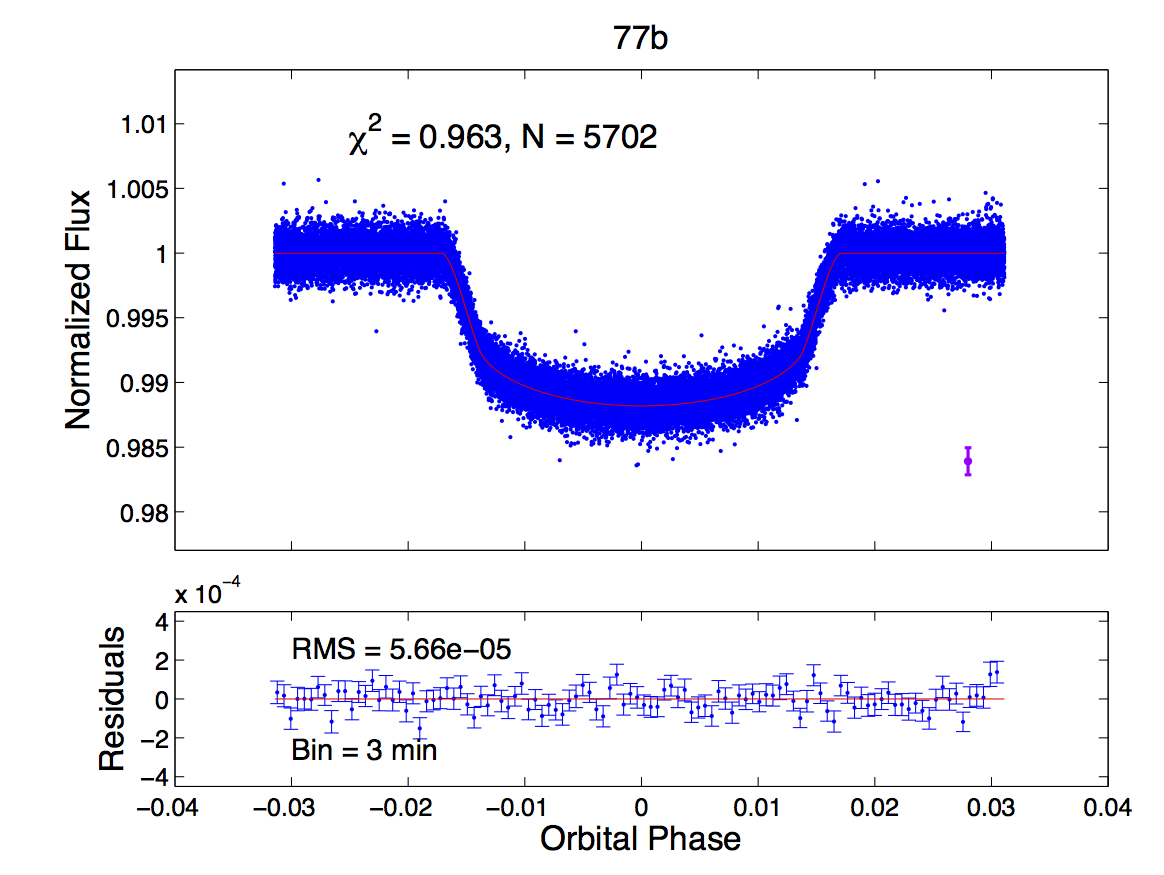}
}
\subfloat[]{
  \includegraphics[width=59mm]{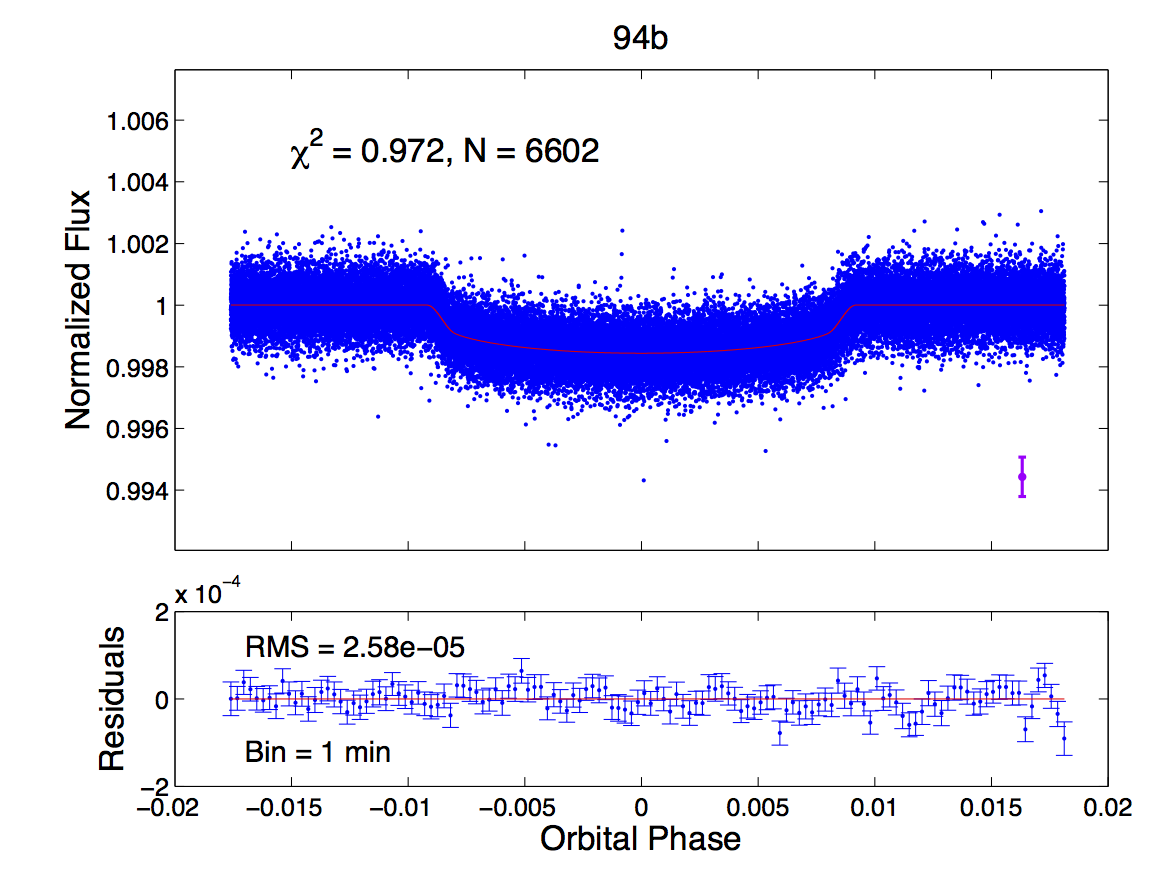}
}\\
\subfloat[]{
  \includegraphics[width=59mm]{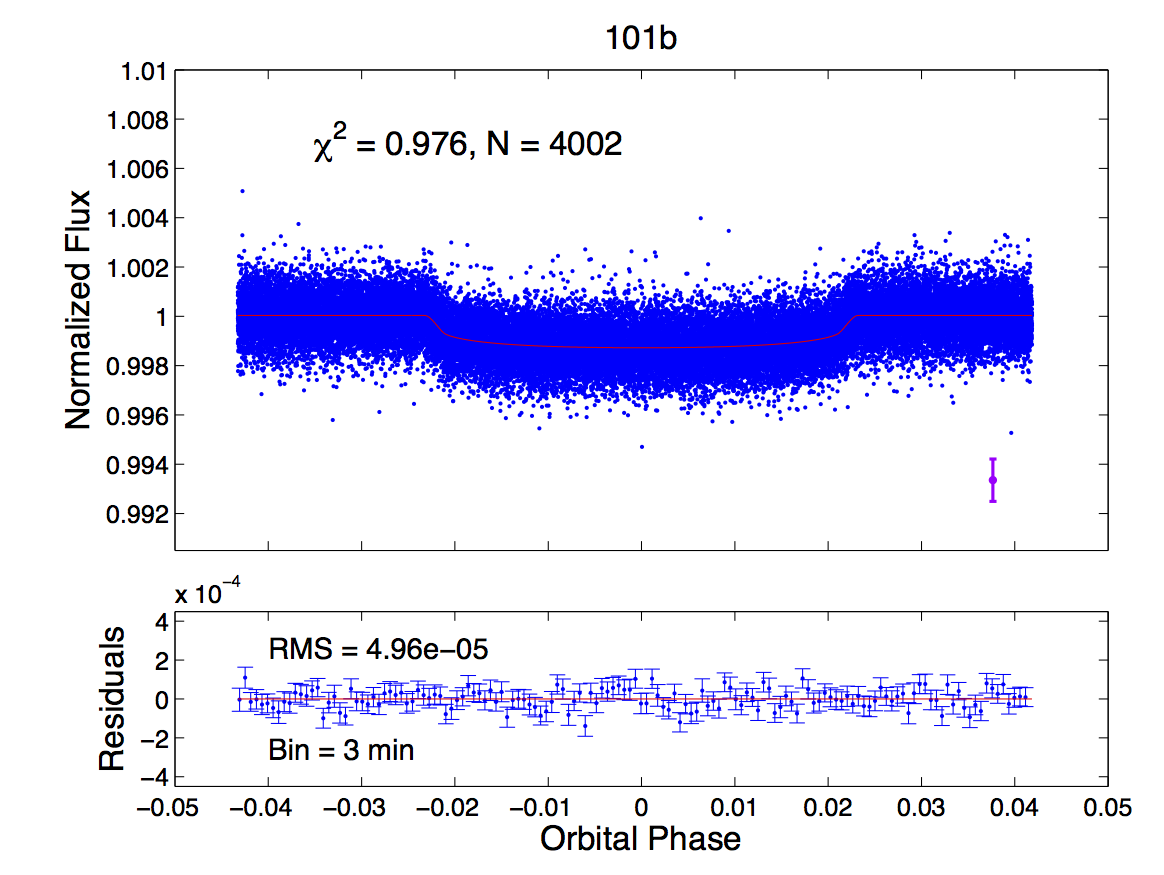}
}
\subfloat[]{
  \includegraphics[width=59mm]{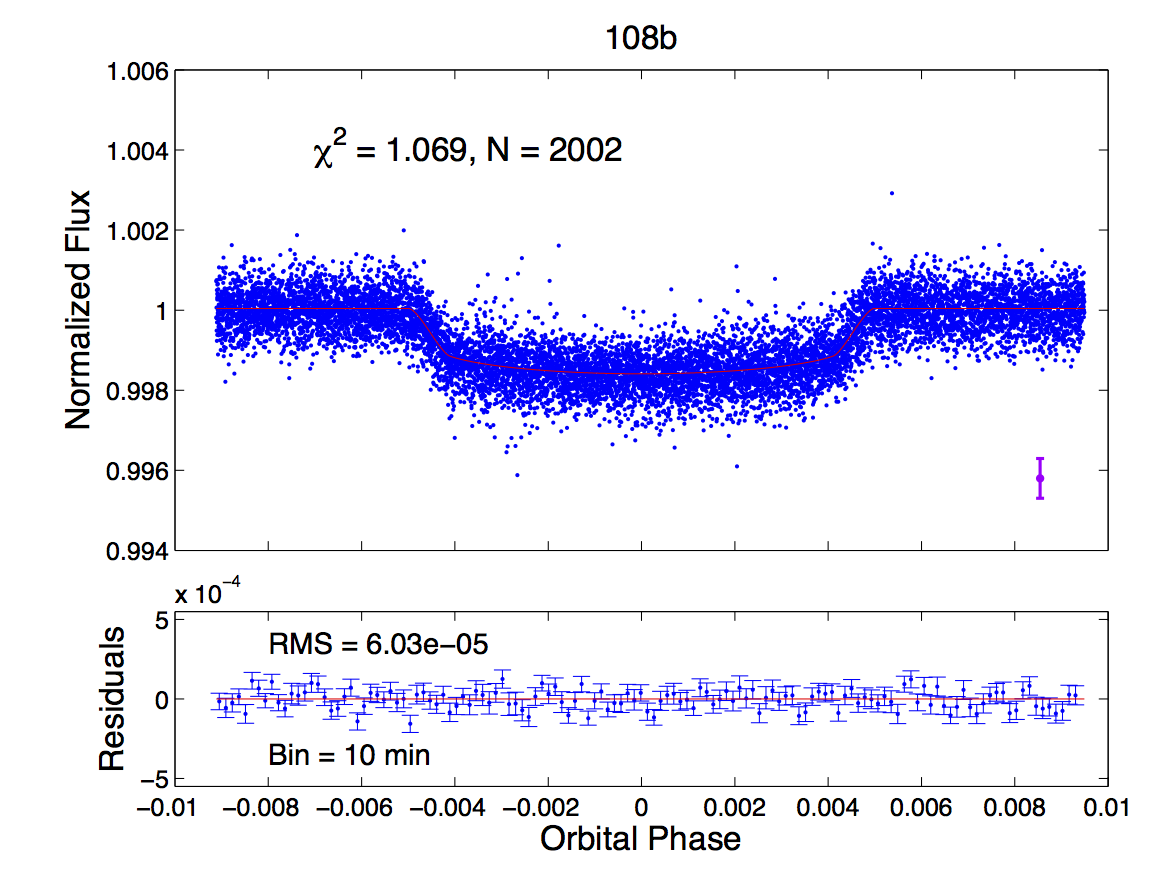}
}
\subfloat[]{
  \includegraphics[width=59mm]{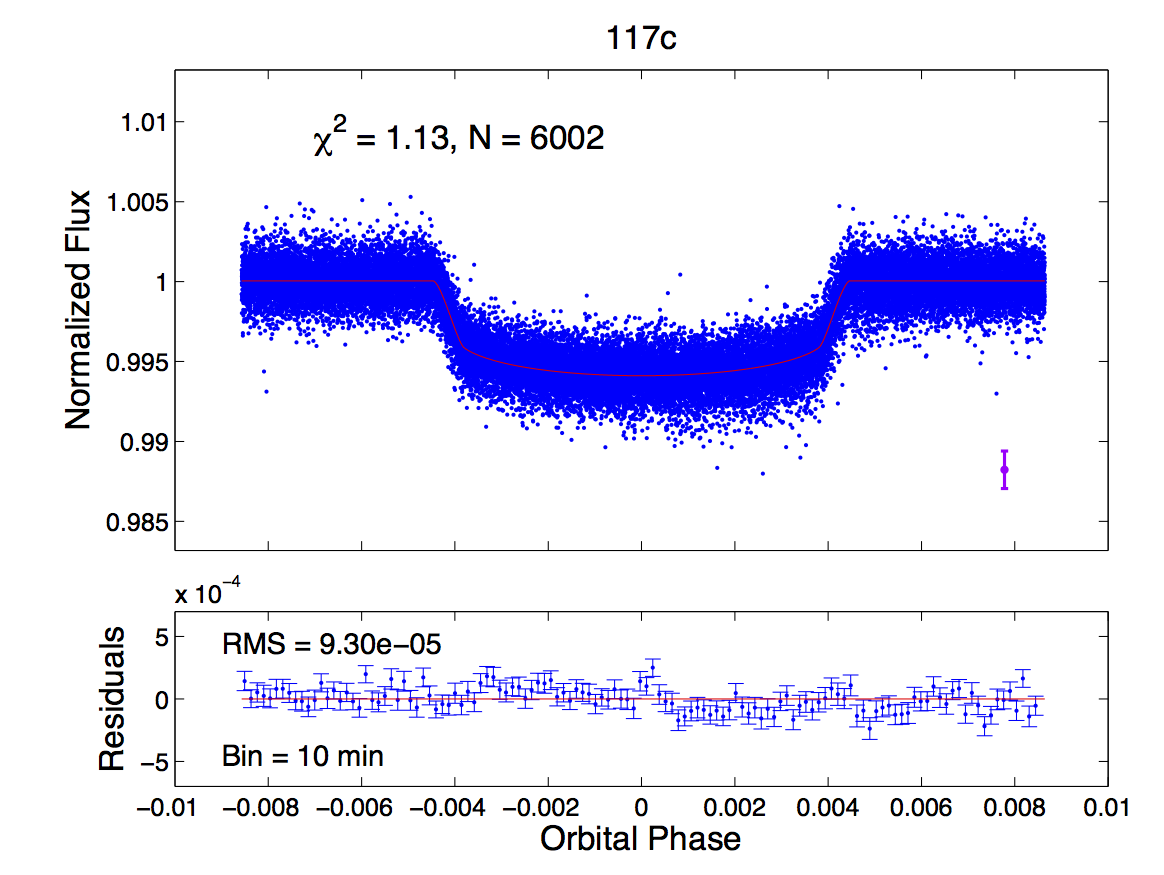}
}\\
\subfloat[]{
  \includegraphics[width=59mm]{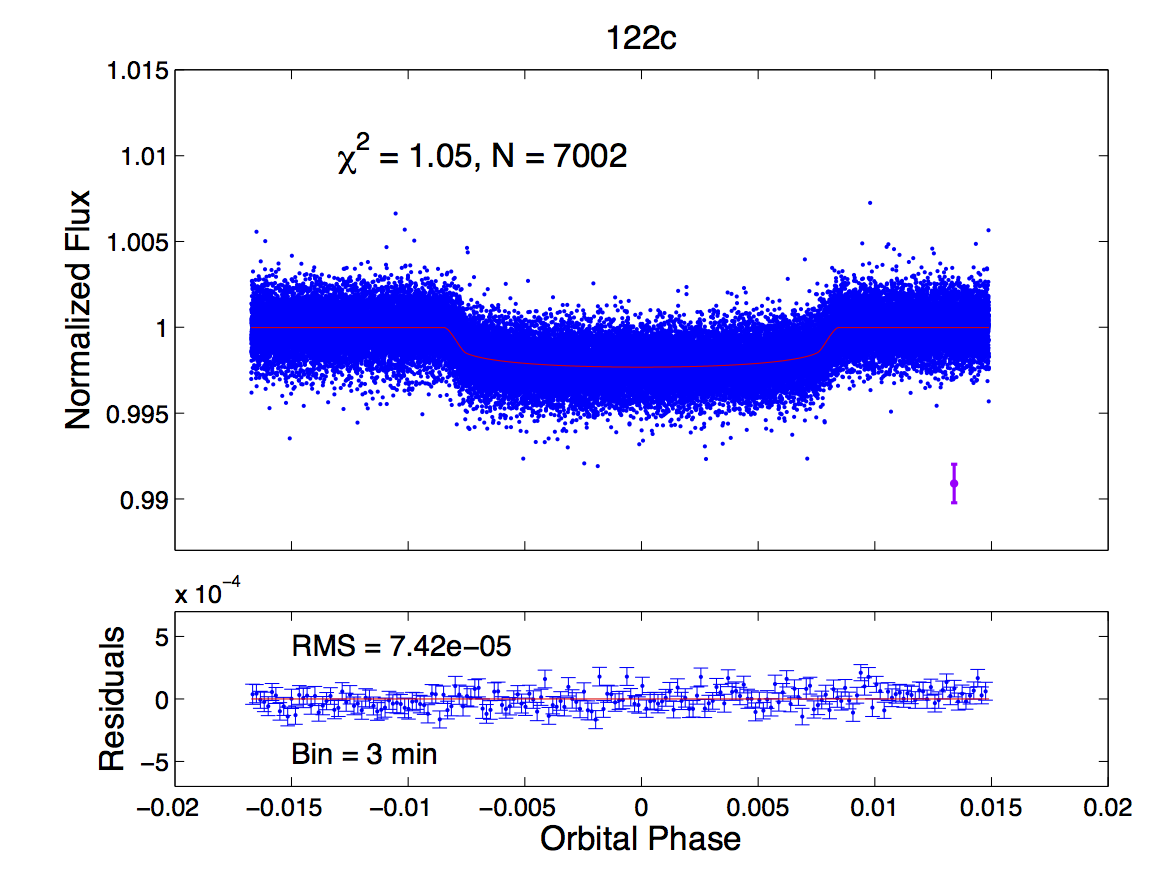}
}
\subfloat[]{
  \includegraphics[width=59mm]{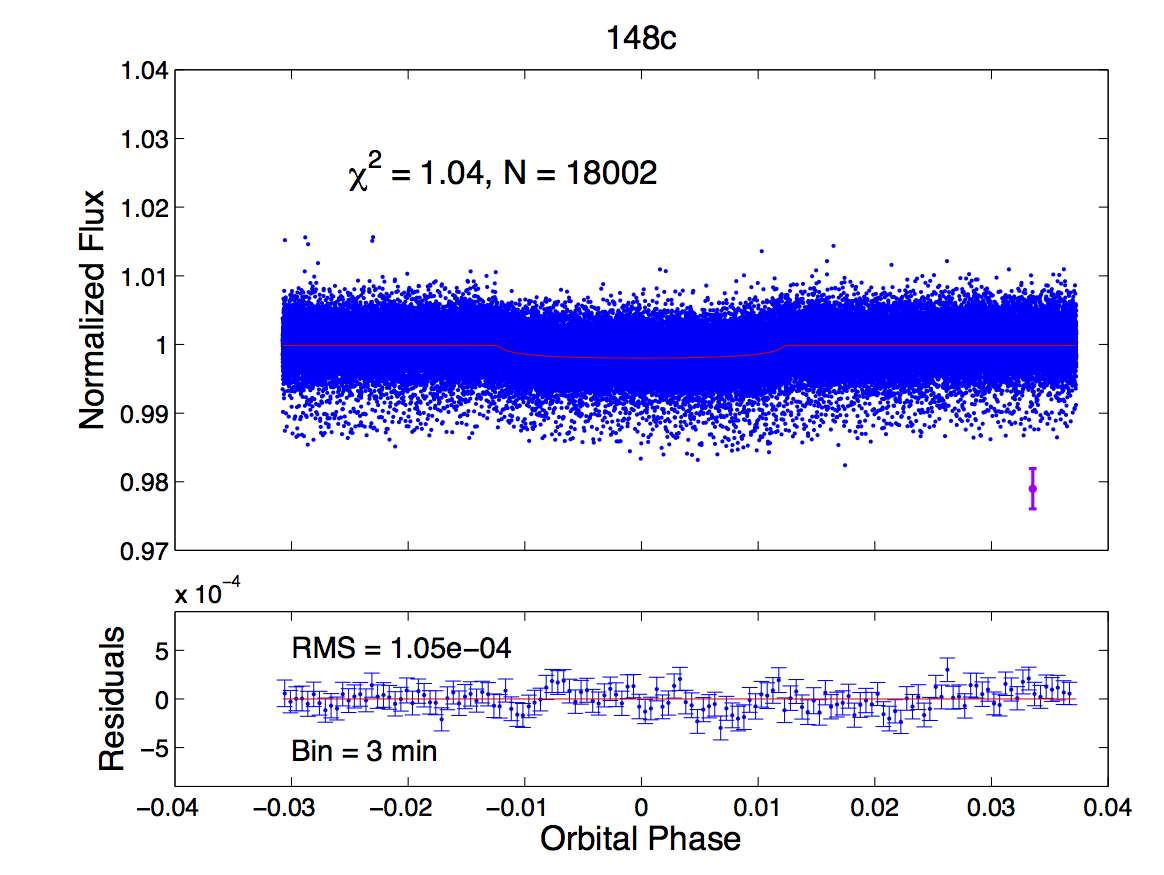}
}
\subfloat[]{
  \includegraphics[width=59mm]{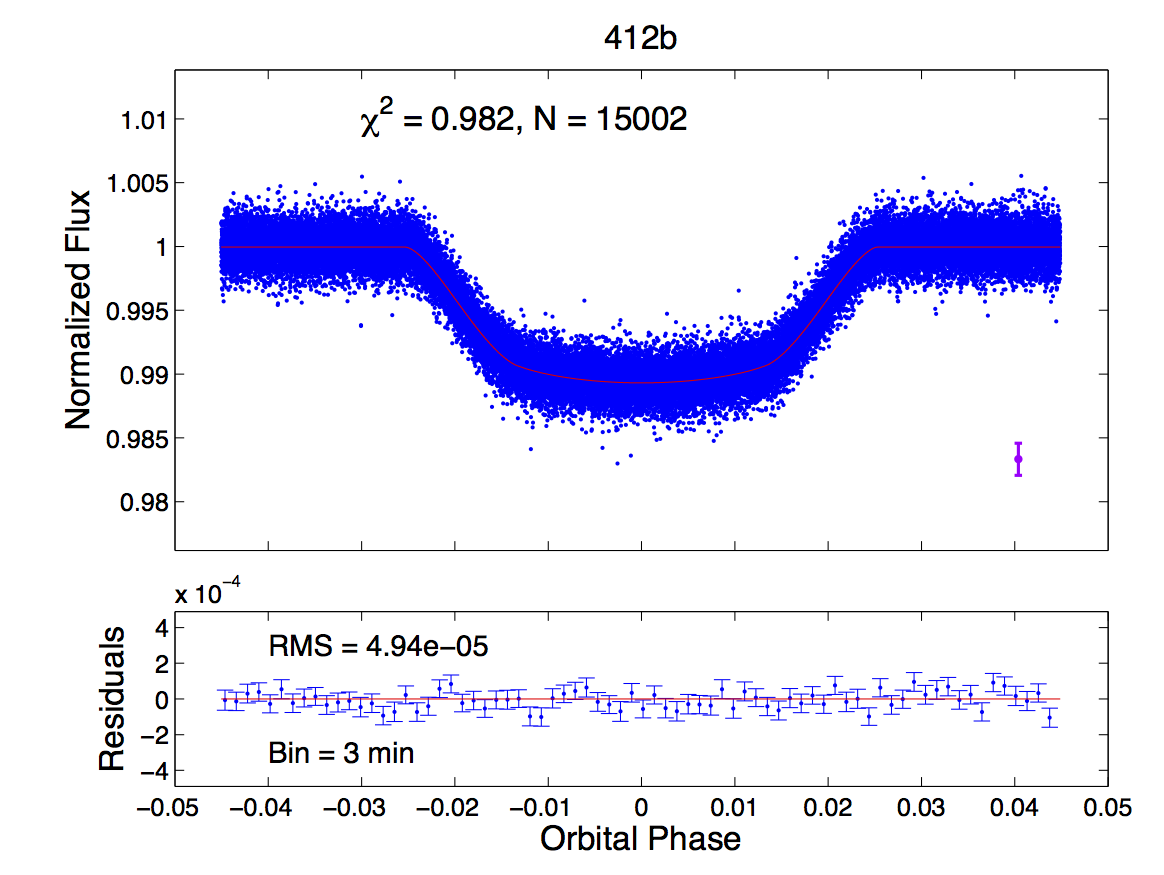}
}\\
\end{adjustwidth}
%\vspace{-1.9mm}
\caption{Continued from previous page}
\end{figure}

\renewcommand{\tabcolsep}{6pt}
\begin{table}[ht]
\captionsetup{justification=centering}
\centering
%\caption{Physical parameters found for ringless fits}
\caption{Physical parameters found for ringless fits\label{params}}
\begin{tabular}{p{1.4cm}|p{2.95cm}|p{1.6cm}|p{1.6cm}|p{2.8cm}|p{1.6cm}}
\hline\hline
\textit{Kepler} name &$R_p$ (stellar radii)&$\gamma_1$&$\gamma_2$&$a$ (stellar radii)&$i$ ($^{\circ}$)\\
\hline
4b&0.0255&0.3570&0.2823&5.5582&95.1251\\[0.1cm]
5b&0.0803&0.3702&0.1090&6.4078&91.5488\\[0.1cm]
6b&0.0949&0.4933&0.0789&7.4813&91.4879\\[0.1cm]
7b&0.0835&0.4073&0.1153&6.6119&94.9736\\[0.1cm]
8b&0.0964&0.4094&0.0771&6.8141&96.1487\\[0.1cm]
12b&0.1199&0.4232&0.0837&7.9815&91.6415\\[0.1cm]
14b&0.0457&0.3714&0.1427&7.4017&94.6129\\[0.1cm]
15b&0.1057&0.5561&0.0002&9.8952&94.0340\\[0.1cm]
17b&0.1364&0.4430&0.0436&5.6886&91.8789\\[0.1cm]
41b&0.1036&0.4905&0.0791&5.1185&97.7321\\[0.1cm]
43b&0.0873&0.4253&0.1225&6.9370&95.4968\\[0.1cm]
44b&0.0835&0.5634&0.0005&7.0327&95.1964\\[0.1cm]
75b&0.1206&0.4928&0.1072&36.9133&90.0086\\[0.1cm]
77b&0.1001&0.5025&0.0971&9.6839&92.2142\\[0.1cm]
94b&0.0378&0.6835&0.0029&14.4239&92.4944\\[0.1cm]
101b&0.0341&0.2986&0.3989&6.0979&95.1086\\[0.1cm]
108b&0.0406&0.5646&0.0000&22.7241&91.9159\\[0.1cm]
117c&0.0717&0.4259&0.0732&36.4522&90.5104\\[0.1cm]
122c&0.0450&0.3711&0.2266&17.7374&91.4752\\[0.1cm]
148c&0.0371&0.5000&0.4998&13.1271&90.0516\\[0.1cm]
412b&0.1061&0.5263&0.0202&4.8855&99.4487\\[0.1cm]
\end{tabular}
%\label{table:params}
\end{table}

\section{CONSTRAINTS ON POSSIBLE RINGS}
\subsection{Fiducial Rings and Theoretical Concerns}

Even if a transit lightcurve can be adequately described by a ringless planet, it is impossible to rule out the presence of any possible optically thick ring using only the lightcurve data, because for any ring that is sufficiently close to edge-on, or with $R_1$ sufficiently close to $R_2$, or sufficiently close to face-on with $R_1$ sufficiently close to $R_p$, the ringed planet's lightcurve will be indistinguishable from that of a ringless planet.  However, it is possible to rule out rings of certain sizes and orientations.  For each of the 21 transit lightcurves, fits were performed to determine if the data were consistent with rings of certain sizes and orientations.  There were 8 fiducial size/orientation combinations that were tested, consisting of 3 different binary parameters: an obliquity of $10^{\circ}$ vs. $45^{\circ}$, a season of $0^{\circ}$ vs. $45^{\circ}$, and a ring size of ``Saturn" vs. ``Half-Saturn".  ``Saturn" refers to proportions given by $\frac{R_1}{R_p}=1.239$ and $\frac{R_2}{R_p}=2.269$ (approximately equivalent to the relative size of Saturn's main rings), while ``Half-Saturn" refers to proportions given by $\frac{R_1}{R_p}=1.195$ and $\frac{R_2}{R_p}=1.6345$ (in which the distances from the planet surface to both the inner and outer edges of the ring have been halved relative to Saturn).  Each of these 8 fiducial rings was given a letter name by which they will be referred to for the rest of this paper.  The name and profile of each of the 8 ringed planets are shown in Figure \ref{rv}, with the planet's orbital motion being parallel to the horizontal axis.  Also shown in each panel is the letter for each ring size/orientation.

\begin{figure}
\captionsetup[subfigure]{labelformat=empty}
\centering
\vspace{-6mm}
\subfloat[]{
  \includegraphics[width=59mm]{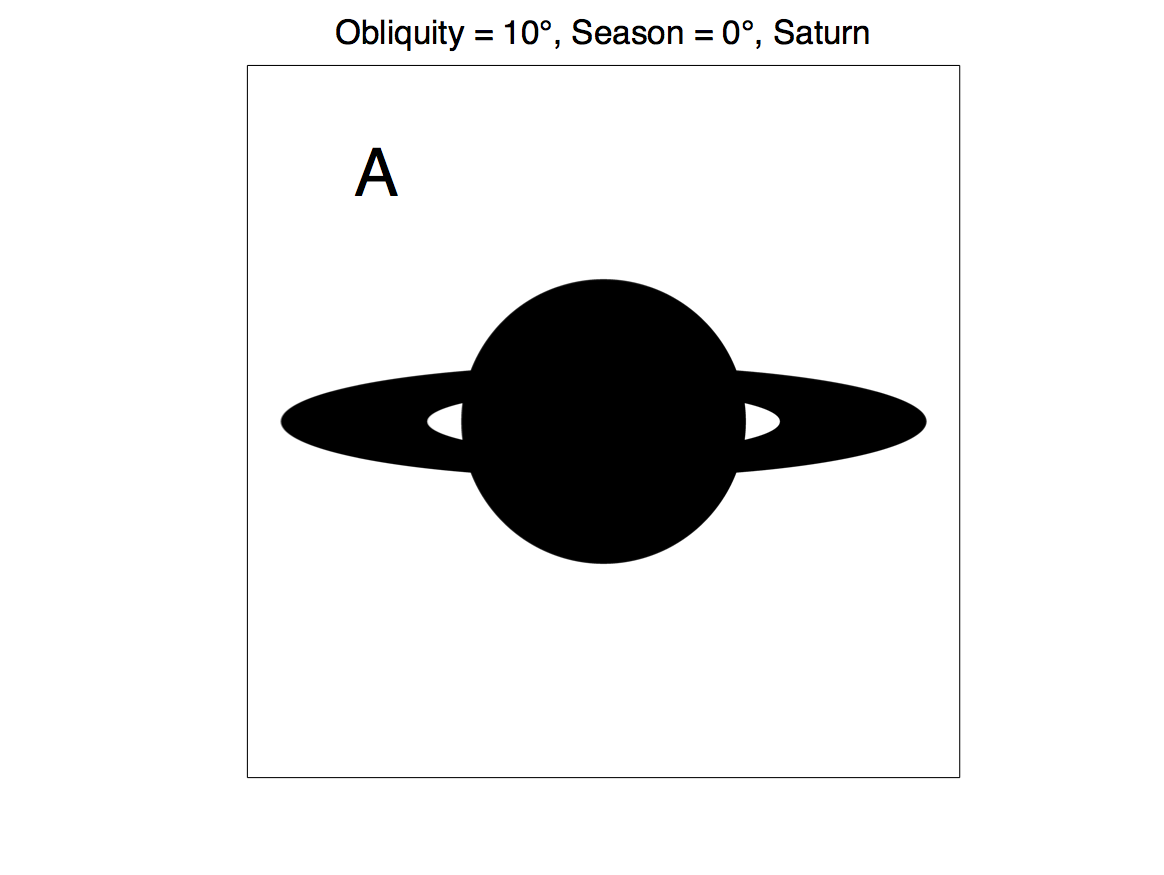}
}
\subfloat[]{
  \includegraphics[width=59mm]{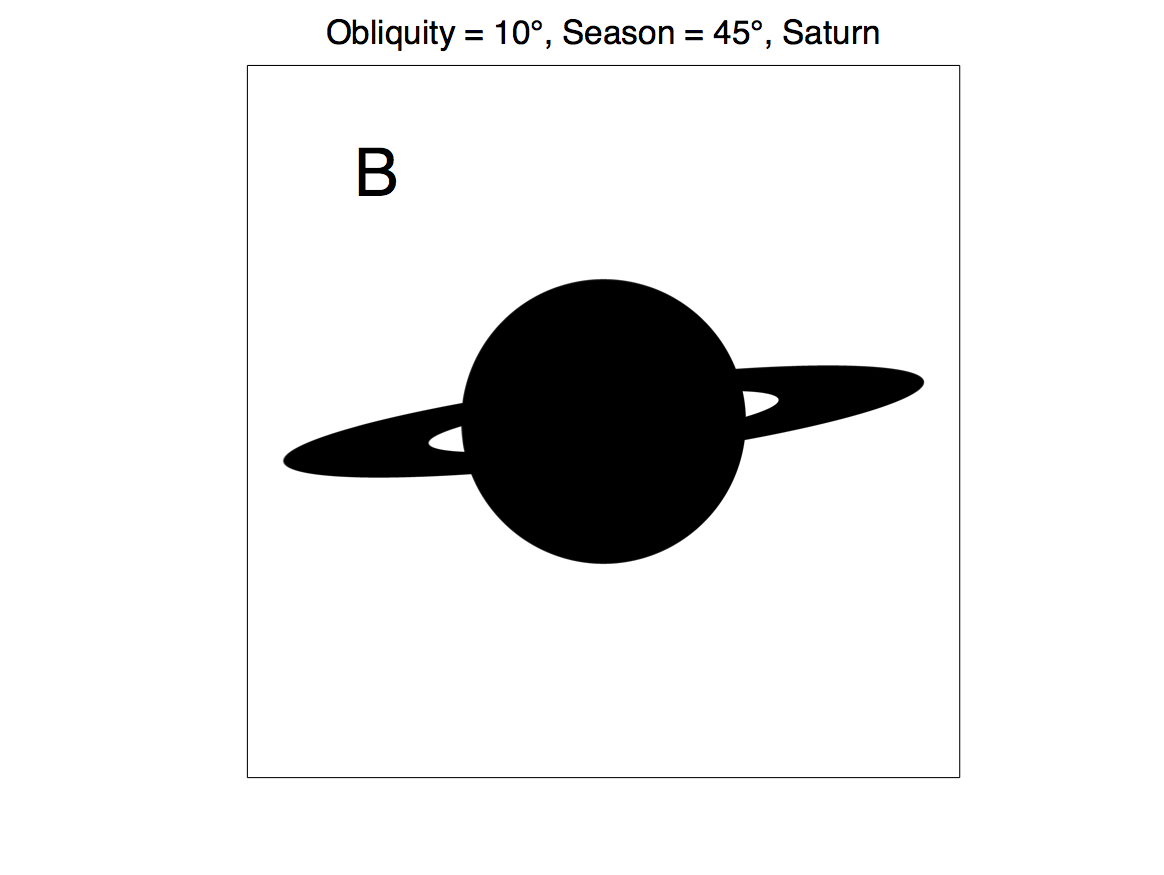}
}\\[-5mm]
\subfloat[]{
  \includegraphics[width=59mm]{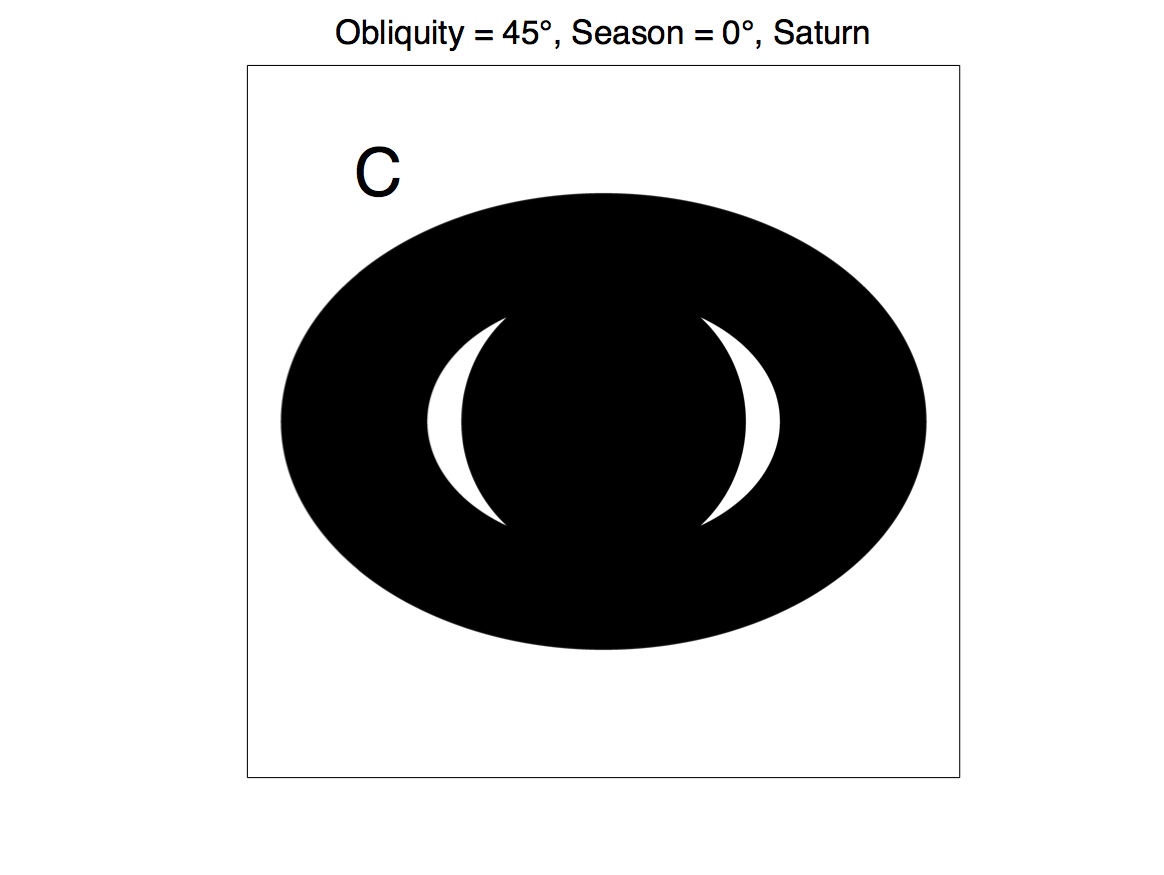}
}
\subfloat[]{
  \includegraphics[width=59mm]{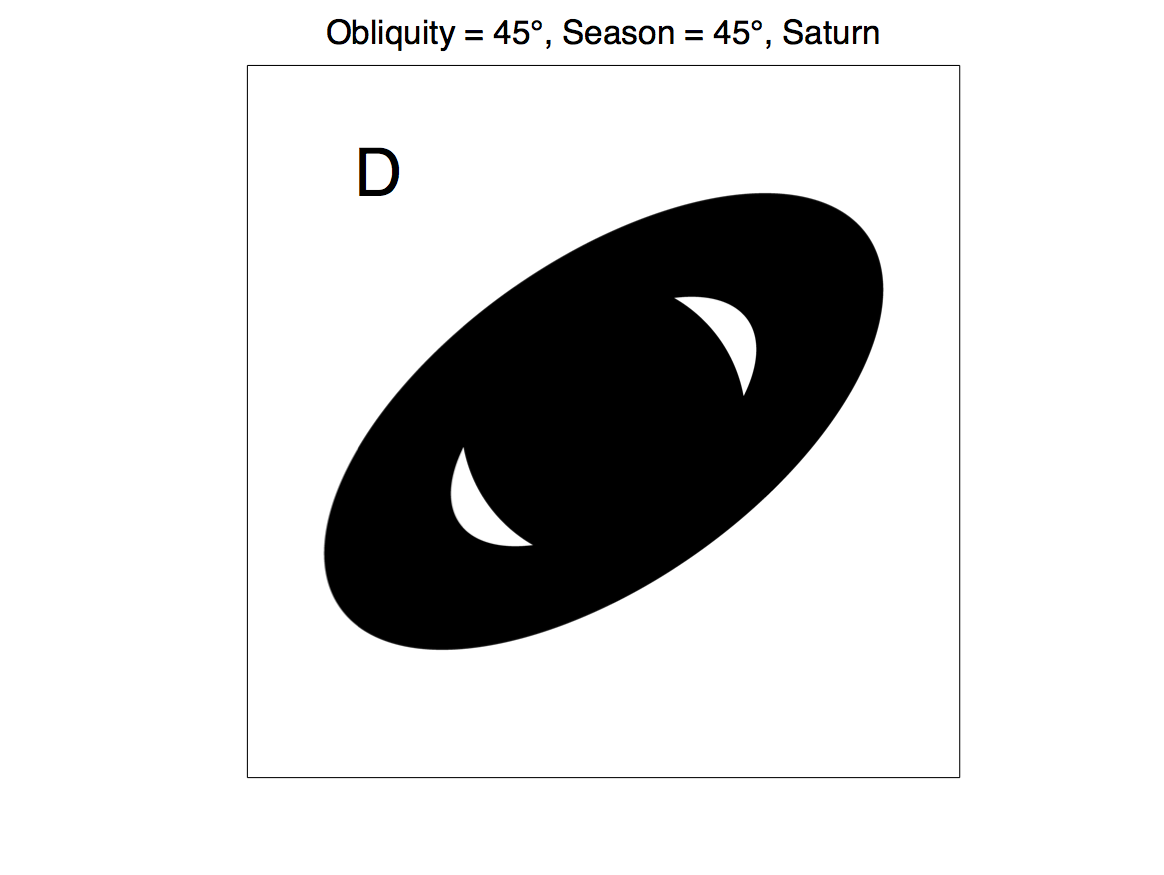}
}\\[-5mm]
\subfloat[]{   % ???
  \includegraphics[width=59mm]{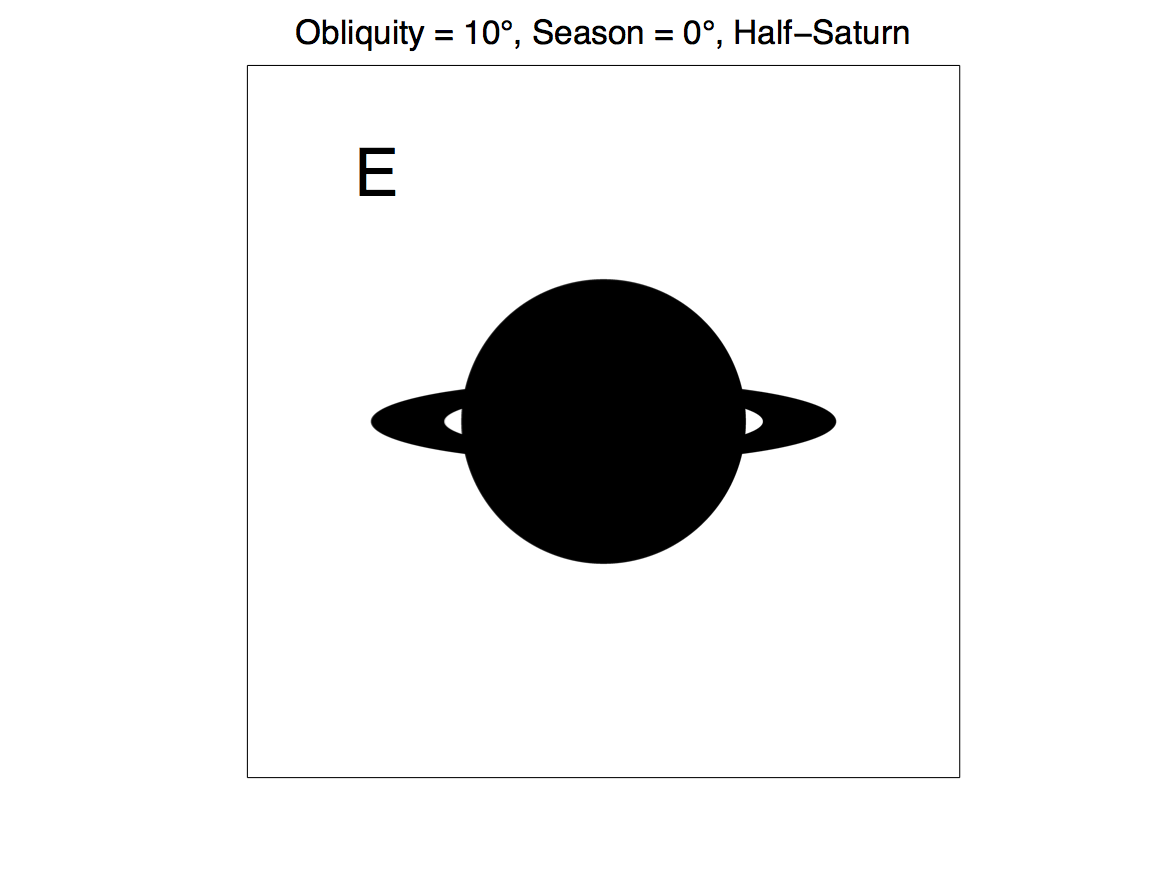}
}
\subfloat[]{
  \includegraphics[width=59mm]{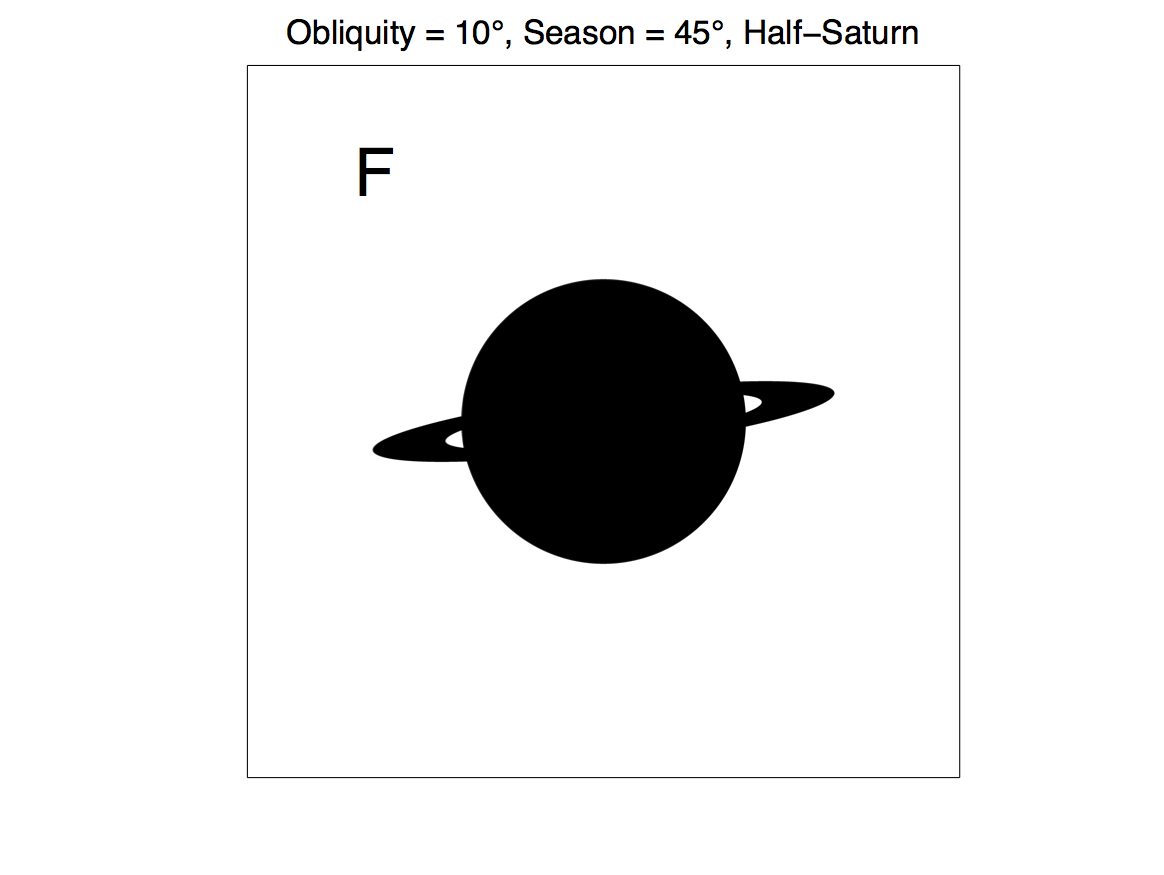}
}\\[-5mm]
\subfloat[]{   % ???
  \includegraphics[width=59mm]{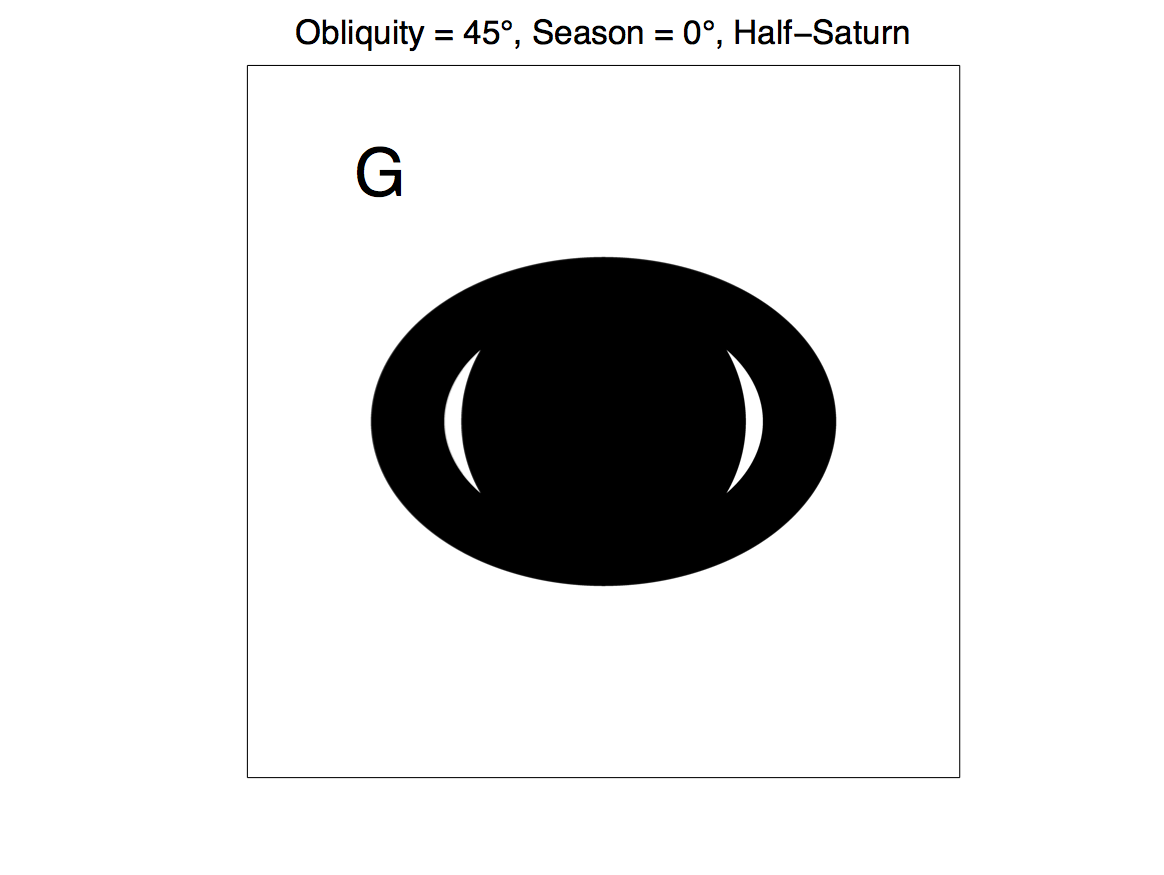}
}
\subfloat[]{
  \includegraphics[width=59mm]{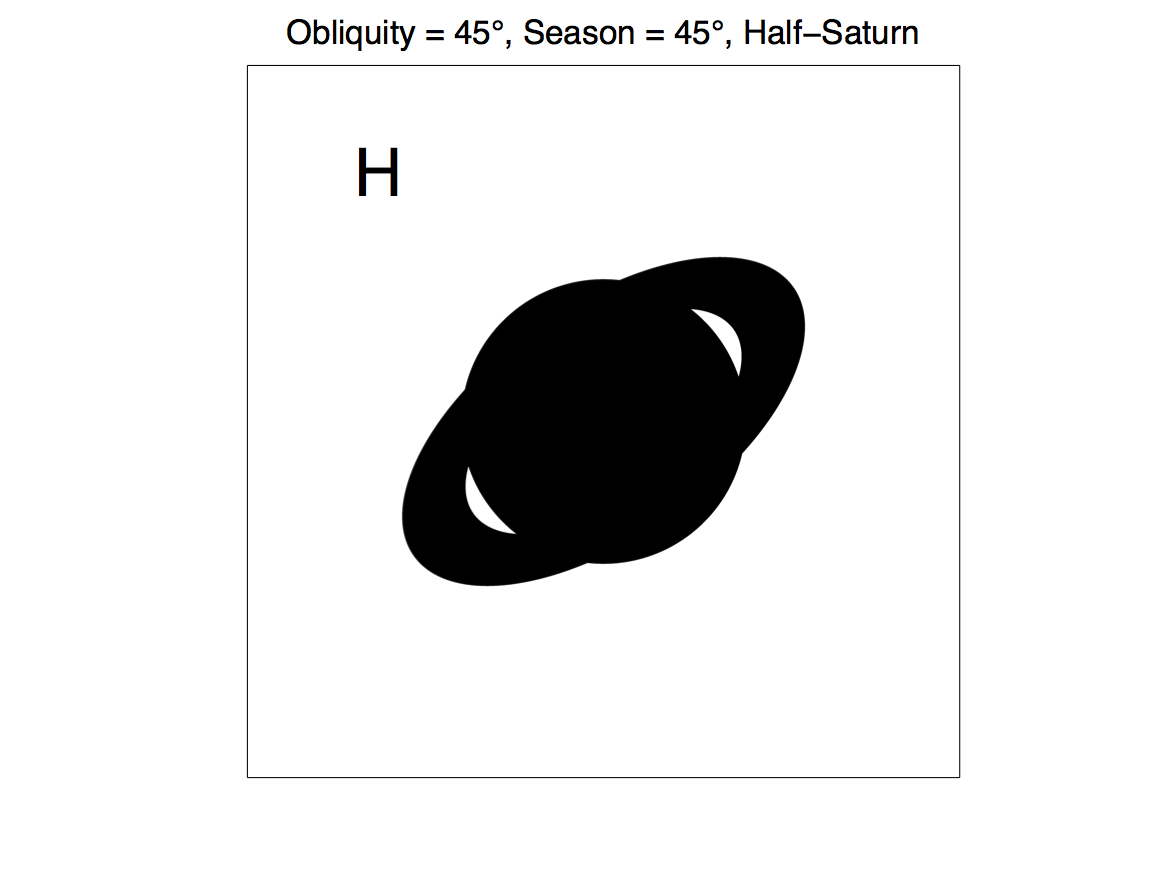}
}
\vspace{-4mm}
\caption{Profiles of 8 different fiducial ringed planets used for fits.  The orientation and size of each ring is given in the title to each panel, while the letter name of each ring is given in the upper left of each panel.  The ringed planet's orbital motion across the stellar profile in each panel is from left to right, parallel to the horizontal axis.\label{rv}}
\end{figure}

There are several theoretical issues to consider when trying to determine how plausible such rings would be about the planets studied in this paper.  One such issue is whether such obliquities are realistic for hot Jupiters.  The axial tilts of most hot Jupiters are expected to be small.  However, the largest possible offset in inclination from $90^{\circ}$ for a transiting planet, given by arcsin$(R_*/a)$, where $R_*$ is the stellar radius, is roughly $5^{\circ}$-$10^{\circ}$ for a hot Jupiter around a Sun-like star.  Such an inclination is sufficient to give a planet an obliquity of $10^{\circ}$, or close to that, even for $\varepsilon=0^{\circ}$.  Thus, $10^{\circ}$ represents an obliquity that is small enough to be relatively common, yet large enough to be potentially detectable.

An obliquity of $45^{\circ}$, on the other hand, requires a large value of $\varepsilon$, which requires that the planet settles into a high-obliquity Cassini state, namely Cassini state 2.  To determine what values of $\varepsilon$ are predicted, we compare $-g/\alpha$ to $(-g/\alpha)_{crit}$.  Here, $g$ is the nodal precession frequency; $\alpha$ is the spin precessional constant, defined as
\begin{equation}
\alpha=\frac{3}{2}\left(\frac{C-A}{C}\right)\left(\frac{n^2}{\omega}\right)
\end{equation}
where $C>A=B$ are the (axisymmetric) planet's principle moments of inertia, $n$ is the orbital mean motion, and $\omega$ is the planet's rotational angular frequency; and $(-g/\alpha)_{crit}$ is defined as
\begin{equation}
(-g/\alpha)_{crit}=(\textrm{sin}^{2/3}(I)+\textrm{cos}^{2/3}(I))^{-3/2}
\end{equation}
where $I$ is the inclination of the orbit normal with respect to the angular momentum of the system, including other bodies ($I=0^{\circ}$ for a coplanar orbit).  In Cassini state 2, for $-g/\alpha\ll(-g/\alpha)_{crit}$, $\varepsilon\rightarrow90^{\circ}$, while for $-g/\alpha\gg(-g/\alpha)_{crit}$, $\varepsilon\rightarrow I$ \citep{c66,wh05,f07}.  This implies two ways of attaining $\varepsilon\approx 45^{\circ}$: $I\approx 45^{\circ}$ and $-g/\alpha>(-g/\alpha)_{crit}$, or $I<45^{\circ}$ and $-g/\alpha\approx(-g/\alpha)_{crit}=\mathcal{O}(1)$.  For nodal precession caused by the stellar gravitational quadrupole moment, $-g/\alpha\ll1$ for a typical hot Jupiter, and thus an obliquity of $45^{\circ}$ is unattainable; however, the presence of other bodies in the system may enhance the nodal precession such that $-g/\alpha$ is of order unity or more, making an obliquity of $45^{\circ}$ possible.

To determine the orbital stability of such rings, we take $R_2=2.269\times R_p$ (Saturn-sized) and we introduce the Hill radius, $R_H$, for a planet on a nearly circular orbit, defined as
\begin{equation}
R_H=a\left(\frac{M_p}{3M_*}\right)^{1/3}=\left(\frac{GP^2\rho_p}{9\pi}\right)^{1/3}
\end{equation}
where $a$ is the planet's semi-major axis, $M_p$ is the planet's mass, $M_*$ is the host star's mass, $G$ is the graviational constant, $P$ is the planet's orbital period, and $\rho_p$ is the planet's density.  Of the 17 planets studied in this paper with measured densities, all have $R_2<R_H$, indicating they could host Saturn-sized rings that are dynamically stable provided that the ring particles are on retrograde orbits, while 10 of the 17 planets have $R_2<0.49\times R_H$, indicating they could host stable Saturn-sized rings that are prograde, as well as retrograde \citep{hamilton,vnw}.

To examine how such rings might have formed, we consider the Roche radius, $R_R$, for a planet with a large, self-gravitating, synchronously rotating satellite, defined as 
\begin{equation}
R_R=2.45R_p\left(\frac{\rho_p}{\rho_s}\right)^{1/3}
\end{equation}
where $\rho_s$ is the density of the satellite.  Assuming a rocky satellite with a density of 3 g/cm$^3$, only 5 of the 17 surveyed planets with measured densities could potentially host Saturn-sized rings within their respective Roche radii.  Rings extending beyond a planet's Roche radius are theoretically possible, though their formation would have to be explained by something other than tidal disintegration of a satellite by the host planet.  Furthermore, such rings would not be long-lived, as the ring particles would eventually coalesce into satellites.

\subsection{Ringed fits}

For each fit, five parameters were left free to float: the overall size of the planet/ring system, the planet's semimajor axis, the planet's inclination, and the two coefficients for stellar limb-darkening.  The phase shift and normalization that were free to float for the ringless fits were here fixed to their values determined from the ringless fits for the sake of faster-running computations, as integrating a ringed lightcurve is more computationally expensive than integrating a ringless lightcurve.  If a given lightcurve could not be adequately described by a ring of a certain size and orientation, then such a ring, if it were to exist around a transiting exoplanet, would yield a transit lightcurve that would be inconsistent with a ringless fit.  Because each of the 21 planets were found to be consistent with a ringless model, the possibility of such a ring could then be eliminated.

Due to the large number of ringed fits performed in total ($21 \times 8 = 168$), the fits are not all included in this paper.  The fits for \textit{Kepler}-15b, however, are included as a representative sample, and are shown in Figure \ref{15b}.  For each ringed fit, the residuals were plotted in bins of the same size as was done with the ringed fits to make clear any possible ring signal.  The amplitude of the residuals (amp.), i.e. the largest residual in absolute magnitude, is also indicated for each fit.  The physical parameters from each fit are shown in Table \ref{params15}.  Note that $R_p$ only indicates the size of the planet and not of the entire planet/ring profile, and therefore changes significantly with the area of the ring relative to that of the planet, as the overall profile area is largely determined by the transit lightcurve depth.

\begin{figure}
\captionsetup[subfigure]{labelformat=empty}
\centering
\vspace{-6mm}
\subfloat[]{
  \includegraphics[width=59mm]{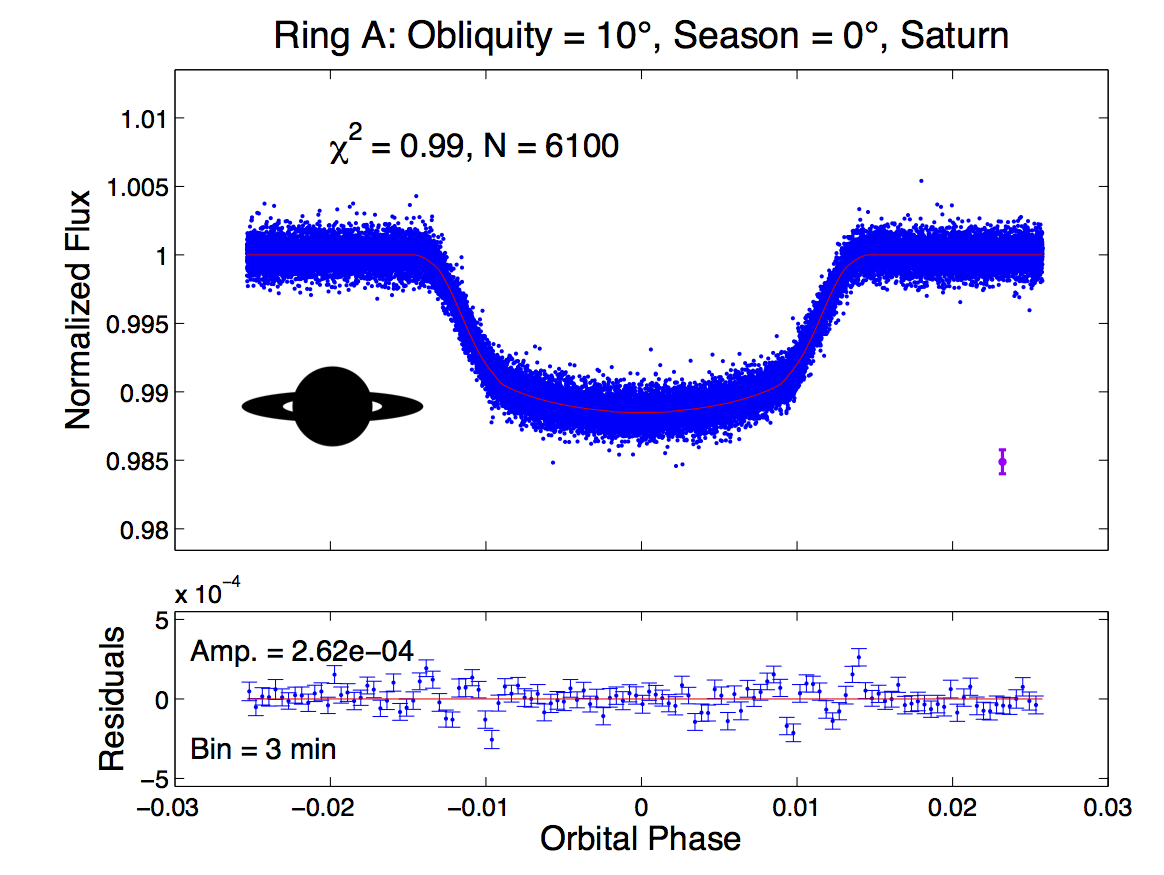}
}
\subfloat[]{
  \includegraphics[width=59mm]{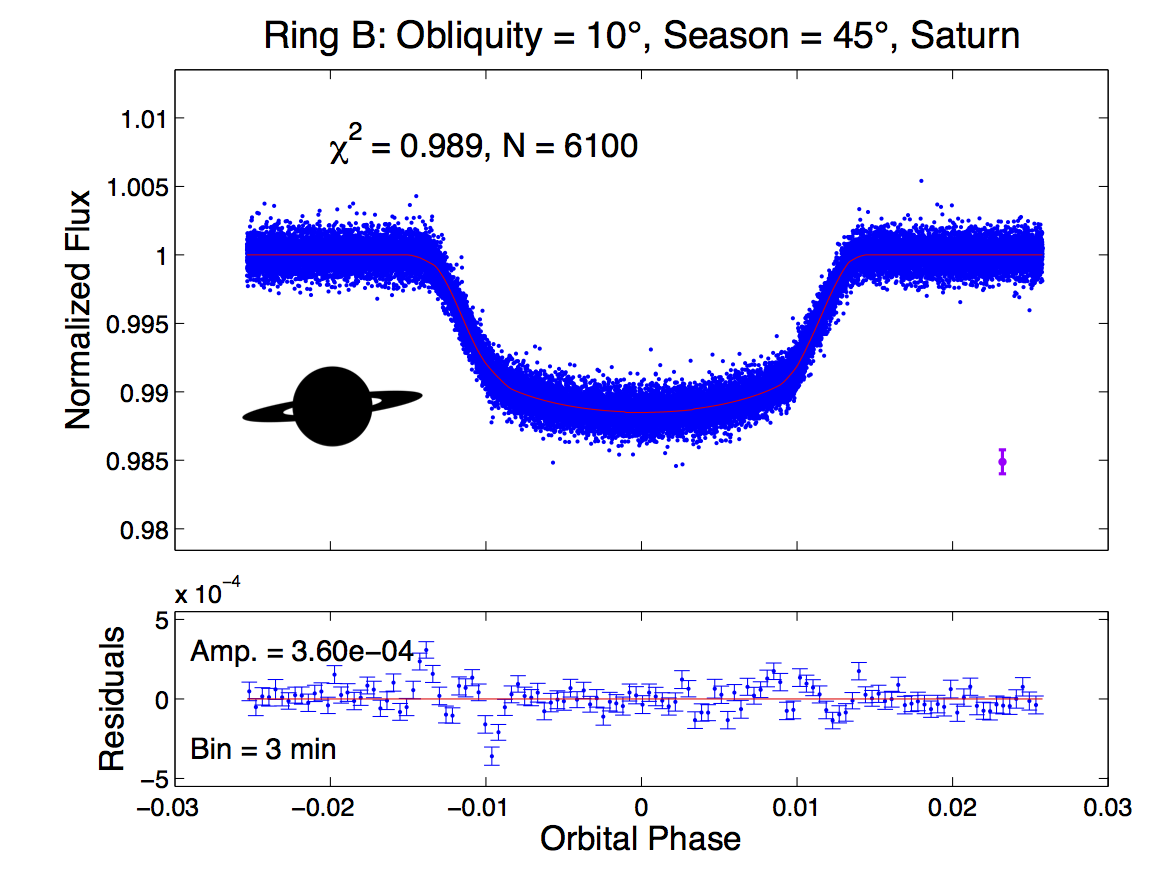}
}\\[-5mm]
\subfloat[]{
  \includegraphics[width=59mm]{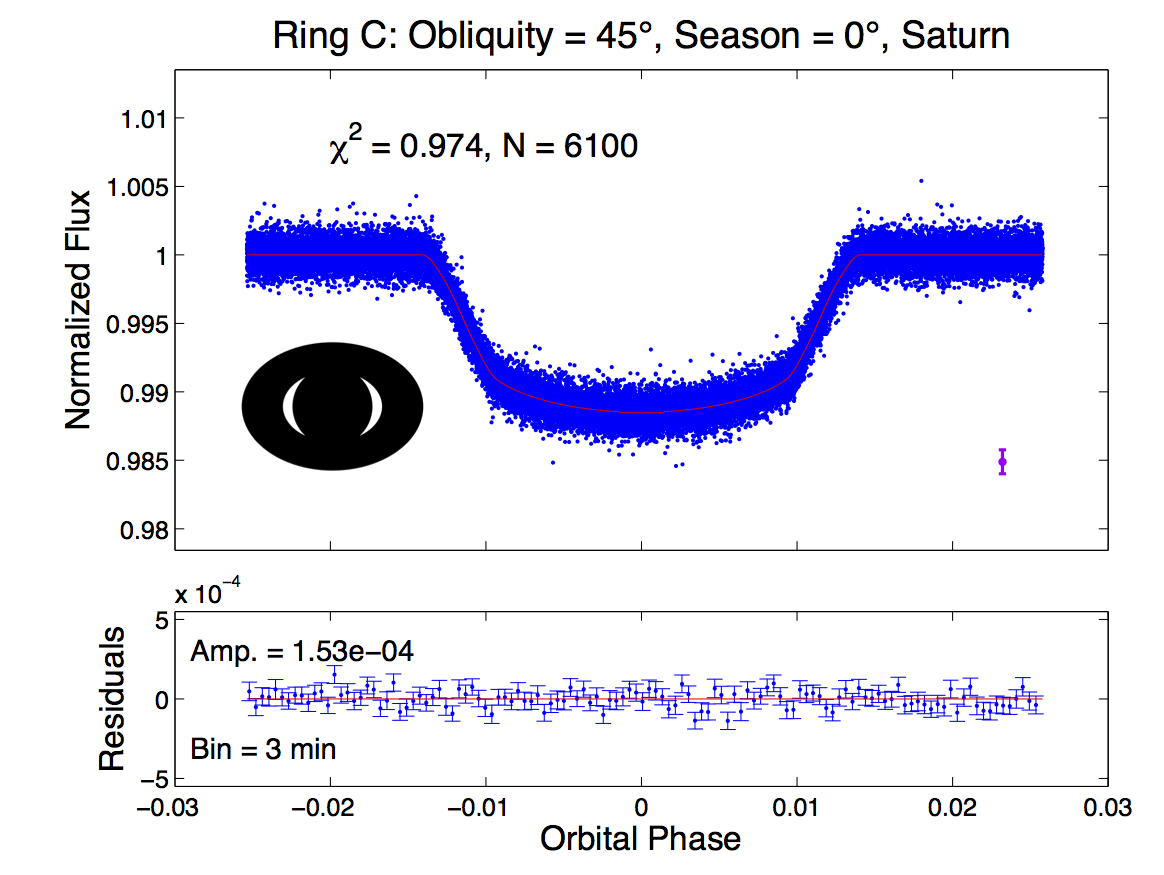}
}
\subfloat[]{
  \includegraphics[width=59mm]{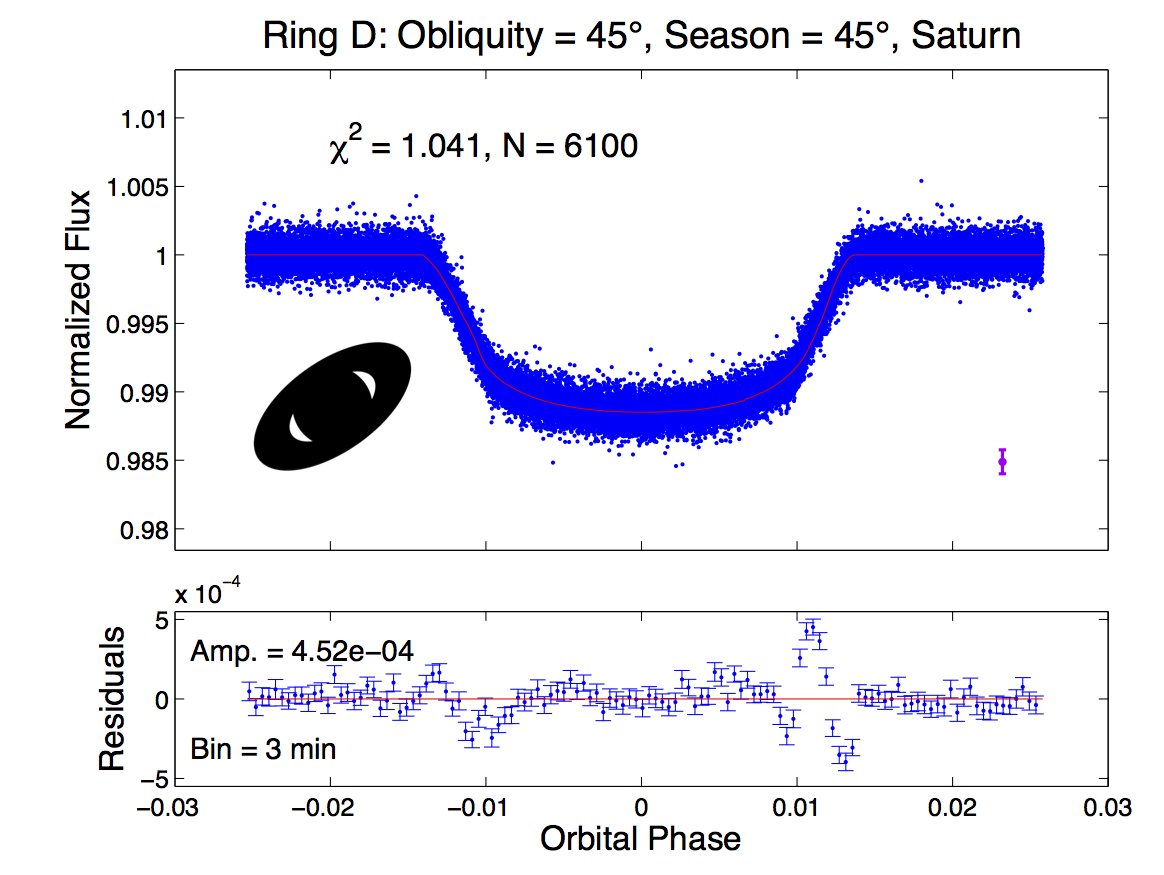}
}\\[-5mm]
\subfloat[]{   % ???
  \includegraphics[width=59mm]{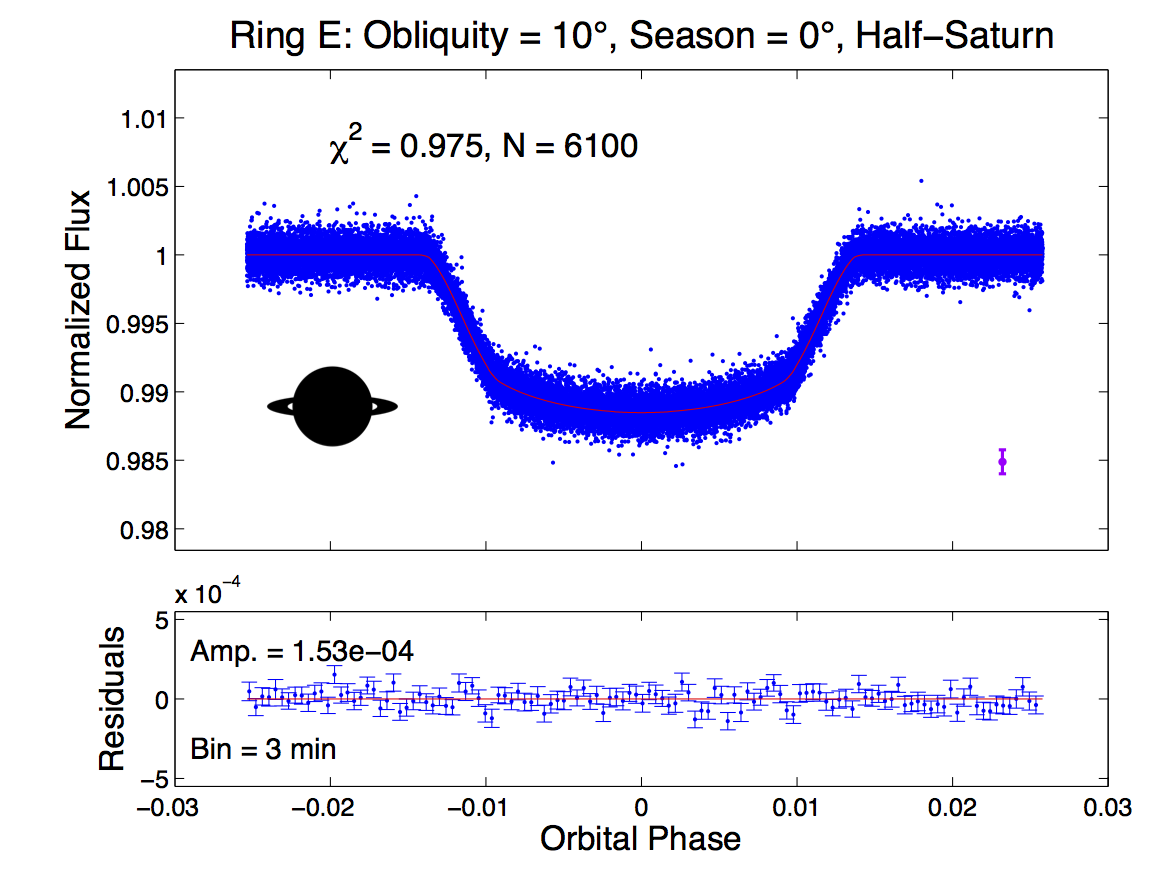}
}
\subfloat[]{
  \includegraphics[width=59mm]{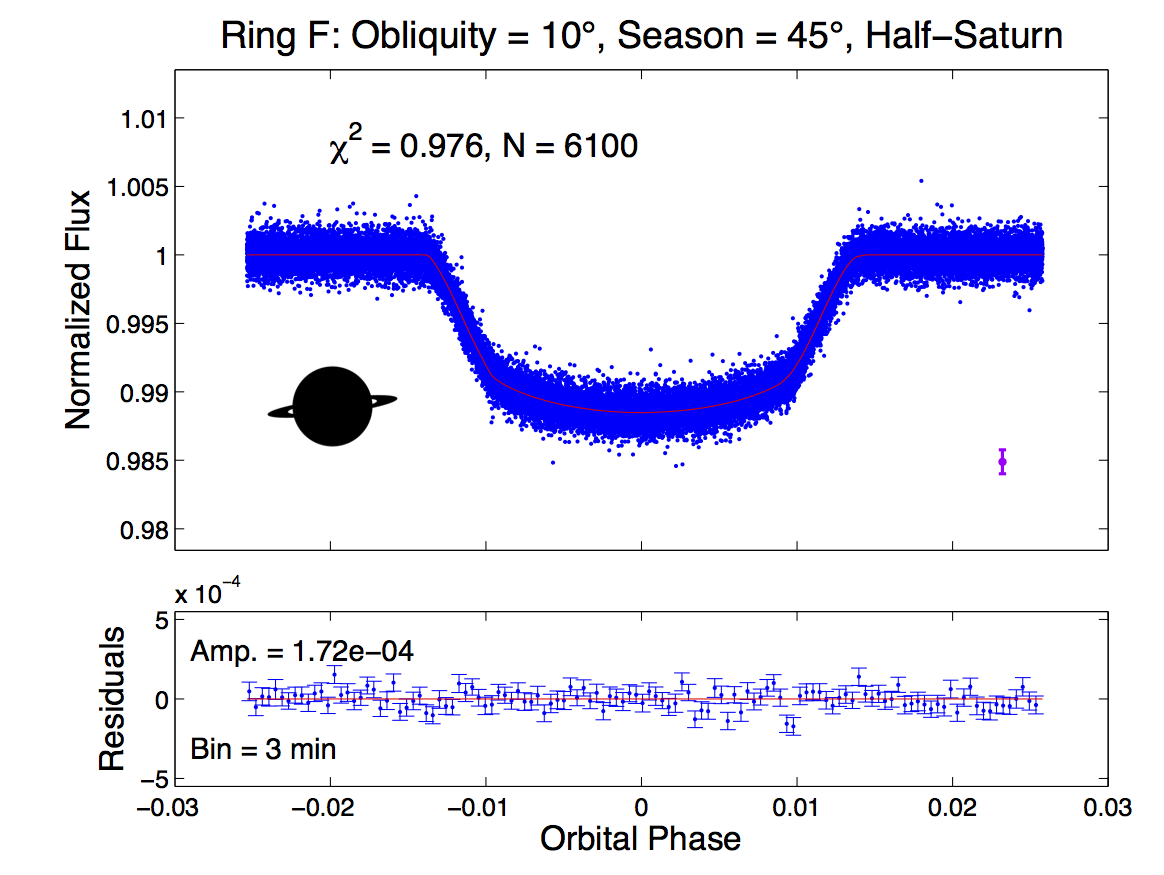}
}\\[-5mm]
\subfloat[]{   % ???
  \includegraphics[width=59mm]{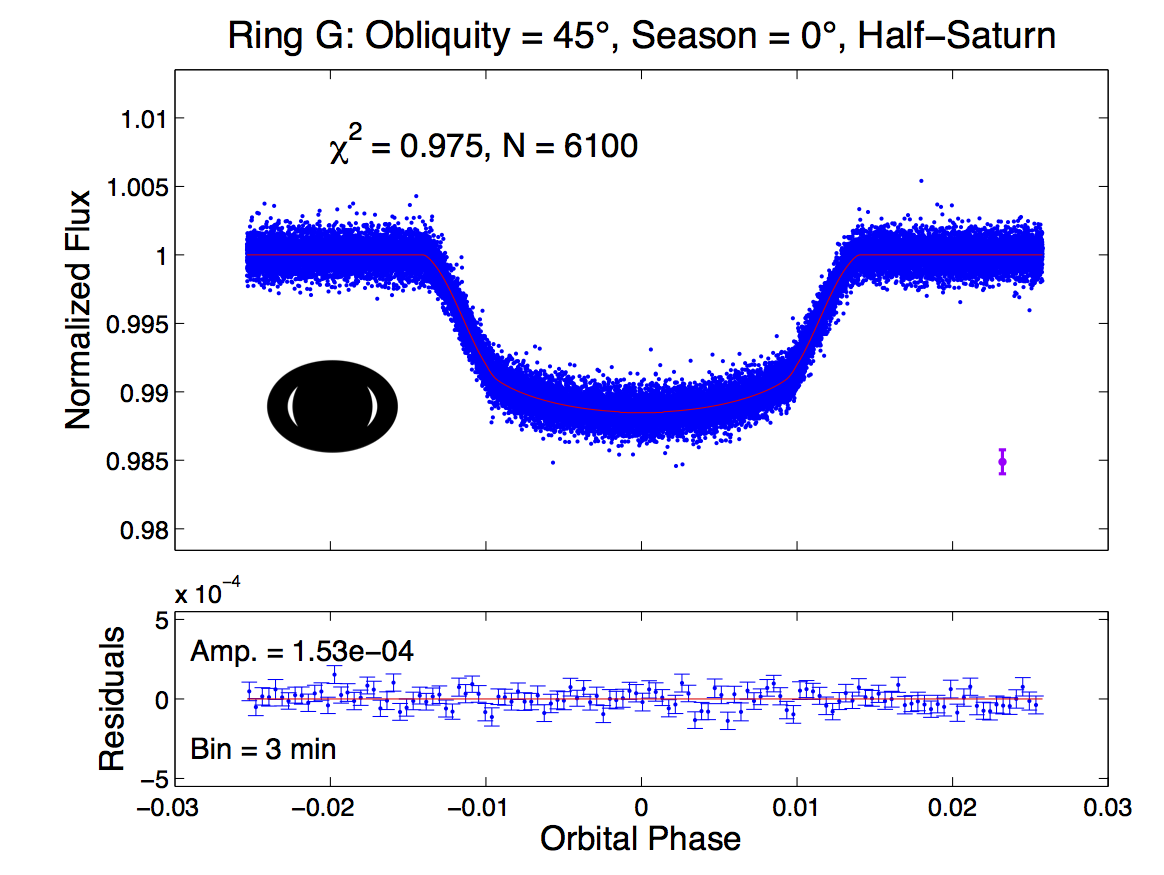}
}
\subfloat[]{
  \includegraphics[width=59mm]{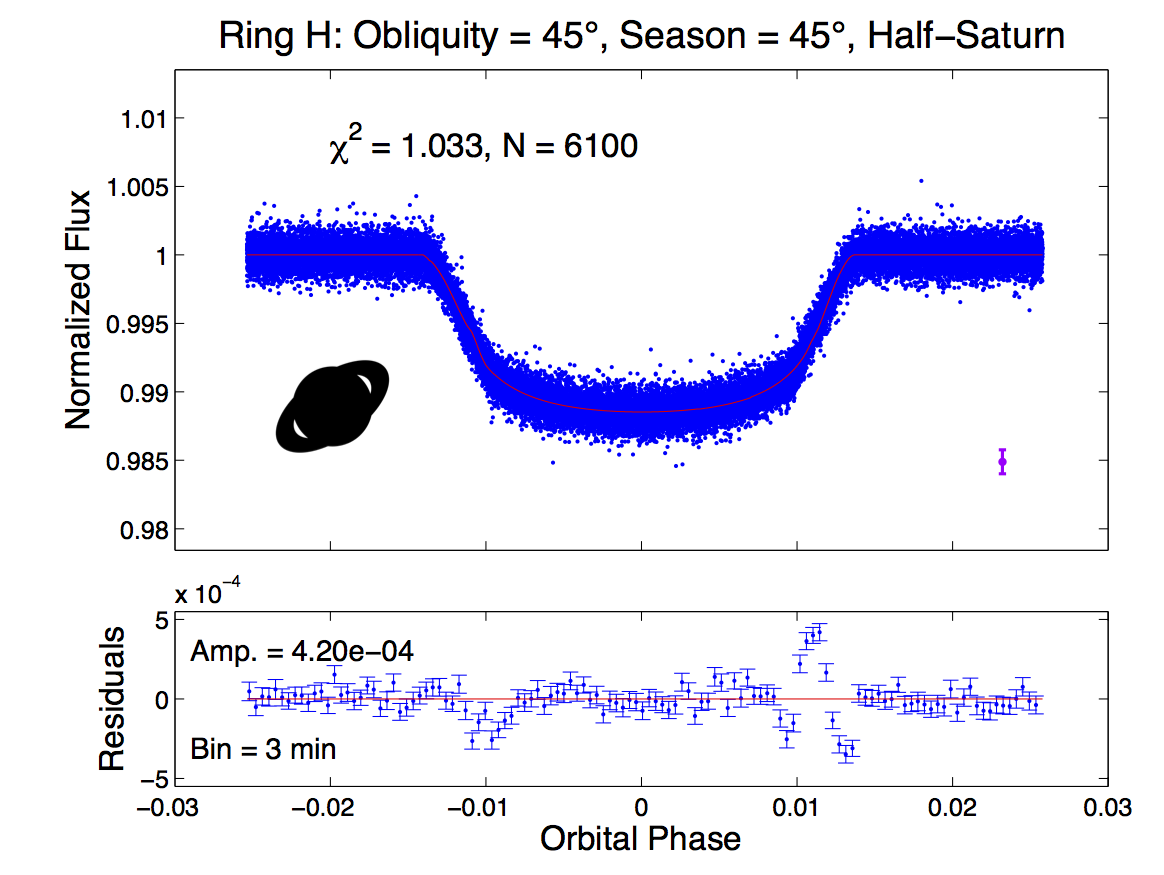}
}
\vspace{-4mm}
\caption{Fits of 8 fiducial ring models to \textit{Kepler}-15b.  The different ring models are referred to by their letter names given in each panel.  In the upper plot of each panel, the blue points represent the phase-folded lightcurve, and the red curve represents the fit.  Individual error bars are omitted for clarity, and the purple error bar in the lower right indicates the average error about an artificial data point.  The profile of the ringed planet is also included.  In the lower plot of each panel is a plot of the residuals (data minus fit) in blue, with bin sizes given by the text, and a red line at $y=0$.\label{15b}}
\end{figure}

\begin{table}[ht]
\begin{adjustwidth}{-1.25cm}{}
\captionsetup{justification=centering}
\centering
\caption{Physical parameters found for ringed fits to \textit{Kepler}-15b\label{params15}}
\begin{tabular}{p{1cm}|p{2.95cm}|p{1.6cm}|p{1.6cm}|p{2.8cm}|p{1.6cm}|p{2.8cm}}
\hline\hline
Ring&$R_p$ (stellar radii)&$\gamma_1$&$\gamma_2$&$a$ (stellar radii)&$i$ ($^{\circ}$)&$b$ (stellar radii)\\
\hline
A&0.0870&0.5125&0.0001&11.1175&87.0224&0.5775\\[0.cm]
B&0.0917&0.5203&0.0002&10.7642&93.2751&-0.6150\\[0.cm]
C&0.0560&0.5040&0.0615&10.2323&93.7010&-0.6605\\[0.cm]
D&0.0620&0.2449&0.7551&11.2572&92.5081&-0.4926\\[0.cm]
E&0.0988&0.5527&0.0062&10.2297&86.2967&0.6607\\[0.cm]
F&0.1008&0.5565&0.0002&10.1135&86.1939&0.6713\\[0.cm]
G&0.0772&0.5203&0.0462&10.1527&93.7677&-0.6672\\[0.cm]
H&0.0827&0.2036&0.7964&11.0322&92.7242&-0.5243\\
\end{tabular}
\end{adjustwidth}
\end{table}

As can be seen  in Figure \ref{15b}, there are several ring configurations (especially those with an obliquity and season of $45^{\circ}$ each) that can be ruled out on the basis of an obvious signal present in the residuals.  However, all 8 fits yield reasonable $\chi^2$ values (calculated over the set of phases in which the planet was in or near ingress or egress, as was the case for the ringless fits).  Thus, another statistical method needs to be used to determine if a potential ring could be eliminated.  The method used in this paper is the following: a potential ring would have been detected if the amplitude of the residual to the best fit with that ring exceeds 5 times the RMS of the residuals for the ringless fit.  As such, the RMS of the residuals for the ringless fit and the amplitude of the residuals for each of the ringed fits are shown in Table \ref{amp}, in which a red background of a given cell indicates a ring that can be ruled out ($Amp. > 5\times RMS$).

Furthermore, to directly test the goodness-of-fit of a ringed fit versus that of a ringless fit, a $\Delta \chi^2$ statistic for each ringed fit was calculated, defined as the $\chi^2$ for the ringed fit minus the $\chi^2$ for the ringless fit.  A negative $\Delta \chi^2$ therefore corresponds to a ringed fit that improved upon the ringless fit.  The $\Delta \chi^2$ statistic for each ringed fit is shown in Table \ref{dc2}, with the same red backgrounds corresponding to Table \ref{amp}, as defined by the residual amplitudes.  As can be seen, the larger $\Delta \chi^2$ values in a given row correlate very strongly with the red backgrounds, indicating that the two statistical measures yield largely equivalent results.  However, the $\Delta \chi^2$ method can distinguish between residuals due to actual ring signals and those due to other effects.

\begin{table}[ht]
\begin{adjustwidth}{-1.6cm}{}
\captionsetup{justification=centering}
\centering
\caption{Amplitudes of residuals to ringed fits and ring detectability for 21 \textit{Kepler} planets\label{amp}}
%\begin{tabular}{l|l|l|l|l|l|l|l|l|l}
\begin{tabular}{p{1.2cm}|p{1cm}|p{1.5cm}|p{1.5cm}|p{1.5cm}|p{1.5cm}|p{1.5cm}|p{1.5cm}|p{1.5cm}|p{1.45cm}}
%\begin{tabaular}{p{1cm}|p{1cm}|p{1cm}|p{1cm}|p{1cm}|p{1cm}|p{1cm}|p{1cm}|p{1cm}|p{1cm}}
\hline\hline
\textit{Kepler} Name&RMS$\newline\times10^5$&A-Amp.$\newline\times10^5$&B-Amp.$\newline\times10^5$&C-Amp.$\newline\times10^5$&D-Amp.$\newline\times10^5$&E-Amp.$\newline\times10^5$&F-Amp.$\newline\times10^5$&G-Amp.$\newline\times10^5$&H-Amp.$\newline\times10^5$\\
\hline
4b&1.54&4.12&4.12&4.12&4.12&4.12&4.12&4.12&4.12\\
5b&2.36&\cellcolor{red!25}48.4&\cellcolor{red!25}37.8&\cellcolor{red!25}26.8&\cellcolor{red!25}24.4&\cellcolor{red!25}15.0&11.1&\cellcolor{red!25}24.9&\cellcolor{red!25}15.9\\
6b&2.45&\cellcolor{red!25}59.6&\cellcolor{red!25}46.6&\cellcolor{red!25}38.8&\cellcolor{red!25}35.4&\cellcolor{red!25}17.5&9.80&\cellcolor{red!25}31.6&\cellcolor{red!25}22.1\\
7b&2.92&\cellcolor{red!25}23.6&\cellcolor{red!25}24.6&7.93&\cellcolor{red!25}16.7&10.7&11.7&8.02&\cellcolor{red!25}24.6\\
8b&2.75&\cellcolor{red!25}16.6&\cellcolor{red!25}23.8&7.97&\cellcolor{red!25}40.6&7.97&9.63&7.97&\cellcolor{red!25}42.4\\
12b&3.62&\cellcolor{red!25}102&\cellcolor{red!25}79.4&\cellcolor{red!25}68.5&\cellcolor{red!25}63.4&\cellcolor{red!25}34.6&\cellcolor{red!25}21.7&\cellcolor{red!25}56.5&\cellcolor{red!25}39.6\\
14b&1.61&6.55&6.50&6.21&\cellcolor{red!25}9.24&6.21&6.21&6.21&\cellcolor{red!25}9.90\\
15b&5.40&26.2&\cellcolor{red!25}36.0&15.3&\cellcolor{red!25}45.2&15.3&17.2&15.3&\cellcolor{red!25}42.0\\
17b&4.11&\cellcolor{red!25}130&\cellcolor{red!25}105&\cellcolor{red!25}85.4&\cellcolor{red!25}84.1&\cellcolor{red!25}37.5&\cellcolor{red!25}26.9&\cellcolor{red!25}76.3&\cellcolor{red!25}55.0\\
41b&5.98&25.2&22.7&16.7&\cellcolor{red!25}44.6&15.7&15.3&16.3&\cellcolor{red!25}37.2\\
43b&2.41&\cellcolor{red!25}16.7&\cellcolor{red!25}23.2&6.71&\cellcolor{red!25}30.0&6.72&7.11&6.71&\cellcolor{red!25}27.6\\
44b&9.96&31.9&31.7&26.6&46.7&28.4&28.4&28.5&42.0\\
75b&22.4&110&96.8&71.0&67.8&67.8&67.8&67.8&67.8\\
77b&5.66&\cellcolor{red!25}51.7&\cellcolor{red!25}45.7&25.4&20.6&21.0&15.4&19.4&15.1\\
94b&2.58&9.52&9.52&9.52&9.52&9.52&9.52&9.52&9.52\\
101b&4.96&14.0&14.0&13.9&13.9&13.9&13.9&13.9&13.9\\
108b&6.03&14.3&14.1&15.5&14.1&16.0&15.0&15.5&14.1\\
117c&9.30&37.1&33.2&24.8&25.8&25.6&25.5&23.7&25.6\\
122c&7.42&21.1&21.1&21.1&21.1&21.1&21.1&21.1&21.1\\
148c&10.5&30.0&30.0&30.0&30.0&30.0&30.0&30.0&30.0\\
412b&4.94&14.3&24.5&10.5&\cellcolor{red!25}55.5&12.1&19.4&10.5&\cellcolor{red!25}55.5\\
\end{tabular}
\end{adjustwidth}
\end{table}

As can be seen in Table \ref{amp}, there are many possible rings which would have been detected, and can thus be rejected.  However, there are only 2 planets surveyed in which each of the 8 different rings studied would have been detected, while there are 9 planets for which none of the rings would have been detected.  For some planets, it is the fiducial rings that are largest in area (A and B) that yield the largest signals, while for others, it is the rings with the most ``tilt" in the vertical direction (D and H, the two for which $\Omega=\phi=45^{\circ}$) that yield the largest signals; which effect dominates is determined by constraints on the planet's impact parameter, which partially determines the asymmetry of a ringed planet's transit lightcurve.  It is also noted that the signals from Saturn-sized rings are generally larger than those from Half-Saturn-sized rings with the same orientation, as is likely to be expected.  However, there are some instances (particularly \textit{Kepler}-7b's H-ring as compared to its D-ring) in which the Half-Saturn ring is larger.  We are not certain what precisely causes this phenomenon, but suspect it may have to do with the discontinuity in slope between the outer edge of ring H and the planet's surface in the ringed planet's profile, which is not present for ring D (see Figure \ref{rv}).  Furthermore, it is important to remember that there are many other possible rings not captured by the 8 fiducial rings in this paper: rings with different proportions between $R_1$, $R_2$, and $R_p$, rings with different orientations, rings that are not optically thick, rings that are in fact composed of multiple smaller rings, as is the case with Saturn, etc.

\begin{table}[ht]
\begin{adjustwidth}{-1cm}{}
\captionsetup{justification=centering}
\centering
\caption{$\Delta \chi^2$ values for ringed fits and ring detectability for 21 \textit{Kepler} planets\label{dc2}}
\begin{tabular}{p{1.2cm}|p{1.5cm}|p{1.5cm}|p{1.5cm}|p{1.5cm}|p{1.5cm}|p{1.5cm}|p{1.5cm}|p{1.45cm}}
\hline\hline
\textit{Kepler} Name&A-$\Delta \chi^2$&B-$\Delta \chi^2$&C-$\Delta \chi^2$&D-$\Delta \chi^2$&E-$\Delta \chi^2$&F-$\Delta \chi^2$&G-$\Delta \chi^2$&H-$\Delta \chi^2$\\
%\textit{Kepler} Name&A-$\Delta\Chi^2$&B-Amp.$\newline\times10^5$&C-Amp.$\newline\times10^5$&D-Amp.$\newline\times10^5$&E-Amp.$\newline\times10^5$&F-Amp.$\newline\times10^5$&G-Amp.$\newline\times10^5$&H-Amp.$\newline\times10^5$\\
\hline
4b&5.08&2.08&0.302&-0.239&0.819&0.334&0.487&0.861\\
5b&\cellcolor{red!25}2268&\cellcolor{red!25}1154&\cellcolor{red!25}746&\cellcolor{red!25}666&\cellcolor{red!25}207&70.9&\cellcolor{red!25}547&\cellcolor{red!25}287\\
6b&\cellcolor{red!25}3719&\cellcolor{red!25}1897&\cellcolor{red!25}1159&\cellcolor{red!25}1177&\cellcolor{red!25}269&83.6&\cellcolor{red!25}861&\cellcolor{red!25}478\\
7b&\cellcolor{red!25}538&\cellcolor{red!25}460&11.2&\cellcolor{red!25}285&44.4&25.4&12.8&\cellcolor{red!25}435\\
8b&\cellcolor{red!25}95.5&\cellcolor{red!25}166&-3.53&\cellcolor{red!25}897&-2.51&23.5&2.08&\cellcolor{red!25}827\\
12b&\cellcolor{red!25}5985&\cellcolor{red!25}3299&\cellcolor{red!25}1964&\cellcolor{red!25}2012&\cellcolor{red!25}436&\cellcolor{red!25}167&\cellcolor{red!25}1396&\cellcolor{red!25}730\\
14b&49.5&42.4&0.0933&\cellcolor{red!25}100&1.48&1.87&-1.47&\cellcolor{red!25}118\\
15b&96.3&\cellcolor{red!25}89.6&-2.53&\cellcolor{red!25}405&3.45&10.7&2.69&\cellcolor{red!25}360\\
17b&\cellcolor{red!25}6755&\cellcolor{red!25}3223&\cellcolor{red!25}3130&\cellcolor{red!25}3358&\cellcolor{red!25}561&\cellcolor{red!25}226&\cellcolor{red!25}2454&\cellcolor{red!25}1406\\
41b&52.0&56.3&10.9&\cellcolor{red!25}206&5.84&5.31&9.36&\cellcolor{red!25}174\\
43b&\cellcolor{red!25}95.0&\cellcolor{red!25}109&3.43&\cellcolor{red!25}203&4.36&6.53&2.98&\cellcolor{red!25}191\\
44b&13.0&10.0&-0.301&24.1&0.653&0.136&1.21&21.7\\
75b&65.1&35.6&29.7&28.0&9.60&4.64&24.1&13.0\\
77b&\cellcolor{red!25}402&\cellcolor{red!25}241&41.0&34.5&50.7&30.6&26.5&7.61\\
94b&5.75&3.18&0.384&3.72&0.796&0.496&0.346&4.24\\
101b&1.39&0.908&-0.321&0.663&0.271&0.930&-0.133&1.58\\
108b&-0.990&-1.72&-0.215&14.2&0.546&-0.934&-0.335&8.62\\
117c&83.3&48.9&12.4&10.8&8.53&4.92&11.5&2.19\\
122c&10.3&6.66&-0.138&-0.945&2.12&1.37&-0.123&-0.506\\
148c&-3.55&-3.88&-3.53&-2.73&-0.546&-0.467&-3.48&-0.713\\
412b&8.02&66.8&-1.08&\cellcolor{red!25}572&16.4&26.5&-6.09&\cellcolor{red!25}588\\
\end{tabular}
\end{adjustwidth}
\end{table}

\section{CONCLUSIONS}

The search for rings around extrasolar planets presented in this paper came up empty-handed, with no evidence for any rings around 21 well-characterized \textit{Kepler} planets.  However, certain classes of rings around some of the surveyed planets were rejected, providing some information on the occurrence rate of rings in other planetary systems.  Moreover, the methodology used for searching for rings is sound, as these rejected rings would have been detected with the quantity and quality of data currently available from \textit{Kepler}.  There are still many possible rings that could exist around some of these planets without being currently detectable.  Furthermore, it is important to remember that the planets studied in this paper are different from any of the planets in our Solar System that host rings, with much tighter orbits.  It may well be the case that rings are not common except about planets with periods comparable to those of the outer planets in our Solar System, in which case the search for extrasolar rings using transit photometry will be unsuccessful for at least as long as it takes for exoplanet transit observations to accumulate and improve until there is high-quality data for transiting planets with periods of several years.  However, we remain optimistic that future work in this manner will help us answer the question of what kinds of rings exist, and with what prevalance, around planets outside of our Solar System.

We thank Eugene Chiang and Dan Fabrycky for insightful discussions.  Furthermore, we thank the NASA Kepler Mission, the Mikulski Archive for Space Telescopes (MAST), and the NASA Exoplanet Archive for providing resources that made this research possible, as well as the MIT UROP office for providing support for undergraduates conducting research.

\clearpage


\begin{thebibliography}{30}
\expandafter\ifx\csname natexlab\endcsname\relax\def\natexlab#1{#1}\fi


\bibitem[Barker 
\& O'Connell(1975)]{boc75} Barker, B.~M., \& O'Connell, R.~F.\ 1975, \prd, 12, 329

\bibitem[Barnes 
\& Fortney(2003)]{bf03} Barnes, J.~W., \& Fortney, J.~J.\ 2003, \apj, 588, 545

\bibitem[Barnes 
\& Fortney(2004)]{bf} Barnes, J.~W., \& Fortney, J.~J.\ 2004, \apj, 616, 1193

\bibitem[Bodenheimer et al.(2001)]{bodenheimer01} Bodenheimer, P., 
Lin, D.~N.~C., \& Mardling, R.~A.\ 2001, \apj, 548, 466

\bibitem[Borucki et al.(2010)]{borucki} Borucki, W.~J., Koch, 
D., Basri, G., et al.\ 2010, Science, 327, 977 

\bibitem[Braga-Ribas et al.(2014)]{chariklo} Braga-Ribas, F., 
Sicardy, B., Ortiz, J.~L., et al.\ 2014, \nat, 508, 72

\bibitem[Carter 
\& Winn(2010)]{cw} Carter, J.~A., \& Winn, J.~N.\ 2010, \apj, 709, 1219

\bibitem[Colombo(1966)]{c66} Colombo, G.\ 1966, \aj, 71, 
891

\bibitem[{{Cranmer} {et~al.}(2014){Cranmer}, {Bastien}, {Stassun}, \& {Saar}}]{flicker} Cranmer, S.~R., 
Bastien, F.~A., Stassun, K.~G., \& Saar, S.~H.\ 2014, \apj, 781, 124

\bibitem[de Pater \& Lissauer(2001)]{dpl01}
de Pater, I., \& Lissauer, J.~J.\ 2001, Planetary Sciences, Cambridge University Press, Cambridge, UK.

\bibitem[Fabrycky et al.(2007)]{f07} Fabrycky, D.~C., 
Johnson, E.~T., \& Goodman, J.\ 2007, \apj, 665, 754

\bibitem[Gilliland et al.(2010)]{gilliland} Gilliland, R.~L., 
Jenkins, J.~M., Borucki, W.~J., et al.\ 2010, \apjl, 713, L160

\bibitem[Goldreich 
\& Soter(1966)]{gs} Goldreich, P., \& Soter, S.\ 1966, \icarus, 5, 375

\bibitem[Gray(1959)]{g59}
Gray, A.\ 1959, A Treatise on Gyrostatics and Rotational Motion, Dover Publications, New York, NY.

\bibitem[Guillot et al.(1996)]{guillot} Guillot, T., Burrows, 
A., Hubbard, W.~B., Lunine, J.~I., \& Saumon, D.\ 1996, \apjl, 459, L35

\bibitem[Hamilton 
\& Burns(1991)]{hamilton} Hamilton, D.~P., \& Burns, J.~A.\ 1991, \icarus, 92, 118

\bibitem[Heller(2014)]{h14} Heller, R.\ 2014, \apj, 787, 14

\bibitem[Hubbard 
\& Marley(1989)]{hm89} Hubbard, W.~B., \& Marley, M.~S.\ 1989, \icarus, 78, 102

\bibitem[Jenkins et al.(2010)]{jc} Jenkins, J.~M., 
Caldwell, D.~A., Chandrasekaran, H., et al.\ 2010, \apjl, 713, L87

\bibitem[Kalas et al.(2008)]{k08} Kalas, P., Graham, J.~R., 
Chiang, E., et al.\ 2008, Science, 322, 1345

\bibitem[Kenworthy 
\& Mamajek(2015)]{km} Kenworthy, M.~A., \& Mamajek, E.~E.\ 2015, arXiv:1501.05652

\bibitem[Kipping(2013)]{kip} Kipping, D.~M.\ 2013, \mnras, 
435, 2152

\bibitem[Kipping et al.(2014)]{k14} Kipping, D.~M., Torres, 
G., Buchhave, L.~A., et al.\ 2014, \apj, 795, 25

\bibitem[Laplace(1805)]{l1805} Laplace, P.~S.\ 1805, M\'{e}canique C\'{e}leste, Vol. 4 (Paris: Courcier)

%\bibitem[Levrard et 
%al.(2007)]{l07} Levrard, B., Correia, A.~C.~M., Chabrier, G., et al.\ 2007, \aap, 462, L5

\bibitem[Magic et 
al.(2015)]{stagger} Magic, Z., Chiavassa, A., Collet, R., \& Asplund, M.\ 2015, \aap, 573, AA90

\bibitem[{{Mamajek} {et~al.}(2012)}]{mamajek} Mamajek, E.~E., 
Quillen, A.~C., Pecaut, M.~J., et al.\ 2012, \aj, 143, 72 

\bibitem[Mandel 
\& Agol(2002)]{ma} Mandel, K., \& Agol, E.\ 2002, \apjl, 580, L171

\bibitem[Mandushev et al.(2007)]{hotjup} Mandushev, G., 
O'Donovan, F.~T., Charbonneau, D., et al.\ 2007, \apjl, 667, L195

\bibitem[Murray \& Dermott(1999)]{md99}
Murray, C.~D., \& Dermott, S.~F.\ 1999, Solar System Dynamics, Cambridge University Press, New York, NY.

\bibitem[Schlichting 
\& Chang(2011)]{warm} Schlichting, H.~E., \& Chang, P.\ 2011, \apj, 734, 117

\bibitem[Tremaine et al.(2009)]{ttn09} Tremaine, S., Touma, 
J., \& Namouni, F.\ 2009, \aj, 137, 3706

\bibitem[Vieira Neto 
\& Winter(2001)]{vnw} Vieira Neto, E., \& Winter, O.~C.\ 2001, \aj, 122, 440

%\bibitem[Ward(1975)]{w75} Ward, W.~R.\ 1975, Science, 189, 
%377

\bibitem[Winn 
\& Holman(2005)]{wh05} Winn, J.~N., \& Holman, M.~J.\ 2005, \apjl, 628, L159

\bibitem[Zuluaga et al.(2015)]{z15} Zuluaga, J.~I., 
Kipping, D.~M., Sucerquia, M., \& Alvarado, J.~A.\ 2015, \apjl, 803, L14 





\end{thebibliography}
\end{document}